\documentclass[aps,prd,twocolumn,preprintnumbers,showpacs,superscriptaddress,nofootinbib,floatfix,10pt]{revtex4-2}
\pdfoutput=1
\usepackage{textcomp}
\usepackage[english]{babel}
\usepackage{amsmath,amssymb,amsbsy,booktabs}
\usepackage{bm}
\usepackage{color}
\usepackage{array}
\usepackage{amstext}
\usepackage{graphicx}
\usepackage{amsfonts}
\usepackage{dcolumn}
\usepackage{rotating}
\usepackage{epstopdf}
\usepackage{esint}
\usepackage{url}
\usepackage[colorlinks=true,
linkcolor=blue,
filecolor=blue,
anchorcolor=blue,
urlcolor=blue,
citecolor=blue
]{hyperref}

\usepackage{array}
\usepackage{booktabs}
\usepackage{multirow}

\usepackage[utf8]{inputenc}

\usepackage{xcolor,colortbl}

\usepackage[T1]{fontenc} 

\usepackage{cancel}
\usepackage{slashed}
\usepackage{hhline}
\usepackage{subfigure} 
\usepackage{enumerate}
\usepackage{times}
\usepackage[version=4]{mhchem}

\DeclareMathOperator{\tr}{Tr}

\allowdisplaybreaks

\begin{document}

\title{\Large \bf \boldmath Probing new physics with polarization components of the tau lepton in quasielastic $e^- p \to \Lambda_c \tau^-$ scattering process}

\author{Xin-Shuai Yan}
\email{yanxinshuai@htu.edu.cn}
\affiliation{Institute of Particle and Nuclear Physics, Henan Normal University, Xinxiang, Henan 453007, China}

\author{Liang-Hui Zhang}
\email{zhanglianghui@stu.htu.edu.cn}
\affiliation{Institute of Particle and Nuclear Physics, Henan Normal University, Xinxiang, Henan 453007, China}

\author{Qin Chang}
\email{changqin@htu.edu.cn (corresponding author)}
\affiliation{Institute of Particle and Nuclear Physics, Henan Normal University, Xinxiang, Henan 453007, China}

\author{Ya-Dong Yang}
\email{yangyd@mail.ccnu.edu.cn}
\affiliation{Institute of Particle and Nuclear Physics, Henan Normal University, Xinxiang, Henan 453007, China}
\affiliation{Institute of Particle Physics and Key Laboratory of Quark and Lepton Physics~(MOE),\\
	Central China Normal University, Wuhan, Hubei 430079, China}

\begin{abstract}
		
Kinematics restricts the ability of rare charm decays to explore the charged lepton-flavor-violation processes mediated 
by the quark-level $c\to u \ell \tau$ transition. To fill the gap, we propose exploring new physics (NP) through the quasielastic 
scattering process $e^-p\to \tau^-\Lambda_c$ and the polarization of the $\tau$ lepton.  
As analyzing modes for the $\tau$ polarization, we consider the decays $\tau^-\to \pi^-\nu_{\tau}$, $\tau^-\to \rho^-\nu_{\tau}$, and $\tau^- \to \ell^-\bar{\nu}_{\ell}\nu_{\tau}$, and show that the $\tau$ polarization components can be extracted from 
analyzing the kinematics of the $\tau$ visible decay products.	
In the framework of a general low-energy effective Lagrangian, we then perform a detailed analysis of the polarization 
components in various aspects and scrutinize possible NP signals. 
With one upcoming experimental setup, we finally demonstrate promising event rate can be expected for the cascade process 
and, even in the worst-case scenario---no signals are observed at all---it can still provide a competitive potential for constraining the NP, 
compared with those from the high-$p_T$ dilepton invariant mass tails at high-energy colliders.	
\end{abstract}

\pacs{}

\maketitle

\section{Introduction} 
\label{sec:intro}

The observation of neutrino oscillations has established that lepton flavor symmetry, 
an accidental symmetry in the Standard Model (SM), is explicitly broken. 
If neutrinos get their masses through Yukawa interactions with the SM Higgs, 
the expected rates for charged lepton flavor violation (cLFV) enabled by the neutrino oscillations are  
suppressed by $G^2_Fm^4_{\nu}\sim 10^{-50}$, 
making them practically unobservable in current experiments~\cite{Calibbi:2017uvl,Ardu:2022sbt}. 
Thus observation of cLFV would clearly point to the existence of new physics (NP) beyond the SM. 

Among the various cLFV processes 
(see Refs.~\cite{Calibbi:2017uvl,Ardu:2022sbt,Altmannshofer:2022hfs,Guadagnoli:2022oxk,Davidson:2022jai} for recent reviews), 
the ones mediated by the $c\to u \ell \ell^\prime$ transitions at the quark level 
have received appreciable attention in recent years. 
These processes have been extensively studied in rare charm decays~\cite{E653:1995rpz,Lees:2011hb,deBoer:2015boa,Aaij:2015qmj,LHCb:2017yqf, DeBoer:2018pdx,Bause:2019vpr,BaBar:2020faa,Bause:2020xzj,LHCb:2020car,Gisbert:2020vjx, Golz:2021imq} 
and also been explored by analyzing the high-$p_T$ dilepton invariant mass tails in the processes $pp\to \ell \ell'$~\cite{ATLAS:2018mrn,Angelescu:2020uug,Descotes-Genon:2023pen}. 
Recently, we have proposed studying them through a low-energy scattering experiment, which turns out very
complementary in searching for NP to the rare charm decays and the high-$p_T$ dilepton invariant mass tails~\cite{Lai:2021sww}. 

Despite the growing efforts above, few can probe the cLFV processes induced by 
the $c\to u \ell \tau$ transitions (with $\ell=e,\mu$).  
Start with the rare charm decays. 
The largest accessible phase space for semileptonic $D$-meson decays is 
given by $m_{D^+}-m_{\pi^0}\simeq 1.735$~GeV, which is smaller than the $\tau$-lepton mass, 
rendering the semitauonic $D$-meson decays kinematically forbidden; the same conclusion also holds for the charmed-baryon decays.
Meanwhile, due to $m_{D^0}<m_{\tau}+m_{\mu}$, the purely tauonic $D$-meson decay $D^0\to \tau \mu$ is 
kinematically forbidden, too.\footnote{Note that $\tau\to D^0\mu$ is also 
kinematically forbidden due to $m_{\tau}<m_{D^0}+m_{\mu}$.} 
Although the decay $D^0\to \tau e$ is kinematically allowed, it has not been detected in experiments 
partially because of its very narrow phase space. 
Given that the new proposed low-energy scattering process involves only the leptons $e$ and $\mu$~\cite{Lai:2021sww}, 
analyzing the high-$p_T$ dilepton invariant mass tails becomes the only way to explore the cLFV processes induced by 
the $c\to u \ell \tau$ transitions for the moment. 
In the framework of  a general low-energy effective Lagrangian 
(denoted by $\mathcal{L}_{\text{eff}}$ as introduced in Eq.~\eqref{eq:Leff}), 
since the high-$p_T$ dilepton invariant mass tails cannot  pinpoint all the possible NP Dirac structures~\cite{Angelescu:2020uug}, 
clearly other new processes and observables, particularly the 
low-energy ones, are badly needed; this is still true even if the decay $D^0\to \tau e$ is measured in the future, 
since the purely leptonic $D$-meson decays are known to be only sensitive to the axial and pseudo-scalar four-fermion operators of the $\mathcal{L}_{\text{eff}}$~\cite{Fleischer:2019wlx,Becirevic:2020rzi,Leng:2020fei,Colangelo:2021dnv}.

In this paper, we will propose the quasielastic (QE) scattering process $e^-p\to \tau^-\Lambda_c$ mediated by the 
quark-level $e^- u \to \tau^- c$ transition. This process is free from the kinematic problem that the semitauonic charmed-hadron decays encounter 
and covers all the effective operators of the $\mathcal{L}_{\text{eff}}$. 
Unfortunately, even with the purely tauonic $D$-meson decays and the high-$p_T$ dilepton invariant mass analyses, there are 
still not enough observables to fully pinpoint all the NP Dirac structures and determine the corresponding  
Wilson coefficients (WCs). We will thus also consider the polarization of the produced $\tau$ lepton, which 
involve all the effective operators of the $\mathcal{L}_{\text{eff}}$, 
and can fill the gap (at least partially), though they are more difficult to measure than the cross sections.

Since the $\tau$ lepton is very short-lived and decays weakly, its polarization is revealed through its ensuing decay distributions. 
As analyzing modes for the $\tau$ polarization, we shall make use of its four dominant decays,
$\tau^-\to \pi^-\nu_{\tau}$, $\tau^-\to \rho^-\nu_{\tau}$, and $\tau^- \to \ell^-\bar{\nu}_{\ell}\nu_{\tau}$ (with $\ell=e,\mu$), 
which together account for more than 70\% of the total $\tau$ decay width~\cite{Workman:2022ynf}.
Since the decay products of these modes contain at least one undetected neutrino, 
we shall focus on the visible final-state kinematics and integrate out all the variables that cannot be directly measured. 
We will show that the three polarization components of the $\tau$ can be extracted from the analyses of the kinematics of 
the visible decay product (i.e., $\pi^-$, $\rho^-$, and $\ell^-$). 

The QE scattering process we propose can be explored through a fixed target experiment. 
Thanks to the advances in technologies of electron beams and proton 
targets---note the ongoing scientific program of 12 GeV Continuous Electron Beam 
Accelerator Facility (CEBAF) at Jefferson Lab (JLab) and its potential 
upgrades~\cite{Arrington:2021alx}---promising event rates can be expected for the scattering process in various NP scenarios,  
if it is measured with a properly selected experimental setup, together with the constraints from 
the high-$p_T$ dilepton invariant mass tails as input.
On the other hand, even in the worst-case scenario—no signals are observed at all—the QE scattering process can still provide 
competitive constraints with respect to those obtained from the high-$p_T$ dilepton invariant mass tails.

We conclude this section by outlining the content of our paper. 
We start, in Sec.~\ref{sec:models}, with a brief introduction of our theoretical framework, including the 
most general low-energy effective Lagrangian, cross section, spin density matrices, as well as the kinematics and form factors.
In such a framework, we show in Sec.~\ref{sec:extract} how to extract the polarization 
components of the $\tau$ in the QE scattering process. 
In Sec.~\ref{sec:Numresults}, we perform comprehensive phenomenology analyses of the polarization 
components in various NP scenarios and explore possible impacts from the form factors. 
Based on the currently available experimental constraints, we evaluate in Sec.~\ref{sec:Prospect} the prospect 
for discovering NP through the low-energy QE scattering experiment in various aspects.
Finally, we collect our main conclusions in Sec.~\ref{sec:con}, and relegate further details on the spin density matrices 
and explicit expressions of the various observables to the appendixes.

\section{Theoretical framework}
\label{sec:models}

\subsection{Low-energy effective Lagrangian}

The idea of using effective Lagrangian for studying lepton-flavor-violation processes, such as the lepton-flavor
changing decays, lepton-flavor conversion, and neutrinoless double beta decay, 
at the quark and hadronic level has been well developed in 
Refs.~\cite{Vergados:1985pq,Bernabeu:1993ta,Hirsch:1995zi,Faessler:1997db,Hisano:1995cp,Faessler:1996ph,
	Faessler:1998qv,Faessler:1999jf,Kosmas:2001mv,Kosmas:2001sx,Kuno:1999jp,Faessler:2004jt,Faessler:2004ea,Faessler:2005hx,
	Gutsche:2009vp,Gutsche:2011bi,Gonzalez:2013rea,Black:2002wh} 
and widely used in a series of papers (see, e.g., Refs.~\cite{Kitano:2002mt,Cirigliano:2009bz,Abada:2014kba,Crivellin:2017rmk,Davidson:2018zuo,
	Gninenko:2018num,Davidson:2018kud,Dib:2018rpy}). 
For the scattering process $e^- p\to \tau^- \Lambda_c$ (or $e^- u \to \tau^- c$ at the quark level) in this work, 	
the general low-energy effective Lagrangian can be written as (see e.g., Refs.~\cite{Gninenko:2018num,Dib:2018rpy,Fuentes-Martin:2020lea,Angelescu:2020uug})
\begin{align}\label{eq:Leff}
	\mathcal{L}_{\text{eff}}=\sum_{\alpha}\frac{g_{\alpha}}{v^2}\mathcal{O}_{\alpha}\,,
\end{align}
where $v=(\sqrt{2}G_F)^{-1/2}$ is the electroweak vacuum expectation value, $\mathcal{O}_{\alpha}$ is the semi-leptonic operator listed 
in Table~\ref{tab:Operator}, and $g_{\alpha}$ is the corresponding effective WC.\footnote{
If the effective hadronic-level Lagrangian is also evoked, its WCs can be connected to those at the quark level through 
the on-mass-shell matching condition~\cite{Faessler:1996ph,Faessler:1998qv,Faessler:2004jt,Faessler:2004ea,Faessler:2005hx,
	Gutsche:2009vp,Gutsche:2011bi,Gonzalez:2013rea,Dib:2018rpy}.} 
Note that the tensor operators with mixed quark and lepton chiralities vanish due to Lorentz invariance. In addition, the 
operators $\mathcal{O}^{LR}_{S}$ and $\mathcal{O}^{RL}_{S}$ are neither generated in the Standard Model effective field 
theory due to gauge invariance~\cite{Fuentes-Martin:2020lea}, nor in certain ultra-violet models, such as the leptoquark models~\cite{Dorsner:2016wpm,Mandal:2019gff}.   
Finally, this framework is only applicable up to an energy scale of $\mathcal{O}(m_b)$, with $m_b$ being the bottom-quark mass, above which new degrees of freedom would appear.

\begin{table}[t]
	\renewcommand*{\arraystretch}{1.6}
	\tabcolsep=0.1cm
	\centering
	\caption{Operator $\mathcal{O}_{\alpha}$ of $\mathcal{L}_{\text{eff}}$ in Eq.~\eqref{eq:Leff}, where $P_{R,L}\!=\!(1\pm \gamma_5)/2$ denote the right- and left-handed projectors, and $\sigma^{\mu \nu}\!= \! i[\gamma^{\mu},\gamma^{\nu}]/2$ is the antisymmetric tensor. And coeff is short for coefficient.}
	\begin{tabular}{cccc}
		\hline \hline
		Coeff. &  Operator & Coeff. &  Operator  \\
		\hline		
		$g^{LL}_V$&$(\bar{\tau}\gamma_{\mu}P_{L}e)(\bar{c}\gamma^{\mu}P_{L}u)$ & $g^{RR}_V$&$(\bar{\tau}\gamma_{\mu}P_{R}e)(\bar{c}\gamma^{\mu}P_{R}u)$\\   
		$g^{LR}_V$&$(\bar{\tau}\gamma_{\mu}P_{L}e)(\bar{c}\gamma^{\mu}P_{R}u)$&
		$g^{RL}_V$&$(\bar{\tau}\gamma_{\mu}P_{R}e)(\bar{c}\gamma^{\mu}P_{L}u)$\\ 
		$g^{LL}_S$&$(\bar{\tau}P_{L}e)(\bar{c}P_{L}u)$&
		$g^{RR}_S$&$(\bar{\tau}P_{R}e)(\bar{c}P_{R}u)$\\   
		$g^{LR}_S$&$(\bar{\tau}P_{L}e)(\bar{c}P_{R}u)$&
		$g^{RL}_S$&$(\bar{\tau}P_{R}e)(\bar{c}P_{L}u)$\\ 
		$g^{LL}_T$&$(\bar{\tau}\sigma_{\mu\nu}P_{L}e)(\bar{c}\sigma^{\mu\nu} P_{L}u)$ &
		$g^{RR}_T$&$(\bar{\tau}\sigma_{\mu\nu}P_{R}e)(\bar{c}\sigma^{\mu\nu} P_{R}u)$\\                                                                                                  
		\hline \hline
	\end{tabular}	
		\label{tab:Operator}
\end{table}

\subsection{Spin density matrices and polarization vector}
\label{subsec:spin density}

The cascade processes considered in this work can be broken down into the QE scattering ($e^- p\to \Lambda_c \tau^-$) 
and successive decays of the $\tau$ lepton ($\tau^- \to \pi^-\nu_{\tau},\rho^-\nu_{\tau},\ell^-\bar{\nu}_{\ell}\nu_{\tau}$). 
The fully differential cross section can be written as (see e.g., Refs.~\cite{Godbole:2006tq,Haber:1994pe}),
\begin{align}\label{eq:diff cross}
	d\sigma_d&\!=\!\frac{1}{4F}\frac{1}{(k^{\prime2}\!-\!m^2_{\tau})^2\!+\!m^2_{\tau}\Gamma^2_{\tau}}\rho^P_{\lambda\lambda^\prime}
	\rho^D_{\lambda^{\prime}\lambda} d\Phi(k,p;p^\prime,p_d,p_\nu) \nonumber \\[0.02cm]
	&\!=\!\Big[\frac{1}{4F} \rho^P_{\lambda\lambda^\prime}d\Phi(k,p;p^\prime,k^\prime)\Big]
   \Big[\frac{1}{2m_{\tau}\Gamma_{\tau}}\rho^D_{\lambda^{\prime}\lambda}d\Phi(k^\prime;p_d,p_\nu)\Big]\,,
\end{align}
where $d\Phi(k_i,...;p_j, ...)\!\equiv\! \prod_{j} \frac{d^3 \pmb{p}_j}{2E_j(2\pi)^3}(2\pi)^4\delta^4(\sum_i k_i\!-\!\sum_jp_j)$ is 
Lorentz invariant.
To account for the spin average of the initial electron and proton, we have introduced a prefactor $1/4$. 
And to obtain the second equation in Eq.~\eqref{eq:diff cross}, we have applied the narrow-width approximation, 
\begin{align}
	\lim_{\Gamma \to 0^+}\frac{1}{\pi}\frac{m\Gamma}{(p^{2}-m^2)^2+m^2\Gamma^2}=\delta(p^{2}-m^2)\,,
\end{align}
to the $\tau$ propagator $1/[(k^{\prime 2}-m_{\tau}^2)^2+m_{\tau}^2\Gamma_{\tau}^2]$, which is valid due to the much 
smaller decay width of $\tau$ than its mass~\cite{Workman:2022ynf}. Finally, we always bear in mind that for the $\tau^- \!\to\!\ell^-\bar{\nu}_{\ell}\nu_{\tau}$ mode, 
$p_{\bar{\nu}_\ell}$, momentum of the outgoing $\ell$-antineutrino, must be taken into account  
in both the $d\Phi(k,p;p^\prime,p_d,p_\nu)$ and $d\Phi(k^\prime;p_d,p_\nu)$. 

Now several necessary explanations of the various symbols in Eq.~\eqref{eq:diff cross} are in order. 
First, the indices $\lambda,\lambda^\prime=\{1/2,-1/2\}$ characterize the helicity of the $\tau$. Second, 
the flux factor $F$ is $4((p\cdot k)^2-m_e^2m^2_p)^{1/2}$ with $k$ and $p$ denoting the 
four momenta of the initial electron and proton, respectively. 
Third, the four-momenta of $\Lambda_c$ and $\nu_{\tau}$ are correspondingly represented by 
$p^\prime$ and $p_\nu$, while $p_d$ refers to momentum of the visible decay product of the  $\tau$, with $d=\pi,\rho,\ell$ 
corresponding to the four decay channels of the $\tau$.
Finally, $\rho^D$ is the spin density decay matrix of the $\tau$, 
whereas its spin density production matrix is denoted by $\rho^P$. Their explicit expressions and details of the calculation 
procedures can be found in Appendixes~\ref{appendix:tdecay} and \ref{appendixx:tprodt}, respectively.  

The density matrix $\rho^P$ can be expanded in terms of Pauli matrices $\sigma^a$ with the first (second) row and 
column corresponding to the helicity $\lambda=1/2(-1/2)$,
\begin{align}\label{eq:rhoP1}
	\rho^P_{\lambda\lambda^\prime}&=\delta_{\lambda\lambda^\prime} C+\sum_a\sigma^a_{\lambda\lambda^\prime}
	\Sigma^a_P \nonumber \\[0.02cm]
	&=C\left(\delta_{\lambda\lambda^\prime}+ \sum_a\sigma^a_{\lambda\lambda^\prime}P_a\right)\,,
\end{align}
where $P_a=\Sigma^a_P/C$ denote the three components of the $\tau$ polarization 
vector $\mathcal{P}^{\mu}\equiv \sum_a P_a s^{\mu}_a$~\cite{Kong:2023kkd}.  
The explicit expressions of $C$ and $\Sigma^a_P$ can be determined by matching 
the $\rho^P_{\lambda\lambda^\prime}$ in Eq.~\eqref{eq:rhoP1} with that in Eq.~\eqref{eq:rhoP}.

\subsection{Extraction of the $\tau$ polarization components}
\label{sec:extract}

Since the spin density matrices $\rho^{P,D}$ and $d\Phi$ are Lorentz invariant, they can be calculated in any frame of reference.  
We pick the laboratory (Lab) frame---i.e., the initial proton is set to static---to evaluate the terms in the first square 
bracket in Eq.~\eqref{eq:diff cross}, and denote the result as 
\begin{align}\label{eq:sigmaP}
   d\sigma^P_{\lambda,\lambda^\prime}&=\frac{C(q^2)}{64\pi F E m_{p}}\left(\delta_{\lambda\lambda^\prime}
+\sum_a\sigma^a_{\lambda\lambda^\prime}P_a(q^2)\right)dq^2 \nonumber \\[0.02cm]
&=\frac{1}{2}\frac{d\sigma_{s}}{dq^2}\left(\delta_{\lambda\lambda^\prime}
+\sum_a\sigma^a_{\lambda\lambda^\prime}P_a(q^2)\right)dq^2\,,
\end{align}
where $q=k-k^\prime$ and $E$ denotes the electron beam energy. Note that $d\sigma_s/dq^2$ is the unpolarized 
differential cross section of the scattering process $e^- p\to \Lambda_c \tau^-$ in the Lab frame, given by 
\begin{align}\label{eq:cross_sec}
	\frac{d\sigma_{s}}{dq^2}=\frac{C}{32\pi F E m_{p}}\,.
\end{align}

We choose the rest frame of the $\tau$ lepton, on the other hand, to deduce
the terms in the second square bracket in Eq.~\eqref{eq:diff cross}, and get  
\begin{align}\label{eq:sigmaD}
	d\sigma^D_{\lambda^\prime,\lambda}=\mathcal{B}_{d}\frac{d^3\pmb{p}_d}{E_d}\Big[\eta_d \delta_{\lambda^\prime\lambda}-\chi_d (p_{d}\cdot s_b)\sigma^b_{\lambda^\prime\lambda}\Big]\,,
\end{align}
where $\mathcal{B}_{d=\pi,\rho,\ell}$ denote the corresponding branching fractions of 
$\tau^-\to \pi^-\nu_{\tau}$, $\tau^-\to \rho^-\nu_{\tau}$, and $\tau^-\to \ell^-\bar{\nu}_{\ell}\nu_{\tau}$ decays. 
The scalar functions $\eta_d$ and $\chi_d$ read, respectively,
\begin{align}\label{eq:scalar_function}
\eta_{d=\pi,\rho}&\!=\!\frac{m^2_{\tau}}{m^2_{\tau}-m^2_{d}}\frac{\delta\left(m_{\tau}-E_d
	-|\pmb{p}_d|\right)}{2\pi|\pmb{p}_d|}\,,\nonumber\\[0.02cm]
	\eta_{d=\ell}&\!=\!\frac{2\theta(1\!+\! y^2\!-\!x)\theta(x\!-\!2y)}{m^2_{\tau} f(y)\pi}\left[x(3\!-\!2x)\!-\!y^2(4\!-\!3x)\right]\,,\nonumber\\[0.02cm] 
	\chi_{d=\pi,\rho}&\!=\!\alpha_d\frac{2m^3_{\tau}}{(m^2_{\tau}-m^2_{d})^2}\frac{\delta\left(m_{\tau}-E_d
		-|\pmb{p}_d|\right)}{2\pi|\pmb{p}_d|}\,,\nonumber\\[0.02cm]
	\chi_{d=\ell}&\!=\!\frac{4\theta(1+y^2-x)\theta(x-2y)}{m^3_{\tau} f(y)\pi}\Big[1+3y^2-2x\Big]\,,
\end{align}
where $\theta(...)$ denotes the step function, $\alpha_{d=\pi,\rho}=\{1,(m^2_{\tau}-2m^2_{\rho})/(m^2_{\tau}+2m^2_{\rho})\}$, and $x=2(p_{\ell}\cdot k^\prime)/m^2_{\tau}$, $y=m_{\ell}/m_{\tau}$ 
in the $d= \ell$ case. For some details of the calculations, 
we refer the reader to Appendix~\ref{appendixx:phase}.

With the $d\sigma^P$ and $d\sigma^D$ given respectively in Eqs.~\eqref{eq:sigmaP} and \eqref{eq:sigmaD},  
we write the fully differential cross section as
\begin{align}\label{eq:sigma01}
		\frac{d\sigma_d}{dq^2}=\mathcal{B}_{d}\frac{d\sigma_s}{dq^2}\int\frac{d^3\pmb{p}_d}{E_{d}}\Big[\eta_d-\chi_d \sum_a (p_{d}\cdot s_a)P_a\Big]\,.
\end{align}
To discuss the polarization components  $P_a$, we must specify the spin vectors $s_a$. 
To this end, we first define three orthogonal unit vectors as follows:
\begin{align}
	\pmb{n}_L=\frac{\pmb{k}^\prime}{|\pmb{k}^\prime|},\quad 
	\pmb{n}_P=\frac{\pmb{k}^\prime\times \pmb{k}}{|\pmb{k}^\prime\times \pmb{k}|}, \quad
	\pmb{n}_T=\pmb{n}_P\times \pmb{n}_L,
\end{align}
where $\pmb{k}$ and $\pmb{k}^\prime$ are the three-momenta of the electron and the $\tau$ in the Lab frame.  
We then define the longitudinal (L), perpendicular (P), and transverse (T) spin four-vectors of the $\tau$ in its rest frame as  
\begin{align}
	s^{\mu}_L=(0,\pmb{n}_L),\quad s^{\mu}_P=(0,\pmb{n}_P), \quad 
	s^{\mu}_T=(0,\pmb{n}_T)\,.
\end{align}  
Finally, a Lorentz boost from the $\tau$ rest frame to the Lab frame leads to  
\begin{align}
	s^{\mu}_L=\left(\frac{|\pmb{k}^\prime|}{m_{\tau}},\frac{E_{\tau}\pmb{k}^\prime}{m_{\tau}|\pmb{k}^\prime|}\right)\,,
\end{align}
but leaves the perpendicular ($s^{\mu}_P$) and transverse ($s^{\mu}_T$) spin four-vectors unchanged. 
These transformed $s^{\mu}_{a}$ can be used to compute the polarization components $P_a$ in the Lab frame,
as shown in Appendix~\ref{app:numerator}. 


Taking the three unit vectors ($\pmb{n}_T$, $\pmb{n}_P$, $\pmb{n}_L$) as the Cartesian basis ($\pmb{n}_x$, $\pmb{n}_y$, $\pmb{n}_z$)  
in the $\tau$ rest frame, we write $\sum_a (p_{d}\cdot s_a)P_a$ in Eq.~\eqref{eq:sigma01} as 
\begin{align}
	\sum_a (p_{d}\cdot s_a)P_a=&-|\pmb{p}_d| \Big(P_T\sin\theta_d \cos\phi_d \nonumber \\[0.02cm]
	&+P_P\sin\theta_d \sin\phi_d+P_L\cos\theta_d\Big)\,, 
\end{align}
where $\theta_d$ and $\phi_d$ denote the polar and azimuthal angles of $\pmb{p}_d$ in this frame. 
Now integrating over $|\pmb{p}_d|$ in Eq.~\eqref{eq:sigma01}, we obtain 
\begin{align}\label{eq:sigma02}
	\frac{d^3\sigma_d}{dq^2d\Omega_d}=&\frac{\mathcal{B}_d}{4\pi}\frac{d\sigma_{s}}{dq^2}
	\Big[g^d+g^d_D\big(P_{T}(q^2)\sin\theta_d\cos\phi_d \nonumber \\
	&+P_{P}(q^2)\sin\theta_d\sin\phi_d
	+P_{L}(q^2)\cos\theta_d\big)\Big]\,,
\end{align}
where 
\begin{align}
	g^{\pi,\rho}=1\,, \qquad g^{\pi,\rho}_D=\alpha_{\pi,\rho}\,,
\end{align}
and 
\begin{align}
	g^{\ell}&=\frac{2}{f(y)}\int^{1+y^2}_{2y} \sqrt{x^2\!-\!4y^2}\left[x(3\!-\!2x)\!-\!y^2(4\!-\!3x)\right] dx\,, \nonumber \\[0.02cm] 
	g_D^{\ell}&=\frac{2}{f(y)}\int^{1+y^2}_{2y}(x^2-4y^2)\big[1+3y^2-2x\big]dx\,.
\end{align}
One can easily verify $g^{\ell}=1$ for $\ell=e,\mu$. 

From Eq.~\eqref{eq:sigma02}, it is easy to extract the $P_{a}$ in terms of the polar $\theta_d$ and azimuthal $\phi_d$ asymmetries.  
For instance, upon integration on $\phi_d$, one obtains 
\begin{align}
	\frac{d^2\sigma_d}{dq^2d\cos\theta_d}=\frac{\mathcal{B}_d}{2}\frac{d\sigma_{s}}{dq^2}\big[g^d+g^d_DP_{L}(q^2)\cos\theta_d\big]\,,
\end{align}
based on which $P_{L}$ is given by 
\begin{align}\label{eq:ExtractPL}
	P_{L}=\frac{2g^d}{g_D^d}\frac{\int_{0}^{1}d\cos\theta_d \frac{d^2\sigma}{dq^2d\cos\theta_d}-\int_{-1}^{0}d\cos\theta_d \frac{d^2\sigma}{dq^2d\cos\theta_d}}{\int_{0}^{1}d\cos\theta_d \frac{d^2\sigma}{dq^2d\cos\theta_d}+\int_{-1}^{0}d\cos\theta_d \frac{d^2\sigma}{dq^2d\cos\theta_d}}\,. 
\end{align}
Similarly, integrating over $\cos\theta_d$ yields
\begin{align}
	\frac{d^2\sigma_d}{dq^2d\phi_d}=&\frac{\mathcal{B}_d}{2\pi}\frac{d\sigma_{s}}{dq^2}\Big[g^d+\frac{\pi}{4}g^d_D\big(P_{T}(q^2)\cos\phi_d\nonumber\\[0.02cm]
	&+P_{P}(q^2)\sin\phi_d\big)\Big]\,.
\end{align}
One can then extract the $P_P$ and $P_T$, respectively, as
\begin{align}\label{eq:ExtractPP}
	P_{P}&=\frac{2g^d}{g_D^d}\frac{\int_{0}^{\pi} d\phi_d \frac{d^2\sigma}{dq^2d\phi_d}-\int_{\pi}^{2\pi} d\phi_d \frac{d^2\sigma}{dq^2d\phi_d}}{\int_{0}^{\pi} d\phi_d \frac{d^2\sigma}{dq^2d\phi_d}+\int_{\pi}^{2\pi} d\phi_d \frac{d^2\sigma}{dq^2d\phi_d}}\,, \\[0.02cm] \label{eq:ExtractPT}
	P_{T}&=\frac{2g^d}{g_D^d}\frac{ \int_{-\frac{\pi}{2}}^{\frac{\pi}{2}} d\phi_d \frac{d^2\sigma}{dq^2d\phi_d}- \int_{\frac{\pi}{2}}^{\frac{3\pi}{2}} d\phi_d \frac{d^2\sigma}{dq^2d\phi_d}}{\int_{-\frac{\pi}{2}}^{\frac{\pi}{2}} d\phi_d \frac{d^2\sigma}{dq^2d\phi_d}+\int_{\frac{\pi}{2}}^{\frac{3\pi}{2}} d\phi_d \frac{d^2\sigma}{dq^2d\phi_d}}\,.
\end{align}

Clearly such an extraction scheme relies on the reconstruction of the $\tau$ rest frame, which in turn depends on 
detection of the $\tau$ momentum. For the QE scattering in this work, we expect that the $\tau$ three-momentum 
can be determined by detecting the $\Lambda_c$ momentum, contrary   
to the $B$-hadron semitauonic decays~\cite{Hagiwara:2014tsa,Bordone:2016tex,Alonso:2016gym,Ligeti:2016npd,Asadi:2018sym,Alonso:2018vwa,
	Bhattacharya:2020lfm,Hu:2020axt,Nierste:2008qe,Tanaka:2010se,Alonso:2017ktd,Asadi:2020fdo,Penalva:2021gef,Hu:2021emb,
	Penalva:2021wye,Penalva:2022vxy,Li:2023gev}, the (anti-)neutrino-nucleus inclusive scattering ($\nu_{\tau}(\bar{\nu}_{\tau})A_Z\to \tau^{\mp}X$)~\cite{Hernandez:2022nmp,Isaacson:2023gwp}, 
or the electron-ion inclusive collision ($ep\to \tau X$)~\cite{Gonderinger:2010yn,Cirigliano:2021img}, 
in which the $\tau$ three-momentum cannot be determined precisely.   

Finally, for the convenience of later discussions, we also integrate over $\Omega_d$ in Eq.~\eqref{eq:sigma02} and get 
\begin{align}\label{eq:total_cross}
   \frac{d\sigma_d}{dq^2}=\mathcal{B}_d\frac{d\sigma_s}{dq^2}=\frac{\mathcal{B}_dC}{32\pi F E m_{p}}\,,
\end{align} 
where the factor $g^d$ has been removed due to $g^d=1$.

\subsection{Form factors and kinematics of the scattering process}
\label{subsec:Cross section}

The spin density matrix $\rho^D$ is constructed out of the leptonic and hadronic matrix 
elements, i.e., $L^{L,R}_a$ and $H^{L,R}_a$ with $\alpha=S,\, V,\, T$, as shown in Eq.~\eqref{eq:rhoP}.
The former can be calculated straightforwardly, while the latter shall turn to its complex conjugate~\cite{Sobczyk:2019uej,Lai:2021sww,Lai:2022ekw,Kong:2023kkd}, 
which are parametrized by the $\Lambda_c \to N$ transition form factors~\cite{Feldmann:2011xf,Meinel:2017ggx,Das:2018sms}.
Since a scattering process generally occupies a different kinematic range ($q^2<0$) from that of a decay ($q^2>0$), theoretical analyses of the scattering process require an extrapolation of the form factors to negative $q^2$.
Thus, the form-factor parametrizations suitable for our purpose must be analytic in the proper $q^2$ range. 

In our previous analyses of the $\tau$ polarizations in the $\nu_{\tau} n\to \tau^- \Lambda_c$ scattering process~\cite{Kong:2023kkd}, we 
considered several schemes that meet the selection criterion and have been utilized to parametrize the $\Lambda_c \to N$ 
form factors by various models. 
For instance, the dipole parametrization scheme, employed within the MIT bag model (MBM)~\cite{Chodos:1974je,Chodos:1974pn} and the nonrelativistic quark model (NRQM)~\cite{Kokkedee1969}, and the double-pole scheme in the relativistic constituent quark model (RCQM)~\cite{Ivanov:1996fj,Branz:2009cd}. However, only the form factors associated with the matrix element $\langle N|\bar{u}\gamma^{\mu}P_L c|\Lambda_c\rangle$ were calculated in these models. 
The primary scheme we took was initially proposed to parametrize the $B\to \pi$ vector form factor~\cite{Bourrely:2008za}, and has been used in the LQCD calculation of the $\Lambda_c\to N$ transition form factors~\cite{Meinel:2017ggx}. Contrary to other model evaluations, the LQCD calculation not only covers all the form factors, but also provides an error estimation. Thus, we will adopt the latest LQCD results~\cite{Meinel:2017ggx} throughout this work, too. For more details about the form factors in these different models, we refer the reader to 
Refs.~\cite{Sobczyk:2019uej, Lai:2021sww, Kong:2023kkd}.
 
Extrapolating the form factors to positive $q^2$ raises ambiguity about the form factors, which in turn 
induces theoretical uncertainties in predictions.
Indeed, model calculations of the $N\to \Lambda_c$ form factors can significantly affect the predictions of $\Lambda_c$ weak production in neutrino QE scattering processes~\cite{DeLellis:2004ovi,Sobczyk:2019uej} and of the $\tau$ polarizations in the scattering $\nu_{\tau} n\to \tau^- \Lambda_c$~\cite{Kong:2023kkd}. Worse, the uncertainties induced by using the different schemes even dwarf 
that from the error propagation of the form factors~\cite{Sobczyk:2019uej,Kong:2023kkd}. 
Thus, we will also analyze the $\tau$ polarizations in the $e^- p\to \tau^- \Lambda_c$ scattering process in terms of the form factors calculated within the models MBM, NRQM, RCQM, and LQCD in various NP scenarios, and examine if the same observation also applies to this process.

\begin{figure}[t]
	\centering
	\includegraphics[width=0.46\textwidth]{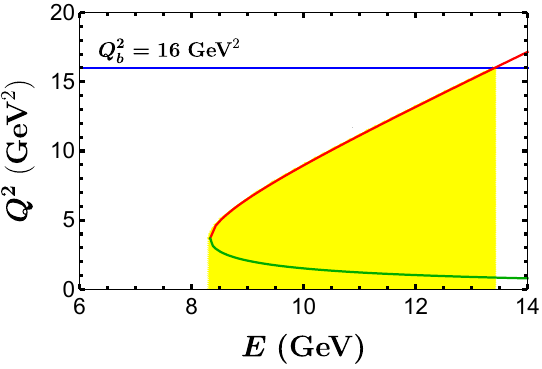}
	\caption{Criteria for selecting the electron beam energy $E$, where the red (green) curve denotes the $E$-$Q^2_{\max(\min)}$ relation given by Eq.~\eqref{eq:LFV_Q2_range}, and the blue line represents the condition $Q^2\leq 16\,\text{GeV}^2$ required by our theoretical framework. The yellow range indicates the eligible $E$.} 
	\label{fig:Eselection} 
\end{figure} 

Let us now turn to the kinematics of the scattering process $e^- p\to \Lambda_c \tau^-$, in which the $q^2$ is bounded by~\cite{Lai:2021sww} 
\begin{align}
	\frac{\alpha-E\sqrt{\lambda}}{m_p+2E} \leq
	q^2 \leq \frac{\alpha +E\sqrt{\lambda}}{m_p+2E}\,,\label{eq:LFV_Q2_range}
\end{align}
where
\begin{align}
	\alpha &\!\equiv \! E(m_{\Lambda_c}^2\!-\! m_p^2\!+\!m^2_{\tau}\!-\! 2m_pE)\!+\!m_p m^2_{\tau}\,, \nonumber \\[0.2cm]
	\lambda &\!\equiv\! m_{\Lambda_c}^4\!+\!(m_p^2\!+\!2m_pE\!-\!m^2_{\tau})^2\!-\!2m_{\Lambda_c}^2(m_p^2\!+\!2m_pE\!+\!m^2_{\tau})\,. \nonumber 
\end{align}
Note that the electron mass has been ignored due to $m_e/E\ll 1$. The condition of Eq.~\eqref{eq:LFV_Q2_range} indicates that the electron beam energy $E$ determines the maximal and minimal values of $Q^2$ ($Q^2=-q^2$), which, in turn, implies that any constraints on $Q^2_{\max}$ and $Q^2_{\min}$ restrict the $E$ selection. 
For instance, condition $Q^2_{\max}=Q^2_{\min}$ indicates a minimal requirement for $E$ ($E\gtrsim 8.33$~GeV);
this can also be visualized in Fig.~\ref{fig:Eselection} by noting the intersection point of the red and green curves that represent the $E$-$Q^2_{\max}$ and $E$-$Q^2_{\min}$ relations, respectively. 
Besides these constraints, we also consider the one from our theoretical framework. Since our analyses are carried out in the framework of $\mathcal{L}_{\text{eff}}$ given by Eq.~\eqref{eq:Leff}, to ensure the validity of our results, we require $Q^2_{\max}$ to not exceed $Q^2_{b}=16\,\text{GeV}^2\approx m_b^2$. Such a requirement, depicted by the blue line in Fig.~\ref{fig:Eselection}, results in an upper bound $E\lesssim 13.41\,\text{GeV}$, provided that the observables one is interested in, such as the total cross section, involve $Q^2_{\max}$. Otherwise, $E$ is not bounded from above, since one can always focus on the lower $Q^2$ range, even though a high $Q^2_{\max}$ is available due to a high $E$.

 \section{Phenomenology analyses}
 \label{sec:Numresults}  
    
\subsection{Observable analyses}
\label{sec:obs}

The factor $C$ in Eq.~\eqref{eq:rhoP1} is connected 
to the unpolarized 
differential cross section  $d\sigma_s$ of the scattering process $e^- p\to \Lambda_c \tau^-$ (cf. Eq.~\eqref{eq:cross_sec}) 
through the following relation 
\begin{align}\label{eq:dsigma}
	2C= |\mathcal{M}|^2=\frac{1}{v^4}\sum g_{\alpha}g^*_{\beta}~\mathcal{A}_{\alpha-\beta}\,,
\end{align}   
where $|\mathcal{M}|^2$ is the amplitude squared without spin average of the initial particles, 
and $\mathcal{A}_{\alpha-\beta}$ with a subscript, e.g., $V_{LL}\!-\!V_{LR}$, represents the reduced amplitude squared that is induced by the interference between the operators $\mathcal{O}^{LL}_V$ and $\mathcal{O}^{LR}_V$.  
Since all of the operators in Table~\ref{tab:Operator} can contribute to the scattering process,
$g_{\alpha}$ and $g^*_{\beta}$ go through all the WCs.

\begin{table}[t!]
	\renewcommand*{\arraystretch}{1.5}
	\centering
	\caption{Non-zero reduced amplitude squared $\mathcal{A}_{\alpha-\beta}$ that contributes to the cross section of the scattering process $e^- p\to \Lambda_c \tau^-$. The subscript $\alpha-\beta$, e.g., $V_{LL}\!-\!V_{LR}$, indicates $\mathcal{A}_{V_{LL}\!-\!V_{LR}}$ is induced by the interference between the operators $\mathcal{O}^{LL}_V$ and $\mathcal{O}^{LR}_V$.}
	\begin{tabular}{lccccc}
		\hline \hline
		&  $(g^{LL}_V)^*$ & $(g^{LR}_V)^*$ & $(g^{LL}_S)^*$ & $(g^{LR}_S)^*$ & $(g^{LL}_T)^*$ \\
		\hline		
		$g^{LL}_V$&$\mathcal{A}_{V_{LL}\!-\!V_{LL}}$ &$\mathcal{A}_{V_{LL}\!-\!V_{LR}}$&$\mathcal{A}_{V_{LL}\!-\!S_{LL}}$ &$\mathcal{A}_{V_{LL}\!-\!S_{LR}}$&$\mathcal{A}_{V_{LL}\!-\!T_{LL}}$ \\   
		$g^{LR}_V$&$\mathcal{A}_{V_{LL}\!-\!V_{LR}}$&$\mathcal{A}_{V_{LR}\!-\!V_{LR}}$&$\mathcal{A}_{V_{LL}\!-\!S_{LR}}$&$\mathcal{A}_{V_{LL}\!-\!S_{LL}}$&$\mathcal{A}_{V_{LR}\!-\!T_{LL}}$\\ 
		$g^{LL}_S$&$\mathcal{A}_{V_{LL}\!-\!S_{LL}}$&$\mathcal{A}_{V_{LL}\!-\!S_{LR}}$&$\mathcal{A}_{S_{LL}\!-\!S_{LL}}$ &$\mathcal{A}_{S_{LL}\!-\!S_{LR}}$&$\mathcal{A}_{S_{LL}\!-\!T_{LL}}$ \\ 
		$g^{LR}_S$&$\mathcal{A}_{V_{LL}\!-\!S_{LR}}$&$\mathcal{A}_{V_{LL}\!-\!S_{LL}}$&$\mathcal{A}_{S_{LL}\!-\!S_{LR}}$&$\mathcal{A}_{S_{LL}\!-\!S_{LL}}$&$\mathcal{A}_{S_{LR}\!-\!T_{LL}}$\\ 
		$g^{LL}_T$&$\mathcal{A}_{V_{LL}\!-\!T_{LL}}$&$\mathcal{A}_{V_{LR}\!-\!T_{LL}}$&$\mathcal{A}_{S_{LL}\!-\!T_{LL}}$&$\mathcal{A}_{S_{LR}\!-\!T_{LL}}$&$\mathcal{A}_{T_{LL}\!-\!T_{LL}}$\\                                                                                                  
		\hline \hline
		&  $(g^{RR}_V)^*$ & $(g^{RL}_V)^*$ & $(g^{RR}_S)^*$ & $(g^{RL}_S)^*$ & $(g^{RR}_T)^*$ \\
		\hline		
		$g^{RR}_V$&$\mathcal{A}_{V_{LL}\!-\!V_{LL}}$ &$\mathcal{A}_{V_{LL}\!-\!V_{LR}}$&$\mathcal{A}_{V_{LL}\!-\!S_{LL}}$ &$\mathcal{A}_{V_{LL}\!-\!S_{LR}}$&$\mathcal{A}_{V_{LL}\!-\!T_{LL}}$ \\   
		$g^{RL}_V$&$\mathcal{A}_{V_{LL}\!-\!V_{LR}}$&$\mathcal{A}_{V_{LR}\!-\!V_{LR}}$&$\mathcal{A}_{V_{LL}\!-\!S_{LR}}$&$\mathcal{A}_{V_{LL}\!-\!S_{LL}}$&$\mathcal{A}_{V_{LR}\!-\!T_{LL}}$\\ 
		$g^{RR}_S$&$\mathcal{A}_{V_{LL}\!-\!S_{LL}}$&$\mathcal{A}_{V_{LL}\!-\!S_{LR}}$&$\mathcal{A}_{S_{LL}\!-\!S_{LL}}$ &$\mathcal{A}_{S_{LL}\!-\!S_{LR}}$&$\mathcal{A}_{S_{LL}\!-\!T_{LL}}$ \\ 
		$g^{RL}_S$&$\mathcal{A}_{V_{LL}\!-\!S_{LR}}$&$\mathcal{A}_{V_{LL}\!-\!S_{LL}}$&$\mathcal{A}_{S_{LL}\!-\!S_{LR}}$&$\mathcal{A}_{S_{LL}\!-\!S_{LL}}$&$\mathcal{A}_{S_{LR}\!-\!T_{LL}}$\\ 
		$g^{RR}_T$&$\mathcal{A}_{V_{LL}\!-\!T_{LL}}$&$\mathcal{A}_{V_{LR}\!-\!T_{LL}}$&$\mathcal{A}_{S_{LL}\!-\!T_{LL}}$&$\mathcal{A}_{S_{LR}\!-\!T_{LL}}$&$\mathcal{A}_{T_{LL}\!-\!T_{LL}}$\\                                                                                                  
		\hline\hline
	\end{tabular}	
	\label{tab:Denominator}
\end{table}

We list in Table~\ref{tab:Denominator} all the non-zero $\mathcal{A}_{\alpha-\beta}$. 
Already, some patterns emerge. First, $\mathcal{A}_{\alpha-\beta}$ vanishes if induced by interference 
between two operators with their lepton currents of opposite chiral properties, equivalently $L^{L}_{\alpha}(\lambda)L^{R*}_{\beta}(\lambda)=0$, 
because the electron mass has been ignored due to $m_e/E\ll 1$. Second, all of the $\mathcal{A}$ are real, leading to 
$\mathcal{A}_{\alpha-\beta}=\mathcal{A}_{\beta-\alpha}$. 
Third, $\mathcal{A}_{S_{LL}\!-\!S_{LL}}$ is equal to $\mathcal{A}_{S_{LR}\!-\!S_{LR}}$.  
Finally, simultaneously flipping the chiral properties of the lepton and quark currents of two operators 
$\mathcal{O}_\alpha$ and $\mathcal{O}_\beta$ 
yields the same reduced amplitude squared $\mathcal{A}_{\alpha-\beta}$, e.g., $\mathcal{A}_{V_{RR}\!-\!V_{RL}}=\mathcal{A}_{V_{LL}\!-\!V_{LR}}$. 
The latter two patterns arise due to the chiral structures of the lepton and quark currents 
and the form-factor paramertrization of the hadronic matrix elements. 
Based on these patterns, we only present one $\mathcal{A}_{\alpha-\beta}$ in Table~\ref{tab:Denominator}, if its duplicates appear. 

\begin{table*}[htpb!]
	\renewcommand*{\arraystretch}{1.5}
	\tabcolsep=0.72cm
	\centering
	\caption{Non-zero reduced amplitude squared $\mathcal{A}^{\tau}_{\alpha-\beta}$ that contains the polarization information of the lepton $\tau$. 
	Note that it has the same subscript with the $\mathcal{A}_{\alpha-\beta}$.}
	\begin{tabular}{lccccc}
		\hline \hline
		&  $(g^{LL}_V)^*$ & $(g^{LR}_V)^*$ & $(g^{LL}_S)^*$ & $(g^{LR}_S)^*$ & $(g^{LL}_T)^*$ \\
		\hline		
		$g^{LL}_V$&$\mathcal{A}^{\tau}_{V_{LL}\!-\!V_{LL}}$ &$\mathcal{A}^{\tau}_{V_{LL}\!-\!V_{LR}}$&$\mathcal{A}^{\tau}_{V_{LL}\!-\!S_{LL}}$ &$\mathcal{A}^{\tau}_{V_{LL}\!-\!S_{LR}}$&$\mathcal{A}^{\tau}_{V_{LL}\!-\!T_{LL}}$ \\   
		$g^{LR}_V$&$\mathcal{A}^{\tau}_{V_{LL}\!-\!V_{LR}}$&$\mathcal{A}^{\tau}_{V_{LR}\!-\!V_{LR}}$&$\mathcal{A}^{\tau}_{V_{LL}\!-\!S_{LR}}$&$\mathcal{A}^{\tau}_{V_{LL}\!-\!S_{LL}}$&$\mathcal{A}^{\tau}_{V_{LR}\!-\!T_{LL}}$\\ 
		$g^{LL}_S$&$\mathcal{A}^{\tau*}_{V_{LL}\!-\!S_{LL}}$&$\mathcal{A}^{\tau*}_{V_{LL}\!-\!S_{LR}}$&$\mathcal{A}^{\tau}_{S_{LL}\!-\!S_{LL}}$ &$\mathcal{A}^{\tau}_{S_{LL}\!-\!S_{LR}}$&$\mathcal{A}^{\tau}_{S_{LL}\!-\!T_{LL}}$ \\ 
		$g^{LR}_S$&$\mathcal{A}^{\tau*}_{V_{LL}\!-\!S_{LR}}$&$\mathcal{A}^{\tau*}_{V_{LL}\!-\!S_{LL}}$&$\mathcal{A}^{\tau}_{S_{LL}\!-\!S_{LR}}$&$\mathcal{A}^{\tau}_{S_{LL}\!-\!S_{LL}}$&$\mathcal{A}^{\tau}_{S_{LR}\!-\!T_{LL}}$\\ 
		$g^{LL}_T$&$\mathcal{A}^{\tau*}_{V_{LL}\!-\!T_{LL}}$&$\mathcal{A}^{\tau*}_{V_{LR}\!-\!T_{LL}}$&$\mathcal{A}^{\tau*}_{S_{LL}\!-\!T_{LL}}$&$\mathcal{A}^{\tau*}_{S_{LR}\!-\!T_{LL}}$&$\mathcal{A}^{\tau}_{T_{LL}\!-\!T_{LL}}$\\                                                     	\hline \hline
		&  $(g^{RR}_V)^*$ & $(g^{RL}_V)^*$ & $(g^{RR}_S)^*$ & $(g^{RL}_S)^*$ & $(g^{RR}_T)^*$ \\
		\hline		
		$g^{RR}_V$&$-\mathcal{A}^{\tau}_{V_{LL}\!-\!V_{LL}}$ &$-\mathcal{A}^{\tau}_{V_{LL}\!-\!V_{LR}}$&$-\mathcal{A}^{\tau*}_{V_{LL}\!-\!S_{LL}}$ &$-\mathcal{A}^{\tau*}_{V_{LL}\!-\!S_{LR}}$&$-\mathcal{A}^{\tau*}_{V_{LL}\!-\!T_{LL}}$ \\   
		$g^{RL}_V$&$-\mathcal{A}^{\tau}_{V_{LL}\!-\!V_{LR}}$&$-\mathcal{A}^{\tau}_{V_{LR}\!-\!V_{LR}}$&$-\mathcal{A}^{\tau*}_{V_{LL}\!-\!S_{LR}}$&$-\mathcal{A}^{\tau*}_{V_{LL}\!-\!S_{LL}}$&$-\mathcal{A}^{\tau*}_{V_{LR}\!-\!T_{LL}}$\\ 
		$g^{RR}_S$&$-\mathcal{A}^{\tau}_{V_{LL}\!-\!S_{LL}}$&$-\mathcal{A}^{\tau}_{V_{LL}\!-\!S_{LR}}$&$-\mathcal{A}^{\tau}_{S_{LL}\!-\!S_{LL}}$ &$-\mathcal{A}^{\tau}_{S_{LL}\!-\!S_{LR}}$&$-\mathcal{A}^{\tau*}_{S_{LL}\!-\!T_{LL}}$ \\ 
		$g^{RL}_S$&$-\mathcal{A}^{\tau}_{V_{LL}\!-\!S_{LR}}$&$-\mathcal{A}^{\tau}_{V_{LL}\!-\!S_{LL}}$&$-\mathcal{A}^{\tau}_{S_{LL}\!-\!S_{LR}}$&$-\mathcal{A}^{\tau}_{S_{LL}\!-\!S_{LL}}$&$-\mathcal{A}^{\tau*}_{S_{LR}\!-\!T_{LL}}$\\ $g^{RR}_T$&$-\mathcal{A}^{\tau}_{V_{LL}\!-\!T_{LL}}$&$-\mathcal{A}^{\tau}_{V_{LR}\!-\!T_{LL}}$&$-\mathcal{A}^{\tau}_{S_{LL}\!-\!T_{LL}}$&$-\mathcal{A}^{\tau}_{S_{LR}\!-\!T_{LL}}$&$-\mathcal{A}^{\tau}_{T_{LL}\!-\!T_{LL}}$\\                                                                 \hline\hline
	\end{tabular}	
	\label{tab:Numerator}
\end{table*}

Similar to $C$, $\Sigma^a_P$ can be written as 
\begin{align}\label{eq:dsigma2}
	2\Sigma^a_P=\frac{1}{v^4}\sum g_{\alpha}g^*_{\beta}~\mathcal{A}^{\tau}_{\alpha-\beta}\,,
\end{align}   
where the reduced amplitude squared $\mathcal{A}^{\tau}_{\alpha-\beta}$ contains the polarization information of 
the $\tau$, and shares the same subscript with the $\mathcal{A}_{\alpha-\beta}$. 
We list in Table~\ref{tab:Numerator} all the non-zero $\mathcal{A}^{\tau}_{\alpha-\beta}$. 
One can see that certain patterns emerging from the $\mathcal{A}_{\alpha-\beta}$ in Table~\ref{tab:Denominator} 
can apply to $\mathcal{A}^{\tau}_{\alpha-\beta}$, too. 
For instance, $\mathcal{A}^{\tau}_{\alpha-\beta}$ also vanishes if induced by interference 
between two operators with their lepton currents of opposite chiral properties, 
equivalently $L^{L}_{\alpha}(\lambda)L^{R*}_{\beta}(\lambda^\prime)=0$,  due to the same reason.
Another example is that the relation $\mathcal{A}^{\tau}_{S_{LL}\!-\!S_{LL}}\!=\!\mathcal{A}^{\tau}_{S_{LR}\!-\!S_{LR}}$ 
also holds in this case.  

Of course, distinct differences exist between the two cases. First, instead of $\mathcal{A}^{\tau}_{\alpha-\beta}\!=\!\mathcal{A}^{\tau}_{\beta-\alpha}$, 
$\mathcal{A}^{\tau}_{\alpha-\beta}\!=\!\mathcal{A}^{\tau*}_{\beta-\alpha}$ 
in general for $\alpha\neq \beta$ (one exception is when both the $\alpha$ and $\beta$ denote the vector operators).
Such a difference arises from the appearance of $\varepsilon_{\{k\}\{k^\prime\}\{s_a\}\{p\}}$, 
which is always accompanied by the imaginary unit $i$. Interestingly enough, the very same term generates a non-zero $P_P$ 
provided that all the WCs are complex. 
For $\alpha=\beta$, on the other hand, $\mathcal{A}^{\tau}_{\alpha-\alpha}$ remains real 
due to missing the $\varepsilon_{\{k\}\{k^\prime\}\{s_a\}\{p\}}$. Based on these arguments, one can see that 
$P_P$ vanishes in a pure-vector scenario, i.e., only 
the vector operators $\mathcal{O}_V$ are activated. And it also vanishes if only one 
WC is turned on in the $\mathcal{L}_{\text{eff}}$, i.e., all other WCs are set to zero.
The second difference is that, contrary to the $\mathcal{A}_{\alpha-\beta}$ case, simultaneously 
flipping the chiral properties of the lepton and quark currents of two operators 
$\mathcal{O}_\alpha$ and $\mathcal{O}_\beta$ now yields $-\mathcal{A}^{\tau}_{\alpha-\beta}$, e.g., $\mathcal{A}^{\tau}_{V_{RR}\!-\!V_{RL}}=-\mathcal{A}^{\tau}_{V_{LL}\!-\!V_{LR}}$. 

\begin{figure}[t]
	\centering
	\includegraphics[width=0.43\textwidth]{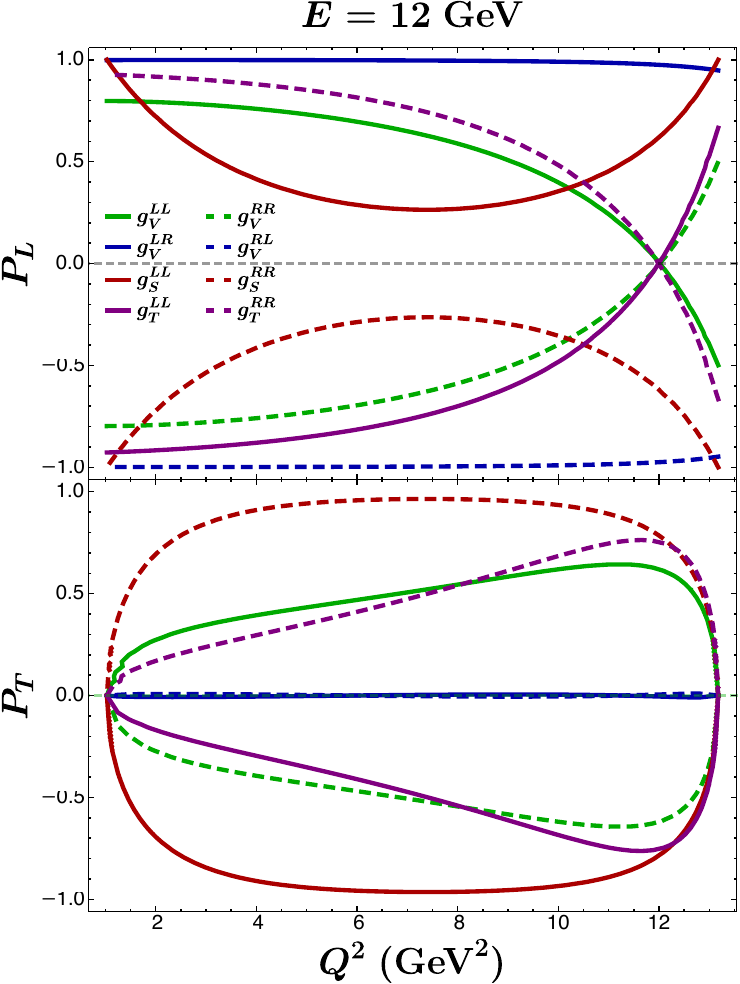}
	\caption{Variations of the polarization $P_L$ and $P_T$ with respect to $Q^2$ in different NP scenarios, where we have set the electron beam energy at $E=12$~GeV.} 
	\label{fig:NonMixCombine} 
\end{figure}

\begin{figure*}[t]
	\centering
	\includegraphics[width=0.96\textwidth]{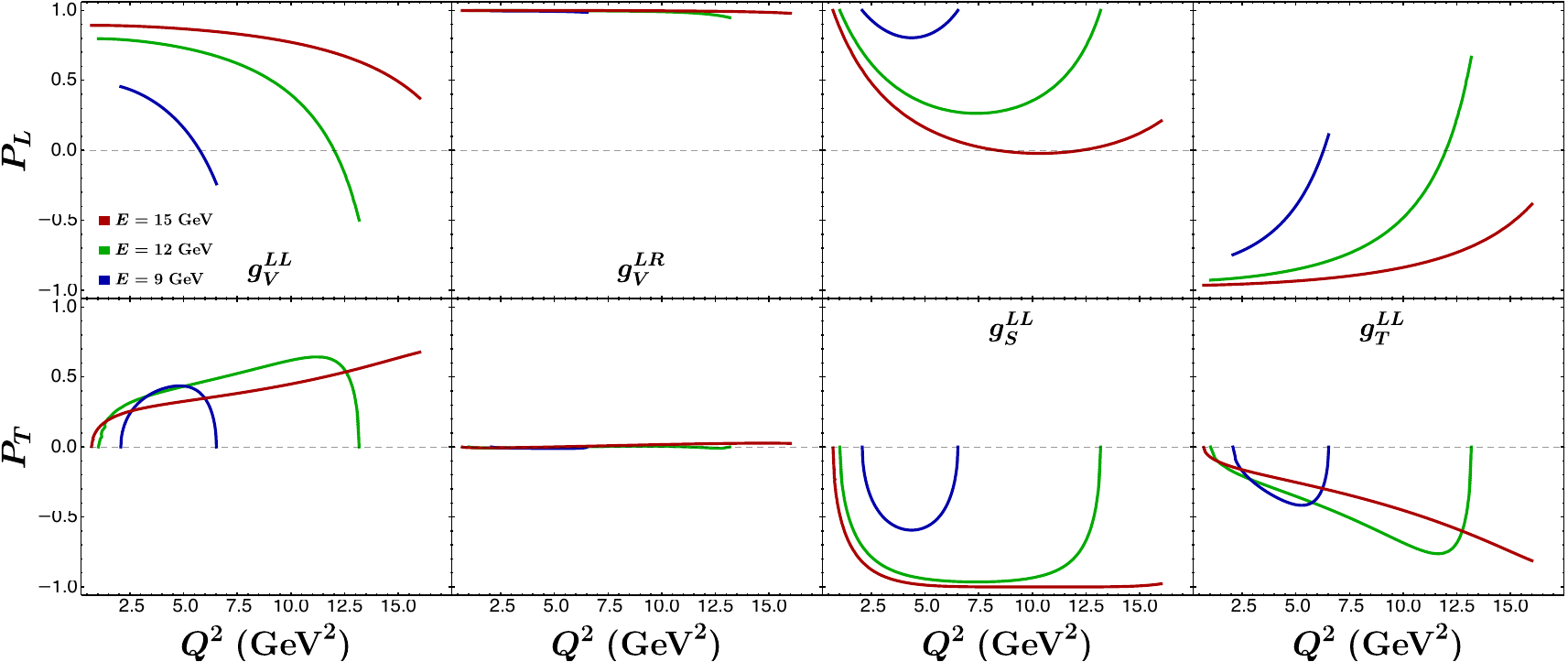}
	\caption{The polarization $P_{L,T}$ in various NP scenarios with three different electron beam energies. Note that the NP scenarios 
		with right-handed lepton current are not presented and neither is the NP $g^{LR}_S$. Also only the $P_{L,T}$ within the kinematic range $Q^2\in[Q^2_{\text{min}}, 16\ \text{GeV}^2]$ are presented for $E=$15 GeV. } 
	\label{fig:CombineDE} 
\end{figure*}

With $C$ and $\Sigma^a_P$ given respectively in Eqs.~\eqref{eq:dsigma} and \eqref{eq:dsigma2}, the polarization components $P_a$ 
are now formulated in terms of the $\mathcal{A}_{\alpha-\beta}$ and $\mathcal{A}^{\tau}_{\alpha-\beta}$ as well as the WCs. 
Since the WCs have not been determined yet, it will be difficult to explore the behavior of $P_{a}$ in general with respect to the kinematics $Q^2$, 
the beam energy $E$, etc. However, if only one operator $\mathcal{O}_\alpha$ in Eq.~\eqref{eq:Leff} is activated at a time, the $P_{a}$ 
become independent of the WC $g_{\alpha}$. Immediately, $P_P$ in these NP scenarios vanishes, as indicated by the patterns arising in Tables~\ref{tab:Denominator} and \ref{tab:Numerator}. In addition, $P_{L,T}$ in the NPs $g^{LL}_V$ and $g^{RR}_V$ differ only 
by a minus sign; the same conclusion also holds for the NP pairs ($g^{LR}_V$, $g^{RL}_V$), ($g^{LL}_S$, $g^{RR}_S$), 
($g^{LR}_S$, $g^{RL}_S$), and ($g^{LL}_T$, $g^{RR}_T$).
Besides the simple case above, the $P_{a}$ can also be independent of the WCs, 
if only $\mathcal{O}^{LL}_S$ and $\mathcal{O}^{LR}_S$ 
($\mathcal{O}^{RR}_S$ and $\mathcal{O}^{RL}_S$) are activated simultaneously. 
To justify this one can check the explicit expressions of 
the corresponding $\mathcal{A}_{\alpha-\beta}$ and $\mathcal{A}^{\tau}_{\alpha-\beta}$ 
in Appendixes~\ref{app:denominator} and \ref{app:numerator}.     
 
The behavior of $P_a$ predicted above in the two simple cases can also be  
justified graphically. Let us focus on the first one and explore in Fig.~\ref{fig:NonMixCombine} 
how the $P_{L,T}$ vary with respect to the $Q^2$ in various NP scenarios (note $P_P=0$). 
For a simple demonstration, we consider the beam energy $E=12$ GeV as a benchmark. 
And since the $P_{L,T}$ induced by the $\mathcal{O}^{LL}_S$ 
and $\mathcal{O}^{LR}_S$ ($\mathcal{O}^{RR}_S$ and $\mathcal{O}^{RL}_S$) are identical, 
only the former are presented, as already argued in the second simple case.
Now one can clearly see that the sign of $P_{L,T}$ indeed flips as one 
simultaneously flips the chiral properties of the lepton and quark currents of an operator---check
the solid and dashed curves in the same color. 
In addition, the lepton $\tau$ produced in the NP $g_V^{LR}$ ($g_V^{RL}$) 
is almost fully polarized along the longitudinal direction, as indicated by the blue (dashed) curves. 
In the NP $g_S^{LL}$ ($g_S^{RR}$) denoted by the red (dashed) curves, 
on the other hand, the $\tau$ is fully polarized in the longitudinal direction only at the maximal and minimal $Q^2$; 
in the middle $Q^2$ regions, say $Q^2\in (5, 9)$ GeV$^2$, it prefers being polarized transversely.   
Finally, the $P_{L,T}$ in the NPs $g_V^{LL}$ and $g_T^{LL}$ ($g_V^{RR}$ and $g_T^{RR}$)
display a roughly opposite pattern and, interestingly, the $P_{L}$ in both scenarios becomes zero at $Q^2\simeq 12$ GeV$^2$. 

Finally, we explore the dependence of $P_{L,T}$ on the electron beam energy $E$. 
Based on the observations made in Fig.~\ref{fig:NonMixCombine}, we shall focus on the NPs $g_V^{LL}$, $g_V^{LR}$, 
$g_S^{LL}$, and $g_T^{LL}$. As shown in Fig.~\ref{fig:CombineDE}, the $P_{L}$ ($P_{T}$) 
in the NP $g_V^{LR}$ remains $+1$ (0) to a very good approximation in the whole allowed $Q^2$ regions. 
Meanwhile, the $P_{L}$ ($P_{T}$) in the NP $g_S^{LL}$ at the corresponding $Q^2_{\text{min}}$ and $Q^2_{\text{max}}$ 
remains $+1$ (0) for $E=9, 12$ GeV; this pattern would also apply to the case of $E=15$ GeV if its corresponding $Q^2_{\text{max}}$ were not excluded by our theoretical framework. On the other hand, the $P_{T}$ in the middle $Q^2$ regions
becomes more favorite than the $P_L$ for the $\tau$ as $E$ increases (note $P_L=0$ 
and $P_T=-1$ at $Q^2\simeq 7.5, 12.5$ GeV$^2$ for $E=15$ GeV). As for the NPs $g_V^{LL}$ and $g_T^{LL}$, 
both the minimal and maximal values of the $P_{L}$ and the minimum of the $P_T$ change vastly as $E$ increases, 
but the roughly opposite pattern between the two NP cases remains.

\subsection{NP model identification}
\label{subsec:model-id}

We considered in Ref.~\cite{Lai:2022ekw} the low-energy polarized scattering process $e^-p\to e^-\Lambda_c$, 
and demonstrated in a model-independent way that 15 NP scenarios constructed only with the vector operators $\mathcal{O}_V$ 
can be effectively disentangled from each other by measuring 4 spin asymmetries $A^e_L$, $A^p_L$, $A^{ep}_{L3}$, and 
$A^{ep}_{L6}$. The key ingredients are the longitudinal polarized electron beam and proton target.  
In this subsection, we will also define 15 NP scenarios in a similar way, 
and show that they can be distinguished from each other by using the longitudinal polarized electron beam
and by measuring the $P_{T}$ of the $\tau$.
 
\begin{table}[t!]
	\renewcommand*{\arraystretch}{1.25}
	\tabcolsep=0.39cm
	\centering
	\caption{The $P_T$ predicated in the cases I-XV with the electron beam left- ($e_L$) and right-handed ($e_R$) polarized, 
		where $c$ is expected to be away from 0 and 0.5 in general; see text for details.}
	\begin{tabular}{lccccc}
		\hline \hline
		&  I & II & III & IV & V \\
		\hline		
		$P_T$($e_L$) & 0.5 & 0 & 0 & 0 & $c$  \\
		$P_T$($e_R$) & 0 & 0 & 0 & $-0.5$ & 0   \\                                                                                     
		\hline \hline
		& VI & VII & VIII & IX & X \\
		\hline		
		$P_T$($e_L$) & 0.5 & 0.5 & 0 & 0 & 0  \\
		$P_T$($e_R$) & 0 & $-0.5$ & 0 & $-0.5$ & $-c$   \\                                                                                     
		\hline \hline	
		& XI & XII & XIII & XIV & XV \\
		\hline		
		$P_T$($e_L$) & $c$ & $c$ & 0.5 & 0 & $c$  \\
		$P_T$($e_R$) & 0 & $-0.5$ & $-c$ & $-c$ & $-c$   \\                                                                                     
		\hline \hline		                                                                                                
	\end{tabular}	
	\label{tab:model-disting}
\end{table}
 
To begin with, let us introduce the new 15 NP scenarios, which are constructed with the 4   
vector operators $\mathcal{O}_V^{LL}$, $\mathcal{O}_V^{LR}$, $\mathcal{O}_V^{RL}$, and $\mathcal{O}_V^{RR}$, 
as well as their possible combinations: 
\begin{enumerate}[(1)]
	\item cases with one vector operator: (I) $\mathcal{O}_V^{LL}$, (II) $\mathcal{O}_V^{LR}$, (III) $\mathcal{O}_V^{RL}$, 
	and (IV) $\mathcal{O}_V^{RR}$;
	
	\item cases with two vector operators: (V) $\mathcal{O}_V^{LL}$ and $\mathcal{O}_V^{LR}$, (VI) $\mathcal{O}_V^{LL}$ and $\mathcal{O}_V^{RL}$, (VII) $\mathcal{O}_V^{LL}$ and $\mathcal{O}_V^{RR}$, (VIII) $\mathcal{O}_V^{LR}$ and $\mathcal{O}_V^{RL}$, (IX) $\mathcal{O}_V^{LR}$ and $\mathcal{O}_V^{RR}$, and (X) $\mathcal{O}_V^{RL}$ and $\mathcal{O}_V^{RR}$;
	
	\item cases with three vector operators: (XI) $\mathcal{O}_V^{LL}$, $\mathcal{O}_V^{LR}$, and $\mathcal{O}_V^{RL}$, (XII) $\mathcal{O}_V^{LL}$, $\mathcal{O}_V^{LR}$, and $\mathcal{O}_V^{RR}$, (XIII) $\mathcal{O}_V^{LL}$, $\mathcal{O}_V^{RL}$, and $\mathcal{O}_V^{RR}$, and (XIV) $\mathcal{O}_V^{LR}$, $\mathcal{O}_V^{RL}$, and $\mathcal{O}_V^{RR}$;
	
	\item cases with four vector operators: (XV) $\mathcal{O}_V^{LL}$, $\mathcal{O}_V^{LR}$, $\mathcal{O}_V^{RL}$, and $\mathcal{O}_V^{RR}$. 
\end{enumerate}
And we assume no scalar and tensor operators are activated. 
  
For a simple demonstration, we once again consider the benchmark beam energy $E=12$~GeV. 
Supposing the scattering experiment is run twice with the electron beam left- ($e_L$) and right-handed ($e_R$) polarized accordingly, 
we collect the complete results for the cases I--XV in Table~\ref{tab:model-disting}.  
And for the sake of simplicity, we focus on $Q^2\simeq 6$ GeV$^2$, so that $P_T\simeq 0.5\, (0)$ for the NP scenario $g^{LL}_V$ ($g^{LR}_V$), 
as shown in Fig.~\ref{fig:NonMixCombine}. 
The constant $c$, depending on $|g_V^{LL}|^2$, $|g_V^{LR}|^2$, and $\text{Re}[g_V^{LL}g_V^{LR*}]$, is 
expected to be away from 0 and 0.5 in general. 
Now based on their results of $P_T$ with $e_L$ and $e_R$, those NP scenarios can be 
divided into nine groups: \{II, III, VIII\}, \{I, VI\}, \{V, XI\}, \{IV, IX\}, \{X, XIV\}, \{VII\}, \{XII\}, \{XIII\}, and \{XV\}. 
Already, the 4 NP scenarios listed accordingly in the last 4 groups can be distinguished from each other and the rest of the NP models.  
To further disentangle the remaining 11 cases, one can simply check the events of the scattering process for the two experiments. 
Take the first group \{II, III, VIII\} for an illustration. 
If one observes events with the $e_L$ beam, 
while none with the $e_R$ beam, the NP scenario is the case II; 
if the observation is completely opposite, it belongs to the case III;
if one observe events in both times, it shall be the case VIII. 
In this way, the NP scenarios II, III, and VIII are disentangled. 
Following the same procedure, one can also disentangle the rest 8 cases. 
 
Let us conclude this subsection by making the following comment. 
In Ref.~\cite{Lai:2022ekw}, seven LQ models $S_1$, $R_2$, $S_3$, $U_3$, $\tilde{V}_2$, $\tilde{U}_1$ and 
$V_2$ were disentangled by using the polarized scattering process $e^-p\to e^-\Lambda_c$; additional processes are 
needed to further distinguish $S_3$ from $U_3$, since they yield the same effective operator $\mathcal{O}_V^{LL}$. 
To achieve this goal, the scalar and tensor contributions 
in the LQ models $S_1$ and $R_2$ were ignored, which was a reasonable assumption, since the 
WCs of the scalar and tensor operators are closed related in the two LQ models
and the former are stringently constrained by the leptonic $D$-meson decays~\cite{Lai:2021sww,Belle:2010ouj}.     
For the WCs of the $\mathcal{L}_{\text{eff}}$ in Eq.~\eqref{eq:Leff}, on the other hand, 
none of them has been constrained more stringently than the others, as to be shown in Sec.~\ref{sec:Prospect}. 
Thus it will be more involved to fully disentangle the LQ models.  
Nonetheless, the LQ models $U_3$ ($S_3$), $\tilde{V}_2$, $\tilde{U}_1$ and $V_2$ are already distinguished through the 
identification mechanism, since they correspond to the cases I, II, IV, and III, respectively. 
Moreover, the scalar LQ models $S_1$ and $R_2$ can be disentangled from the others, 
because they generate a non-zero $P_P$---induced by the interference among the vector, scalar, and tensor 
operators---whereas the others do not.

\subsection{Impacts from the form factors}
\label{sec:obsdiffform}

\begin{figure*}[t]
	\centering
	\includegraphics[width=0.97\textwidth]{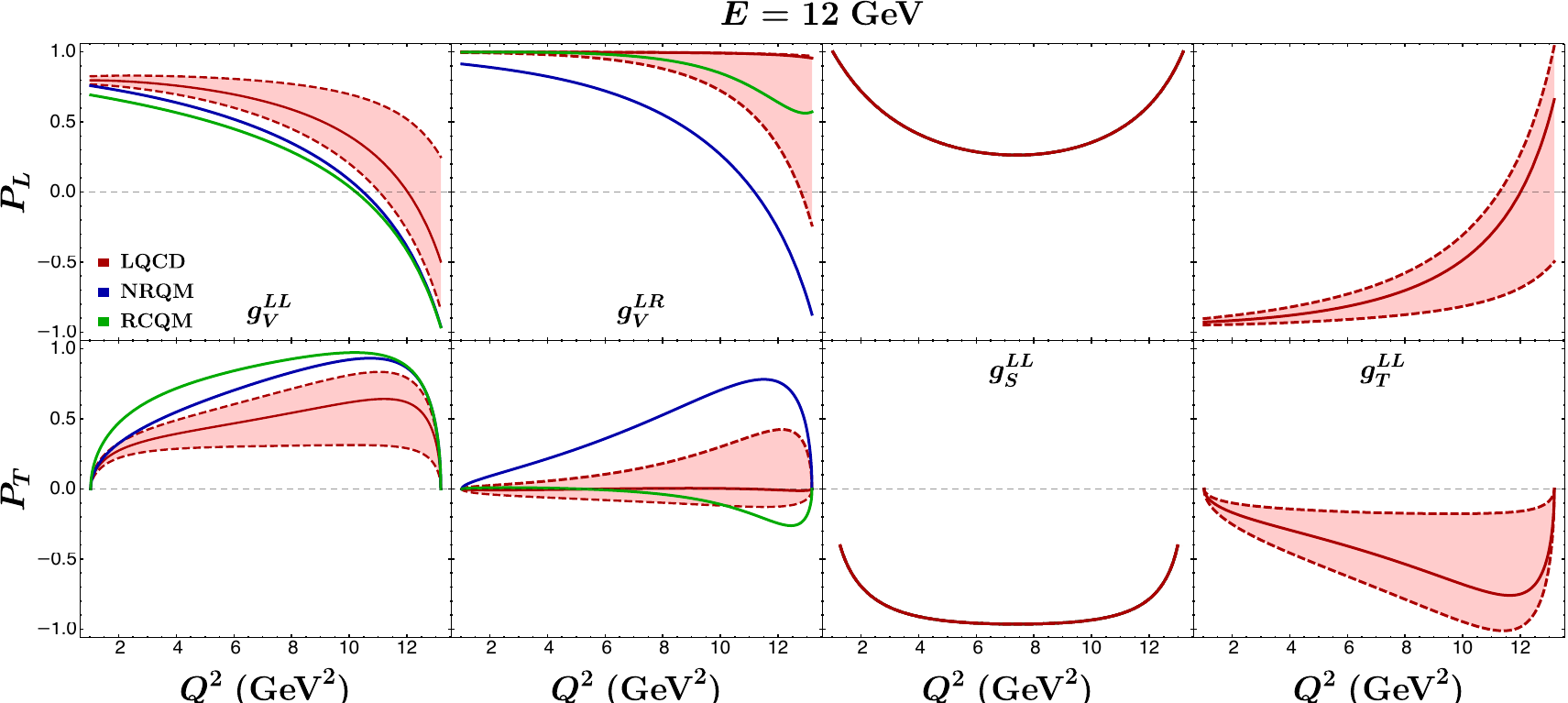}
	\caption{The polarization $P_{L,T}$ as a function of $Q^2$, predicted with the form 
		factors calculated in LQCD (red), NRQM (blue), and RCQM (green), respectively. Note that the 1$\sigma$-level statistical uncertainties of the 
		form factors in LQCD have been propagated to all the observables, as denoted by the outer red regions. In addition, the $P_{L,T}$ denoted by the three colors overlap perfectly for the NP $g^{LL}_S$, while for the NP $g^{LL}_T$ only the LQCD results are presented.} 
	\label{fig:DiffForm} 
\end{figure*}

As mentioned in Sec.~\ref{subsec:Cross section}, the primary results of the form factors we adopt 
are from the LQCD calculations, since they provide us with an error estimation~\cite{Meinel:2017ggx}. 
Yet our calculation has only involved the central values of these inputs so far. 
Given that the uncertainties of the form factors can affect the theoretical predictions significantly~\cite{Sobczyk:2019uej,Kong:2023kkd}, 
they should be taken into account in the analyses of this work, too.  
In addition, the ambiguity arising in extrapolating the form factors to positive $q^2$
with different parametrization schemes can have 
a huge impact on theoretical predictions~\cite{DeLellis:2004ovi,Sobczyk:2019uej,Kong:2023kkd}; 
this impact can even dwarf the one induced by the error propagation of the form-factor uncertainties. 
Since our analysis is based on the same extrapolation, 
we should also examine if the same observation applies to the $P_{a}$ in various NP scenarios in this work.   

As a simple illustration, let us focus again on the first simple case, i.e., only one WC is activated at a time. 
We evaluate in Fig.~\ref{fig:DiffForm} the $P_{L,T}$ with the form factors calculated in LQCD (red), NRQM (blue), and RCQM (green), respectively.\footnote{We do not present the results with the form factors calculated in MBM, because both MBM and NRQM employ the 
dipole form for the $q^2$ dependence of the form factors~\cite{Perez-Marcial:1989sch,Avila-Aoki:1989arc} (see also Refs.~\cite{Sobczyk:2019uej,Kong:2023kkd} for details).} 
And we also take account of the $1\sigma$-level statistical uncertainties of the form factors in the LQCD case. 
Following the same argument made in Fig.~\ref{fig:NonMixCombine}, we only present the NP scenarios $g_V^{LL}$, $g_V^{LR}$, $g_S^{LL}$, 
and $g_T^{LL}$ in this figure. It can be seen that in the case of $g_V^{LL}$, the LQCD always predicts the largest $P_L$ but the smallest $P_T$, 
whereas the situation is completely opposite in the RCQM. For the NP $g_V^{LR}$, on the other hand, the RCQM predicts compatible 
$P_{L,T}$ with the LQCD at the $1\sigma$ level, while the NRQM produces the smallest $P_L$ but the largest $P_T$. 
Contrary to the two NP scenarios above, situation in the NP $g^{LL}_S$ is much simpler. Since both the $P_L$ and $P_T$ are 
independent of the WC $g^{LL}_S$, all of the three models generate the same $P_{L,T}$; for details of the $P_{L,T}$ in this case 
we refer the reader to Appendix~\ref{app:numerator}. 
Finally, for the NP $g^{LL}_T$, only the LQCD results are presented, since the relevant form factors have not been 
calculated in other two models.  
   
Despite the complicated behaviors of the $P_{L,T}$ calculated with various form-factor parametrization schemes 
in the NP scenarios, two overall observations can be made easily. The first one is that 
the uncertainties of the $P_{L,T}$ from the error propagation of the 
statistical uncertainties of the form factors in the LQCD increase along with the $Q^2$---these uncertainties of the $P_{L,T}$ would 
become even larger if the systematical ones were taken into account, too. Such a pattern is closely related to the 
behaviors of the form factors with respect to the $Q^2$~\cite{Kong:2023kkd}. The other observation is that the uncertainties of the $P_{L,T}$   
due to the different schemes overwhelm the ones induced by the error propagation of the form-factor uncertainties. 
Clearly, these two observations are consist with the results of Ref.~\cite{Kong:2023kkd}.

The uncertainties of the $P_{L,T}$ will certainly affect the model identification in Sec.~\ref{subsec:model-id}. 
We have shown that a distinct difference between the predicted $P_{T}$ in the NPs $g^{LL}_V$ and $g^{LR}_V$ at a certain $Q^2$ is 
one of the key ingredients in the success of NP model identification. 
Clearly, such a difference will be blurred when the uncertainties discussed above are taken into account, as shown in Fig.~\ref{fig:DiffForm}. 
We note that the largest possible $P_{T}$ in the NP $g^{LR}_V$ (denoted by the blue curve)  
exceeds the smallest one in the NP $g^{LL}_V$ (denoted by the lower, red dashed curve) at $Q^2\simeq 5$ GeV$^2$, 
indicating that the identification mechanism will fail in the $Q^2>5$ GeV$^2$ regions. Even if one focuses only on the LQCD results, 
the mechanism will also fail at $Q^2\geq 10$ GeV$^2$, 
since the largest possible $P_{T}$ in the NP $g^{LR}_V$ (denoted by the upper, red dashed curve)  
exceeds the smallest one in the NP $g^{LL}_V$ (denoted by the lower, red dashed curve) in this regions. 
The situation will certainly becomes even tougher for this mechanism 
if the systematical uncertainties of the form factors are considered, too.

\section{Prospect and constraints}
\label{sec:Prospect}

We now evaluate the prospect for observing the process $e^- p \to \Lambda_c \tau^-(\to \pi^-\nu_{\tau},\rho^-\nu_{\tau},\ell^-\bar{\nu}_{\ell}\nu_{\tau})$ in the framework of 
the $\mathcal{L}_{\text{eff}}$. As mentioned in Sec.~\ref{subsec:spin density}, 
this cascade process can be broken down into the QE scattering $e^- p\to \Lambda_c \tau^-$
and the successive decays $\tau^- \to \pi^-\nu_{\tau},\rho^-\nu_{\tau},\ell^-\bar{\nu}_{\ell}\nu_{\tau}$. 
Given that the former is a cLFV process, the cascade process would certainly suffer from 
low experimental statistics. To alleviate this problem, we consider all the successive 
decays of the lepton $\tau$ in this process and integrate  
over all the variables in Eq.~\eqref{eq:sigma02}; in other words, we shall consider the total cross section. 
   
We propose searching for the cascade process through a fixed-target experiment, whose event rate is given by
\begin{equation}\label{eq:event}
	\frac{dN}{dt}\!=\!\mathcal{L}\sigma\,.
\end{equation}
The luminosity $\mathcal{L}$ reads $\mathcal{L}=IL\rho_T$, with $I$ being the beam intensity, and 
$L$ and $\rho_T$ denoting the length and the number density of the proton target, 
respectively.
The total cross section $\sigma$ is related to the $\sigma_s$, the unpolarized one of the QE scattering process, 
through the relation $\sigma=\sum_d \mathcal{B}_d\sigma_s$, as indicated by Eq.~\eqref{eq:total_cross}. 

From Eq.~\eqref{eq:event}, it is clear that to perform a concrete event-rate estimation for the scattering experiment, 
we have to choose a suitable experimental setup and assign proper values to the WCs.
For the former, we prefer the 12 GeV CEBAF at 
JLab~\cite{Arrington:2021alx}---specifically, the proposed Solenoidal Large Intensity Device (SoLID)~\cite{JeffersonLabSoLID:2022iod}, 
which is feasible to study subthreshold $J/\psi$ production from liquid hydrogen~\cite{K.Hafidi} 
or deuteron~\cite{Liu:2023htv} (all proton targets). 
We note that the produced $J/\psi$ is reconstructed through its decay into a lepton pair ($e^+e^-$) in these experiments~\cite{K.Hafidi,Liu:2023htv}. 
By analogy, we expect (or assume) the produced $\Lambda_c$ of the cascade process 
can also be reconstructed through its decay products and 
the visible decay products ($\rho,\pi,\ell$) of the $\tau$ can be detected. 
Thus, we choose the same experimental setup as Ref.~\cite{Liu:2023htv}, i.e., the 12 GeV electron beam with intensity 
up to 1.25 $\mu$A and the liquid deuteron target, which together make up the overall luminosity $\mathcal{L}=1.2\times10^{37}$ cm$^{-2}$s$^{-1}$.
With the beam energy $E=12$ GeV, the $\sigma$ can be now written as 
\begin{align}\label{eq:Nsigma}
	\sigma\!=\!&\frac{\sum_d\mathcal{B}_dG_F^2}{32\pi F E m_p}\big\{1.86\left(|g_{V}^{LL}|^2+|g_{V}^{RR}|^2\right)\nonumber\\[0.15cm]
	&+1.20\left(|g_{V}^{LR}|^2+|g_{V}^{RL}|^2\right)
	+13.60\left(|g_{T}^{LL}|^2+|g_{T}^{RR}|^2\right)\nonumber\\[0.15cm]
	&+0.51\left(|g_{S}^{LL}|^2+|g_{S}^{RR}|^2+|g_{S}^{LR}|^2+|g_{S}^{RL}|^2\right) \nonumber\\[0.15cm]
	&\!+\!1.01\text{Re}\left[g_{V}^{LL}g_{S}^{LL*}\!+\!g_{V}^{RR}g_{S}^{RR*}\!+\!g_{V}^{LR}g_{S}^{LR*}\!+\!g_{V}^{RL}g_{S}^{RL*}\right]	\nonumber\\[0.15cm]
	&\!+\!0.56\text{Re}\left[g_{V}^{LL}g_{S}^{LR*}\!+\!g_{V}^{RR}g_{S}^{RL*}\!+\!g_{V}^{LR}g_{S}^{LL*}\!+\!g_{V}^{RL}g_{S}^{RR*}\right]\nonumber\\[0.15cm]
	&+0.65\text{Re}\left[g_{S}^{LL}g_{S}^{LR*}+g_{S}^{RR}g_{S}^{RL*}\right] \nonumber\\[0.15cm]
	&+0.90\text{Re}\left[g_{V}^{LL}g_{V}^{LR*}+g_{V}^{RR}g_{V}^{RL*}\right] \nonumber\\[0.15cm]
	&-4.39\text{Re}\left[g_{V}^{LL}g_{T}^{LL*}+g_{V}^{RR}g_{T}^{RR*}\right]	\nonumber\\[0.15cm]
	&+0.36\text{Re}\left[g_{V}^{LR}g_{T}^{LL*}+g_{V}^{RL}g_{T}^{RR*}\right]\nonumber\\[0.15cm]
	&-1.98\text{Re}\left[g_{S}^{LL}g_{T}^{LL*}+g_{S}^{RR}g_{T}^{RR*}\right] \nonumber\\[0.15cm]   
	&+0.38\text{Re}\left[g_{S}^{LR}g_{T}^{LL*}+g_{S}^{RL}g_{T}^{RR*}\right]\big\}\!\times\!10^3\ \text{GeV}^6\,.
\end{align}  
For the general expression of the $d\sigma/dq^2$, we refer the reader to Appendix~\ref{app:denominator}.

We now turn to the values of the WCs. 
Thus far, available constraints on the WCs are solely set by the analyses of high-$p_T$ dilepton invariant mass tails 
in $pp\to \tau^\pm e^\mp$~\cite{Angelescu:2020uug}. 
Taking account of the renormalization group (RG) running effects neglected in 
Ref.~\cite{Angelescu:2020uug},\footnote{Detailed discussion about the RG effects for the $\ell u\to \ell^\prime c$ 
process in the framework of the $\mathcal{L}_{\text{eff}}$ can be found in Ref.~\cite{Lai:2021sww}.} and 
assuming only one WC contributes to the process at a time, 
we list in Table~\ref{table:constraint} the upper bounds at the $90\%$ confidence level (CL). 
For an optimistic event-rate estimation, we take those upper bounds for the WCs. 
Note that since $O_S^{LR}$ and $O_S^{RL}$ were not included in the $\mathcal{L}_{\text{eff}}$ 
in Ref.~\cite{Angelescu:2020uug}, no constraints on their WCs $g^{LR}_S$ and $g^{RL}_S$ has been set yet. 

  \begin{table}[t]
	\renewcommand\arraystretch{1.6} 
	\tabcolsep=0.07cm
	\centering
	\caption{Summary of the upper bounds on the WCs $g$ from the high-$p_T$ dilepton invariant mass tails (at $90\%$ CL)
		and the $e^- p \to \Lambda_c \tau^-(\to \pi^-\nu_{\tau},\rho^-\nu_{\tau},\ell^-\bar{\nu}_{\ell}\nu_{\tau})$
		scattering processes in the framework of the $\mathcal{L}_{\text{eff}}$ by Eq.~\eqref{eq:Leff}. Note that a common factor $10^{-3}$ has been factored out, and constraints on $g^{LR}_S$ and $g^{RL}_S$ have not been set by the high-$p_T$ dilepton invariant mass tails~\cite{Angelescu:2020uug}.}
	\begin{tabular}{ccccc}
		\hline\hline
		Processes &$\big|g^{LL,RR}_V\big|$ &$\big|g^{LR,RL}_V\big|$ & $\big|g^{LL,RR}_S\big|$& $\big|g^{LL,RR}_T\big|$  \\ \hline
		$pp\to \tau^\pm e^\mp$~\cite{Angelescu:2020uug} & 5.27& 5.27 & 11.1 &2.11 \\
		$e^-p\to\Lambda_c\tau^-(\to d \nu_{\tau})$ & 1.40& 1.74 & 2.67&0.52 \\
		\hline \hline
	\end{tabular}
	\label{table:constraint} 
\end{table} 

\begin{table}[t]
	\renewcommand\arraystretch{1.6} 
	\tabcolsep=0.55cm
	\centering
	\caption{Summary of the event-rate estimations for the $e^- p \to \Lambda_c \tau^-(\to \pi^-\nu_{\tau},\rho^-\nu_{\tau},\ell^-\bar{\nu}_{\ell}\nu_{\tau})$ processes in the first simple case  
	discussed in Sec.~\ref{sec:obs}. Note that we have set $\sum_d \mathcal{B}_d=70\%$ and assumed no event is observed after 1 year's running of 
	the experiment. And the event rate is given in units of $N$/yr.}
	\begin{tabular}{llll}
		\hline\hline
		$g^{LL,RR}_V $ &$g^{LR,RL}_V$ & $g^{LL,RR}_S$& $g^{LL,RR}_T$  \\ \hline
		14.16 & 9.13 & 17.22&16.59 \\
		\hline \hline
	\end{tabular}
	\label{table:event} 
\end{table}

Supposing a run time of 1 year (yr) and 100\% detecting efficiency of the visible final particles, 
we evaluate the expected event rates in units of number 
per year ($N$/yr) for the first simple case discussed in Sec.~\ref{sec:obs}. The 
final results are listed in Table~\ref{table:event}. 
It can be seen that comparable event rates can be expected for the NP $g_V^{LL,RR}$, $g^{LL,RR}_S$, and $g^{LL,RR}_T$. 
Compared with the NP $g_V^{LL,RR}$, the predicted event rate for the NP $g_V^{LR,RL}$, on the other hand, is slightly smaller, 
though its WC takes the same value as the $g_V^{LL,RR}$. 
The underlying reason is their different reduced amplitude squared $\mathcal{A}$, as explicitly shown in Eq.~\eqref{eq:Nsigma} 
and in Eq.~\eqref{eq:amplitude2} for general cases.   
In fact, this is an unique character of the QE scattering process mediated by $\ell u\to \ell^{(\prime)} c$ 
in our framework; one can see that the constraints on the WCs $g_V^{LL,RR}$ and $g_V^{LR,RL}$ set by the charmed-hadron weak decays and the high-$p_T$ dilepton invariant mass tails are always the same (see, e.g., Refs.~\cite{Lai:2021sww,Lai:2022ekw}). 

The expected event rates in Table~\ref{table:event}, though non-zero, are still not big enough to 
extract the $P_a$ of the $\tau$ with good statistics. One way out is to consider the facilities with higher luminosity. 
We note that even higher luminosity ($10^{38}$ to $10^{39}$ cm$^{-2}$s$^{-1}$) 
can be achieved in Hall C at JLab in the upcoming era~\cite{Arrington:2021alx}. On top of that, both liquid hydrogen and deuterium targets 
and polarized electron beams are available. If our proposed experiment can be conducted in this hall, 
not only can the expected event rate be further enhanced, extraction of the $P_a$ and NP model identification may become promising, too.  

The event rates in Table~\ref{table:event} are what one can expect in the best scenario, 
because they are carried out with the upper limits of the WCs. 
Therefore, it is still possible that no event of the cascade process will be observed 
with our preferred experimental setup in future. 
If such a worst-case scenario indeed happens, our proposed scattering experiment can still set competitive constraints on 
the WCs, as shown in Table~\ref{table:constraint}, 
where we have also assumed a run time of 1 yr and 100\% detecting efficiency of the visible final particles. 
Surely more stringent constraints can be obtained if the experiment can be conducted in the Hall C at JLab in the future. 

However, to fully pinpoint the WCs of the $\mathcal{L}_{\text{eff}}$ in Eq.~\eqref{eq:Leff}, clearly more observables are needed. 
Recall that we had to make the assumption, i.e., only one WC contributes at a time, 
to get the constraints listed in Table~\ref{table:constraint}. 
Even if the purely tauonic $D$-meson decays were measured in the future, they would be still not enough, since they are 
only sensitive to the axial and pseudo-scalar four-fermion operators of the $\mathcal{L}_{\text{eff}}$. 
Since the polarization components $P_a$ of the $\tau$ involve all the effective operators of the $\mathcal{L}_{\text{eff}}$ (see Appendix~\ref{app:numerator}), they can partially fill the gap, though they  
are more difficult to measure than the cross section, as already indicated in Table~\ref{table:event}. 
One may also consider the QE scattering with both the electron beam and proton target polarized,  
since polarized QE scattering processes can also shed light on the NP Dirac structures~\cite{Lai:2022ekw}.    

Finally we would like point out that our predictions in Tables~\ref{table:constraint} and \ref{table:event} 
can be weakened by the non-100\% detecting efficiency of the produced particles, such as the $\Lambda_c$ baryon, 
since it may be hard to keep track of all its decay products. Thus a sophisticated detecting system for these particles is crucial. 
Our results can also be affected significantly by the uncertainties of the form factors. 
As demonstrated in Sec.~\ref{sec:obsdiffform}, they can exert huge influence on the predictions of $P_{L,T}$, 
which in turn affect the efficiency of the NP model identification. By analogy, they will certainly 
have a huge impact on the event-rate estimation and the constraints on the WCs. All this calls for a more concrete form factor 
parametrization scheme and better control of the uncertainties in future LQCD calculations.

\section{Conclusion}
\label{sec:con}

We have proposed searching for NP through the polarization of the $\tau$ lepton in the 
QE scattering process $e^- p \to \Lambda_c \tau^-$. As analyzing modes for the $\tau$ polarization, 
we have considered its four dominant decays 
$\tau^-\to \pi^-\nu_{\tau}$, $\tau^-\to \rho^-\nu_{\tau}$, and $\tau^- \to \ell^-\bar{\nu}_{\ell}\nu_{\tau}$, 
and demonstrated that its three polarization components $P_a$ can be extracted from the 
analyses of the kinematics of its visible decay products.

Working in the framework of a general low-energy effective Lagrangian given by Eq.~\eqref{eq:Leff}, 
we have first performed a detailed analysis of the $P_a$  
and discovered some interesting patterns. 
First, the $P_a$ in a NP scenario with only left-handed lepton currents differ by a minus sign to that in another NP scenario with 
only right-handed lepton currents, as shown in Tables~\ref{tab:Denominator} and \ref{tab:Numerator}. 
Second, due to missing the $\varepsilon_{\{k\}\{k^\prime\}\{s_a\}\{p\}}$, 
the perpendicular component $P_P$ vanishes in the simplest NP scenario, i.e., only one WC is activated at a time, 
and in the pure-vector case, i.e., only the vector operators $\mathcal{O}_V$ are activated. 
Third, the $P_{a}$ can be independent of the WCs in certain NP scenarios, e.g., 
when only $\mathcal{O}^{LL}_S$ and $\mathcal{O}^{LR}_S$ 
($\mathcal{O}^{RR}_S$ and $\mathcal{O}^{RL}_S$) are activated simultaneously. 
Based on these patterns, we have then shown that 15 NP models constructed with only vector operators  
can be distinguished from each other by using the longitudinal polarized electron beam and measuring the transverse component 
$P_T$ of the $\tau$. 
We have also explored the impacts of the form factors in various aspects and 
observed that large uncertainties of the polarization observables arise from using 
the different schemes and dwarf that from the error propagation of the 
form factors, which is consist with our previous result in Ref.~\cite{Kong:2023kkd}, though it was focused on 
the neutrino QE scattering process $\nu_{\tau} n \to \Lambda_c \tau^-$.

We have finally performed a simple event-rate estimation for the cascade process $e^- p\! \to \!\Lambda_c \tau^-(\!\to\! \pi^-\nu_{\tau},\rho^-\nu_{\tau},\ell^-\bar{\nu}_{\ell}\nu_{\tau})$ in the simplest NP scenarios with our preferred experimental setup and 
the upper boundary of the WCs from the analyses of high-$P_T$ dilepton invariant mass tails. 
Although promising event rate can be expected for this process, it remains challenging to 
extract the $P_a$ of the $\tau$ with good statistics, which thus calls for an experimental setup of even higher luminosity.
Nonetheless, even in the worst-case scenario—no signals is observed at all—we have shown in a model-independent 
way that the low-energy scattering process can provide competitive constraints, in comparison with the high-$P_T$ 
dilepton invariant mass tails.

\section*{Acknowledgments}

We are grateful to Dong-Hui Zheng for helpful discussions about the spin density matrices.
This work is supported by the National Natural Science Foundation of China under Grants No. 12135006 and No. 12275067, 
Natural Science Foundation of Henan Province under Grant No. 225200810030, Excellent Youth Foundation of Henan Province 
under Grant No. 212300410010, as well as the Pingyuan Scholars Program under Grand No.5101029470306.  

\appendix

\section{Calculation of the spin density matrices}
\label{appendix:spin density}

\subsection{$\tau$ decay density matrix}
\label{appendix:tdecay}

The helicity amplitude for the $\tau^-\to \pi^- \nu_{\tau}$ decay is given by 
\begin{align}\label{eq:pi}	
	\mathcal{M}_{\lambda}(\tau^-\to\pi^-\nu_{\tau})=i\sqrt{2}G_FV^*_{ud}f_{\pi}p^{\mu}_{\pi}\bar{u}_{\nu_{\tau}}\gamma_{\mu}
	P_Lu_{\tau}(\lambda)\,,
\end{align}
where $f_{\pi}$ is the $\pi$ decay constant and $\lambda=\{1/2,-1/2\}$ characterizes the helicity of the $\tau$. 
With the amplitude above, we compute the spin density decay matrix of the $\tau$ in this 
decay process~\cite{Haber:1994pe},
\begin{align}\label{eq:rhoDpi}
	(\rho^D_{\pi})_{\lambda\lambda^{\prime}}&=\mathcal{M}_{\lambda}\mathcal{M}^*_{\lambda^\prime} \nonumber \\[0.02cm]	&=\frac{|G|^2}{4}\Big[2m^2_{\tau}(m^2_{\tau}\!-\!m^2_{\pi})\delta_{\lambda\lambda^\prime}\!-\!4m^3_{\tau}(p_{\pi}\cdot s^a) \sigma^a_{\lambda\lambda^\prime} \Big]\,,
\end{align}
where $G=i\sqrt{2}G_FV^*_{ud}f_{\pi}$, $\sigma^a$ ($a=1,2,3$) are the Pauli matrices, and 
the subscript $\pi$ in $\rho^D_{\pi}$ indicates the $\tau^-\to \pi^- \nu_{\tau}$ decay channel. 
The spacelike four-vector $s^a_{\mu}$  
are the spin vectors of the $\tau$, which together with the $k^\prime/m_{\tau}$ form 
an orthonormal set~\cite{Haber:1994pe},
\begin{align}
	\frac{k^\prime}{m_{\tau}}\cdot s^a&=0\,, \quad
	s^a\cdot s^{a\prime}=-\delta^{aa^\prime}\,,\nonumber \\[0.02cm] 
	\sum_a s^a_{\mu}\cdot s^a_{\nu}&=-g_{\mu\nu}+\frac{k^{\prime}_{\mu}k^{\prime}_{\nu}}{m^2_{\tau}}\,.
\end{align}  
Note that in computing the density matrix $\rho^D_{\pi}$ in Eq~\eqref{eq:rhoDpi}, as well as the rest of density matrices 
for production and decay of the $\tau$, we always employ the Bouchiat-Michel formulae~\cite{Bouchiat:1958yui,Michel:1959dvg,Haber:1994pe}:
\begin{align}\label{eq:BMformulea}
	u_{\lambda^\prime}(p)\bar{u}_{\lambda}(p)=\frac{1}{2}\left[\delta_{\lambda\lambda^\prime}+\sum_a\gamma_5\slashed{s}^a\sigma^a_{\lambda\lambda^\prime}\right](\slashed{p}+m)\,.
\end{align}

The helicity amplitude for the $\tau^-\to \rho^- \nu_{\tau}$ decay is written as 
\begin{align}
	\mathcal{M}^{\lambda_{\rho}}_{\lambda}(\tau^-\!\to\!\rho^-\nu_{\tau})\!=\! i\sqrt{2}G_FV^*_{ud}f_{\rho}m_{\rho}\varepsilon^{\mu*}_{\rho}(\lambda_{\rho})\bar{u}_{\nu_{\tau}}\gamma_{\mu}P_Lu_{\tau}(\lambda)\,,
\end{align} 
where $f_{\rho}$ and $\varepsilon^{\mu}_{\rho}(\lambda_{\rho})$ represent the decay constant and polarization vectors of the $\rho$ meson. 
Similar to the previous case, we also compute the spin density decay matrix of $\tau$ in this case,
\begin{align}\label{eq:rhoDrho}
	\left(\rho^D_{\rho}\right)_{\lambda\lambda^{\prime}}=&\frac{|G_{\rho}|^2}{2}\Big[ (m^2_{\tau}-m^2_{\rho})(m^2_{\tau}+2m^2_{\rho})\delta_{\lambda\lambda^\prime}\nonumber \\[0.02cm]
	 &-\sum_a 2(m^2_{\tau}-2m^2_{\rho})m_{\tau}(p_{\rho}\cdot s^a)\sigma^a_{\lambda\lambda^\prime}\Big]\,,
\end{align}
with $G_{\rho}=i\sqrt{2}G_FV^*_{ud}f_{\rho}$. 

From Eqs.~\eqref{eq:rhoDpi} and \eqref{eq:rhoDrho}, 
one can easily obtain the decay rates for these two channels, which read respectively as
\begin{align}
	\Gamma_{\tau}^{\pi}&=\Gamma(\tau^-\to\pi^-\nu_{\tau})=\frac{1}{2}\frac{1}{2m_{\tau}}\frac{|\pmb{p_{\pi}}|}{4m_{\tau}\pi}\tr[\rho^D_{\pi}] \nonumber\\[0.02cm]
	&=\frac{G^2_Ff^2_{\pi}|V_{ud}|^2m^3_{\tau}}{16\pi}\left(1-\frac{m^2_{\pi}}{m^2_{\tau}}\right)^2\,, \\[0.02cm]
	\Gamma_{\tau}^{\rho}&=\Gamma(\tau^-\to\rho^-\nu_{\tau})=\frac{1}{2}\frac{1}{2m_{\tau}}\frac{|\pmb{p_{\rho}}|}{4m_{\tau}\pi}\tr[\rho^D_{\rho}] \nonumber\\[0.02cm]
	&=\frac{G^2_Ff^2_{\rho}|V_{ud}|^2m_{\tau}}{16\pi}(m^2_{\tau}+2m^2_{\rho})\left(1-\frac{m^2_{\rho}}{m^2_{\tau}}\right)^2\,.
\end{align}
Based on these two decay rates, we can rewrite $\rho^D_{\pi,\rho}$ universally as 
\begin{align}\label{eq:tau2decay}
	(\rho^{D}_d)_{\lambda\lambda^{\prime}}
	\!=\!16\pi m_{\tau}\Gamma_{\tau}^{d}\left[\eta_d \delta_{\lambda\lambda^\prime}
	\!-\!\sum_a\chi_d (p_{d}\cdot s^a)\sigma^a_{\lambda\lambda^\prime} \right],
\end{align}
where
\begin{align}
	\eta_d=\frac{m^2_{\tau}}{m^2_{\tau}-m^2_{d}}\,, \quad \chi_d=\alpha_d\frac{2m^3_{\tau}}{(m^2_{\tau}-m^2_{d})^2} \,,
\end{align}
with 
\begin{align}
	\alpha_{d=\pi}=1\,, \qquad
	\alpha_{d=\rho}=\frac{m^2_{\tau}-2m^2_{\rho}}{m^2_{\tau}+2m^2_{\rho}}\,,
\end{align}

The helicity amplitude for the $\tau^-\to \ell^- \bar{\nu}_{\ell}\nu_{\tau}$ decays are given by 
\begin{align}
	\mathcal{M}_{\lambda}(\tau^-\!\to\! \ell^- \bar{\nu}_{\ell}\nu_{\tau})=i2\sqrt{2}G_F \bar{u}_{\nu}\gamma^{\mu}P_Lu_{\tau}(\lambda)
	\bar{u}_{\ell}\gamma_{\mu}P_Lv_{\bar{\nu}}\,, 
\end{align}
which leads to the spin density decay matrix of the $\tau$
\begin{align}
	(\rho^D_\ell)_{\lambda\lambda^{\prime}}=& 64G^2_F\Big[(p_{\ell}\cdot p_{\nu_{\tau}})(k^\prime\cdot p_{\bar{\nu}})\delta_{\lambda\lambda^\prime}\nonumber \\
	&-\sum_a m_{\tau}\sigma^{a}_{\lambda \lambda^{\prime}}(p_{\ell}\cdot p_{\nu_{\tau}})(s^a\cdot p_{\bar{\nu}}) \Big]\,.
\end{align}
After integrating over the phase space, we can get the decay rate as 
\begin{align}
	\Gamma_{\tau}^{\ell}&=\Gamma(\tau^-\to \ell^- \bar{\nu}_{\ell}\nu_{\tau})\nonumber \\[0.02cm]
	&=\frac{G^2_F m^5_{\tau}}{192\pi^3}(1-8y^2+8y^6-y^8-24y^4\ln y)\nonumber \\[0.02cm]
	&=\frac{G^2_F m^5_{\tau}}{192\pi^3}f(y)\,,  
\end{align}
where $y=m_{\ell}/m_{\tau}$. Following the same procedure in the previous two cases, we 
rewrite the $\rho^D_\ell$ as 
\begin{align}\label{eq:tau3decay}
	(\rho^D_\ell)_{\lambda\lambda^{\prime}}=&\frac{64\times 192\pi^3\Gamma_{\tau}^{\ell}}{m^5_{\tau} f(y)}\Big[(p_{\ell}\cdot p_{\nu_{\tau}})(k^\prime\cdot p_{\bar{\nu}})\delta_{\lambda\lambda^\prime}\nonumber \\[0.02cm]
	&-\sum_am_{\tau}\sigma^{a}_{\lambda \lambda^{\prime}}(p_{\ell}\cdot p_{\nu_{\tau}})(s^a\cdot p_{\bar{\nu}}) \Big]\,.
\end{align}

\subsection{$\tau$ production density matrix}
\label{appendixx:tprodt}

The amplitude $\mathcal{M}$ of the QE scattering process $e^-p\to \Lambda_c \tau$ can be generically written as
\begin{align}\label{eq:amplitude}
	\mathcal{M}=\frac{1}{v^2}\sum_\alpha\left(H^{L}_{\alpha}L^{L}_{\alpha}+H^{R}_{\alpha}L^{R}_{\alpha}\right)\,,
\end{align}
with $\alpha=S,\, V,\, T$. The hadronic matrix elements $H^{L,R}_a$ are defined, respectively, as
\begin{widetext}
\begin{align}
	H^L_S&=\frac{1}{2}\langle \Lambda_c(p^\prime,s^\prime)|\left[\left(g^{LL}_S+g^{LR}_S\right)\bar{c}u
	+\left(g^{LR}_S-g^{LL}_S\right)\bar{c}\gamma^5u\right]|p(p,s)\rangle\,, \\[0.02cm]
	H^R_S&=\frac{1}{2}\langle \Lambda_c(p^\prime,s^\prime)|\left[\left(g^{RR}_S+g^{RL}_S\right)\bar{c}u
	+\left(g^{RR}_S-g^{RL}_S\right)\bar{c}\gamma^5u\right]|p(p,s)\rangle\,, \\[0.02cm]
	H^{L\mu}_V&=\frac{1}{2}\langle \Lambda_c(p^\prime,s^\prime)|\left[\left(g^{LL}_V+g^{LR}_V\right)\bar{c}\gamma^{\mu} u+\left(g^{LR}_V-g^{LL}_V\right)\bar{c}\gamma^{\mu}\gamma^5u\right]|p(p,s)\rangle\,, \\[0.02cm]
	H^{R\mu}_V&=\frac{1}{2}\langle \Lambda_c(p^\prime,s^\prime)|\left[\left(g^{RR}_V+g^{RL}_V\right)\bar{c}\gamma^{\mu} u+\left(g^{RR}_V-g^{RL}_V\right)\bar{c}\gamma^{\mu}\gamma^5u\right]|p(p,s)\rangle\,,  \\[0.02cm]
	H^{L\mu\nu}_T&=g^{LL}_T\langle\Lambda_c(p^\prime,s^\prime)|\bar{c}\sigma^{\mu\nu} u |p(p,s)\rangle\,, \\[0.02cm]
	H^{R\mu\nu}_T&=g^{RR}_T\langle\Lambda_c(p^\prime,s^\prime)|\bar{c}\sigma^{\mu\nu} u |p(p,s)\rangle\,,
\end{align}
\end{widetext}
which can be evaluated by using the form factors discussed in Refs.~\cite{Lai:2021sww,Lai:2022ekw,Kong:2023kkd}. 
Meanwhile, the leptonic matrix element $L^{L,R}_a$ are correspondingly given by
\begin{align}
	L^L_S&=\bar{u}(k^{\prime},\lambda)P_Lu(k,r)\,,\\[0.02cm]
	L^R_S&=\bar{u}(k^{\prime},\lambda)P_Ru(k,r)\,,\\[0.02cm]
	L^L_{V\mu}&=\bar{u}(k^{\prime},\lambda)\gamma_{\mu} P_Lu(k,r)\,,\\[0.02cm]
	L^R_{V\mu}&=\bar{u}(k^{\prime},\lambda)\gamma_{\mu}  P_Ru(k,r)\,,\\[0.02cm]
	L^L_{T\mu\nu}&=\bar{u}(k^{\prime},\lambda)\sigma_{\mu\nu} P_Lu(k,r)\,,\\[0.02cm]
	L^R_{T\mu\nu}&=\bar{u}(k^{\prime},\lambda)\sigma_{\mu\nu} P_Ru(k,r)\,,
\end{align}
where $r$ and $s$ ($r^\prime$ and $s^\prime$) denote the helicities of initial (final) lepton and baryon, respectively.  
Based on the amplitude $\mathcal{M}$ in Eq.~\eqref{eq:amplitude}, we can write the $\tau$ production spin density matrix as 
\begin{align}\label{eq:rhoP}
	\rho^P_{\lambda,\lambda^{\prime}}&=\sum_{r,s,s^\prime} \mathcal{M}(\lambda) \mathcal{M}^*(\lambda^\prime)\nonumber \\[0.02cm]
	&=\frac{1}{v^4}\sum_{r,s,s^\prime}\sum_{\alpha,\beta}\Big\{H^{L}_{\alpha}H^{L*}_{\beta}L^{L}_{\alpha}(\lambda)L^{L*}_{\beta}(\lambda^\prime)\nonumber \\[0.02cm]
	&\qquad+H^{R}_{\alpha}H^{R*}_{\beta}L^{R}_{\alpha}(\lambda)L^{R*}_{\beta}(\lambda^\prime)\Big\}\,.
\end{align} 
Note that in deriving the second equation in Eq.~\eqref{eq:rhoP}, we have used the following relation, 
\begin{align}
	L^{L}_{\alpha}(\lambda)L^{R*}_{\beta}(\lambda^\prime)=0\,,
\end{align}  
which is valid due to the negligible electron mass in this process.  
To obtain the concrete expression of $\rho^P_{\lambda,\lambda^{\prime}}$, 
one has to deal with $L^{L,R}_{\alpha}(\lambda)L^{L,R*}_{\beta}(\lambda^\prime)$, 
which can be accomplished by using the Bouchiat-Michel formulae in Eq.~\eqref{eq:BMformulea}.

Finally, the density matrix $\rho^P$, as discussed in Sec.~\ref{subsec:spin density}, 
can be expanded in terms of Pauli matrices $\sigma^a$: 
\begin{align}
	\rho^P_{\lambda\lambda^\prime}=C\left(\delta_{\lambda\lambda^\prime}+ \sum_a\sigma^a_{\lambda\lambda^\prime}P_a\right)\,,
\end{align}
where $P_a$ represent the three components of the $\tau$ polarization 
vector $\mathcal{P}$~\cite{Kong:2023kkd}.

\subsection{Phase space integration}
\label{appendixx:phase}

We here provide some useful results and formulae to get the $d\sigma^P$ 
and $d\sigma^D$ in Eqs.~\eqref{eq:sigmaP} and \eqref{eq:sigmaD}, respectively.

First, for the $d\sigma^P$ the relevant phase space integration in the Lab frame is given by~\cite{Lai:2021sww}
\begin{align}
	\int d\Phi(k,p;p^\prime,k^\prime)&=\int \frac{d^3 \pmb{p}^\prime}{2E_{\Lambda_c}(2\pi)^3}\frac{d^3 \pmb{k}^\prime}{2E^\prime(2\pi)^3}\nonumber \\
	&\quad\times (2\pi)^4\delta^4(k\!+\!p-\!p^\prime\!-\!k^\prime)\nonumber \\
	&=\int \frac{1}{(2\pi)^2}\frac{d^3 \pmb{k}^\prime}{2E^\prime}\delta(p^{\prime2}-m^2_{\Lambda_c})\nonumber \\
	&=\frac{dq^2}{16\pi Em_p}\,,
\end{align}
where $q\equiv k-k^\prime=p^\prime-p$, and the $\delta$ function in the second step has been used to get rid of the angular integration of 
$\pmb{k}^\prime$. 

Second, for the $d\sigma^D$ with $d=\pi,\rho$, the involved phase space integration in the $\tau$ rest frame can be written as
\begin{align}
	\int d\Phi(k^\prime;p_d,p_\nu)&=\int \frac{d^3 \pmb{p}_d}{2E_{d}(2\pi)^3}\frac{d^3 \pmb{p}_{\nu}}{2E_{\nu}(2\pi)^3}\nonumber \\
	&\quad\times (2\pi)^4\delta^4(k^\prime-p_d-p_{\nu})\nonumber \\
	&=\frac{1}{(2\pi)^2}\int \frac{d^3 \pmb{p}_d}{2E_{d}}\frac{\delta\left(m_{\tau}-E_d
		-|\pmb{p}_d|\right)}{2|\pmb{p}_d|}\,.
\end{align}
For the $d\sigma^D$ with $d=\ell$, on the other hand, the following equation helps~\cite{Penalva:2021wye}, 
\begin{align}
	&\int \frac{d^3 \pmb{p}_{\nu}}{2|\pmb{p}_{\nu}|}\frac{d^3 \pmb{p}_{\bar{\nu}_\ell}}{2|\pmb{p}_{\bar{\nu}_\ell}|}
	\delta^4(q^\prime-p_{\nu}-p_{\bar{\nu}_\ell})p^{\alpha}_{\nu}p^{\beta}_{\bar{\nu}_\ell} \nonumber \\
	=&\frac{\pi q^{\prime2}}{24}\left(g^{\alpha\beta}+2\frac{q^{\prime\alpha}q^{\prime\beta}}{q^{\prime2}}\right)
	\theta(q^{\prime2})\,,
\end{align}
where $q^\prime=k^{\prime}-p_\ell$ and $\theta(...)$ denotes the step function.
Given $x=2(p_{\ell}\cdot k^\prime)/m^2_{\tau}$ and $y=m_{\ell}/m_{\tau}$ (see Sec.~\ref{sec:extract}), 
the step function $\theta(q^{\prime2})$ can be written as $\theta(1+y^2-x)$. 
It is also important to note that $x=2m_{\tau}E_{\ell}/m^2_{\tau}\geq 2m_{\ell}/m_{\tau}=2y$ in the $\tau$ rest frame;
hence the second step function in the scalar functions $\eta_{d=\ell}$ and $\chi_{d=\ell}$ in Eq.~\eqref{eq:scalar_function}.

\section{Amplitude squared of the QE scattering process}
\label{app:denominator}

The amplitude squared $|\mathcal{M}|^2 $ of the QE scattering process $e^-p\to \tau^-\Lambda_c$ mediated by the $\mathcal{L}_{\text{eff}}$
is connected to the diagonal element of the density matrix $\rho^P$ through the relation $|\mathcal{M}|^2=2C$, 
or equivalently $|\mathcal{M}|^2=\tr[\rho^P]$ as it should be.
With all the operators of $\mathcal{L}_{\text{eff}}$ in Eq.~\eqref{eq:Leff} taken into account,
we here write the $|\mathcal{M}|^2 $ as
\begin{widetext}
	\begin{align}\label{eq:amplitude2}
		v^4|\mathcal{M}|^2=&\left(|g_{V}^{LL}|^2+|g_{V}^{RR}|^2\right)\mathcal{A}_{V_{LL}\!-\!V_{LL}}
		+\left(|g_{V}^{LR}|^2+|g_{V}^{RL}|^2\right)\mathcal{A}_{V_{LR}\!-\!V_{LR}}
		+\left(|g_{T}^{LL}|^2+|g_{T}^{RR}|^2\right)\mathcal{A}_{T_{LL}\!-\!T_{LL}} \nonumber\\[0.15cm]
		&+\left(|g_{S}^{LL}|^2\!+\!|g_{S}^{RR}|^2\!+\!|g_{S}^{LR}|^2\!+\!|g_{S}^{RL}|^2\right)\mathcal{A}_{S_{LL}\!-\!S_{LL}}
		+2\text{Re}\left[g_{S}^{LL}g_{S}^{LR*}+g_{S}^{RR}g_{S}^{RL*}\right]\mathcal{A}_{S_{LL}\!-\!S_{LR}}
		\nonumber\\[0.15cm]
		&+2\text{Re}\left[g_{V}^{LL}g_{V}^{LR*}+g_{V}^{RR}g_{V}^{RL*}\right]\mathcal{A}_{V_{LL}\!-\!V_{LR}}
 		   +2\text{Re}\left[g_{V}^{LL}g_{S}^{LL*}+g_{V}^{RR}g_{S}^{RR*}+g_{V}^{LR}g_{S}^{LR*}+g_{V}^{RL}g_{S}^{RL*}\right]\mathcal{A}_{V_{LL}\!-\!S_{LL}} 	\nonumber\\[0.15cm]
 		&+2\text{Re}\left[g_{V}^{LL}g_{S}^{LR*}\!+\!g_{V}^{RR}g_{S}^{RL*}\!+\!g_{V}^{LR}g_{S}^{LL*}\!+\!g_{V}^{RL}g_{S}^{RR*}\right]\mathcal{A}_{V_{LL}\!-\!S_{LR}}
 		   +2\text{Re}\left[g_{V}^{LL}g_{T}^{LL*}+g_{V}^{RR}g_{T}^{RR*}\right]\mathcal{A}_{V_{LL}\!-\!T_{LL}} 	\nonumber\\[0.15cm]
    	&+2\text{Re}\left[g_{V}^{LR}g_{T}^{LL*}+g_{V}^{RL}g_{T}^{RR*}\right]\mathcal{A}_{V_{LR}\!-\!T_{LL}} 
 		   +2\text{Re}\left[g_{S}^{LL}g_{T}^{LL*}+g_{S}^{RR}g_{T}^{RR*}\right]\mathcal{A}_{S_{LL}\!-\!T_{LL}} \nonumber\\[0.15cm]   
 	    &+2\text{Re}\left[g_{S}^{LR}g_{T}^{LL*}+g_{S}^{RL}g_{T}^{RR*}\right]\mathcal{A}_{S_{LR}\!-\!T_{LL}}\,, 	 
	\end{align}
\end{widetext}
where the various patterns discovered in Table~\ref{tab:Denominator} have been taken into account. 
For the convenience of interested readers and future discussions, we provide the explicit expressions of the $\mathcal{A}$ on the right-hand side of Eq.~\eqref{eq:amplitude2} as 
\begin{widetext}
	\begin{align}
		\mathcal{A}_{V_{LL}\!-\!V_{LL}}=&\frac{m_\tau^2 \left(m_\tau^2\!-\!q^2\right)}{2q^4}\left[f_0^2 (m_{\Lambda_c}\!-\!m_p)^2 s_++g_0^2 (m_{\Lambda_c}+m_p)^2s_-\right]\!-\!\frac{m_\tau^2\left(m_{\Lambda_c}^2-m_p^2\right) }{q^4}		\nonumber\\[0.15cm]
		&\times\!(f_0 f_+\!+\!g_0 g_+)\left[4 E m_p q^2\!+\!\left(m_\tau^2\!-\!q^2\right) \left(m_{\Lambda_c}^2\!-\!m_p^2-q^2\right)\right]
		+\left[\frac{f_+^2 (m_{\Lambda_c}\!+\!m_p)^2}{2q^4 s_+}
		\right.\nonumber\\[0.15cm]
		&\left.+\frac{ g_+^2 (m_{\Lambda_c}\!\!-\!m_p)^2}{2q^4 s_-}\right]\Big{\{}4 m_p^2 q^4 \left(4 E^2\!-\!m_\tau^2+q^2\right)+\left(m_\tau^2\!-\!q^2\right) \left(m_{\Lambda_c}^2-m_p^2-q^2\right) 
		\nonumber\\[0.15cm]
		&\times\!\left[8 E m_p q^2\!+\!m_\tau^2 \left(m_{\Lambda_c}^2\!\!-\!m_p^2\!-\!q^2\right)\right]\Big{\}}
		\!+\!\left(\frac{f_\perp^2}{s_+}\!+\!\frac{ g_\perp^2}{s_-}\right)\Bigg\{8 E^2 m_p^2 q^2\!+\!\left(m_\tau^2\!-\!q^2\right) \nonumber\\[0.12cm]
		&\times\left[2 m_{\Lambda_c}^2 q^2-4 E m_p \left(m_p^2-m_{\Lambda_c}^2+q^2\right)-\left(m_{\Lambda_c}^2-m_p^2\right)^2+2 m_p^2 m_\tau^2-q^4\right]\Bigg\}
		\nonumber\\[0.15cm]
		&-2f_\perp g_\perp \left[4 E m_p q^2+\left(m_\tau^2-q^2\right) \left(m_{\Lambda_c}^2-m_p^2-q^2\right)\right]
		\,,
		\\[0.2cm]
		\mathcal{A}_{V_{LR}\!-\!V_{LR}}=&\frac{m_\tau^2 \left(m_\tau^2\!-\!q^2\right)}{2q^4}\left[f_0^2 (m_{\Lambda_c}\!-\!m_p)^2 s_++g_0^2 (m_{\Lambda_c}+m_p)^2s_-\right]\!-\!\frac{m_\tau^2\left(m_{\Lambda_c}^2-m_p^2\right) }{q^4}		\nonumber\\[0.15cm]
		&\times\!(f_0 f_+\!+\!g_0 g_+)\left[4 E m_p q^2\!+\!\left(m_\tau^2\!-\!q^2\right) \left(m_{\Lambda_c}^2\!-\!m_p^2-q^2\right)\right]
		+\left[\frac{f_+^2 (m_{\Lambda_c}\!+\!m_p)^2}{2q^4 s_+}
		\right.\nonumber\\[0.15cm]
		&\left.+\frac{ g_+^2 (m_{\Lambda_c}\!\!-\!m_p)^2}{2q^4 s_-}\right]\Big{\{}4 m_p^2 q^4 \left(4 E^2\!-\!m_\tau^2+q^2\right)+\left(m_\tau^2\!-\!q^2\right) \left(m_{\Lambda_c}^2-m_p^2-q^2\right) 
		\nonumber\\[0.15cm]
		&\times\!\left[8 E m_p q^2\!+\!m_\tau^2 \left(m_{\Lambda_c}^2\!\!-\!m_p^2\!-\!q^2\right)\right]\Big{\}}
		\!+\!\left(\frac{f_\perp^2}{s_+}\!+\!\frac{ g_\perp^2}{s_-}\right)\Bigg\{8 E^2 m_p^2 q^2\!+\!\left(m_\tau^2\!-\!q^2\right) 	\nonumber\\[0.15cm]
		&\times\left[2 m_{\Lambda_c}^2 q^2-4 E m_p \left(m_p^2-m_{\Lambda_c}^2+q^2\right)-\left(m_{\Lambda_c}^2-m_p^2\right)^2+2 m_p^2 m_\tau^2-q^4\right]\Bigg\}
		\nonumber\\[0.15cm]
		&+2f_\perp g_\perp \left[4 E m_p q^2+\left(m_\tau^2-q^2\right) \left(m_{\Lambda_c}^2-m_p^2-q^2\right)\right]
		\,,
		\\[0.2cm]
		\mathcal{A}_{S_{LL}\!-\!S_{LL}}=&\dfrac{m_\tau^2-q^2 }{2m_c^2}\left[f_0^2 (m_{\Lambda_c}-m_p)^2 s_++g_0^2 (m_{\Lambda_c}+m_p)^2 s_-\right]\,,
		\\[0.2cm]		
		\mathcal{A}_{T_{LL}\!-\!T_{LL}}=&-8\left(\dfrac{h_+^2}{s_+}\!+\!\dfrac{\tilde{h}_+^2}{s_-}\right)\Big\{4 m_\tau^4 m_p^2 \!+ \!m_{\Lambda_c}^4 (q^2\!-\!m_\tau^2)\!+\!2 m_{\Lambda_c}^2 (m_\tau^2\!-\!q^2) (4 E m_p\!+\!m_p^2\nonumber\\[0.15cm]
		&+\!q^2)\!+\!q^2 (4 E m_p\!+\!m_p^2 + q^2)^2\!-\! m_\tau^2 \big[m_p^4\!+\!6 m_p^2 q^2\!+\!q^4\!+\!8 E m_p (m_p^2\!+\! q^2)\big]\Big\}\nonumber\\[0.15cm]
		&+\!16\left[\dfrac{h_\perp^2(m_{\Lambda_c}\!+\!m_p)^2}{s_+q^4}\!+\!\dfrac{\tilde{h}_\perp^2(m_{\Lambda_c}\!-\!m_p)^2}{s_-q^4}\right]
		\Big\{2 m_p (2 E\!+\!m_p) q^4 (2 E m_p\!+\!q^2)\nonumber\\[0.15cm]
		& - m_\tau^2 q^2 (m_p^2 + q^2) (4 E m_p + m_p^2 + q^2) + m_{\Lambda_c}^4 m_\tau^2 (m_\tau^2 - q^2) +m_\tau^4 (m_p^4 + q^4)\nonumber\\[0.15cm]
		& - 2 m_{\Lambda_c}^2 (m_\tau^2 - q^2) \big[m_\tau^2 (m_p^2 + q^2)-2 E m_p q^2\big]\Big\}-\frac{32 m_\tau^2 (m_{\Lambda_c}^2 - m_p^2)}{q^4}\nonumber\\[0.15cm]
		&\times \big[m_{\Lambda_c}^2 (m_\tau^2 - q^2) - m_\tau^2 (m_p^2 + q^2) + 
		q^2 (4 E m_p + m_p^2 + q^2)\big]h_\perp\,\tilde{h}_\perp\,, 
		\\[0.2cm]	
  	   \mathcal{A}_{V_{LL}\!-\!V_{LR}}=&\frac{m_\tau^2 \left(m_\tau^2\!-\!q^2\right)}{2q^4}\left[f_0^2 (m_{\Lambda_c}\!-\!m_p)^2 s_+-g_0^2 (m_{\Lambda_c}+m_p)^2s_-\right]\!-\!\frac{m_\tau^2\left(m_{\Lambda_c}^2-m_p^2\right) }{q^4}		\nonumber\\[0.15cm]
		&\times\!(f_0 f_+\!-\!g_0 g_+)\left[4 E m_p q^2\!+\!\left(m_\tau^2\!-\!q^2\right) \left(m_{\Lambda_c}^2\!-\!m_p^2-q^2\right)\right]
		+\left[\frac{f_+^2 (m_{\Lambda_c}\!+\!m_p)^2}{2q^4 s_+}
		\right.\nonumber\\[0.15cm]
		&\left.-\frac{ g_+^2 (m_{\Lambda_c}\!\!-\!m_p)^2}{2q^4 s_-}\right]\Big{\{}4 m_p^2 q^4 \left(4 E^2\!-\!m_\tau^2+q^2\right)+\left(m_\tau^2\!-\!q^2\right) \left(m_{\Lambda_c}^2-m_p^2-q^2\right) 
		\nonumber\\[0.15cm]
		&\times\!\left[8 E m_p q^2\!+\!m_\tau^2 \left(m_{\Lambda_c}^2\!\!-\!m_p^2\!-\!q^2\right)\right]\Big{\}}
		\!+\!\left(\frac{f_\perp^2}{s_+}\!-\!\frac{ g_\perp^2}{s_-}\right)\left\{8 E^2 m_p^2 q^2\!+\!\left(m_\tau^2\!-\!q^2\right) 	\right.\nonumber\\[0.15cm]
		&\left.\times\left[2 m_{\Lambda_c}^2 q^2-4 E m_p \left(m_p^2-m_{\Lambda_c}^2+q^2\right)-\left(m_{\Lambda_c}^2-m_p^2\right)^2+2 m_p^2 m_\tau^2-q^4\right]\right\}
		\,,
		\\[0.2cm]	
		\mathcal{A}_{V_{LL}\!-\!S_{LL}}=&\frac{m_\tau \left(q^2\!-\!m_\tau^2\right)}{2m_c q^2} \left[f_0^2 (m_{\Lambda_c}\!-\!m_p)^2 s_+\!+\!g_0^2 (m_{\Lambda_c}\!+\!m_p)^2s_-\right]\!+\!\frac{m_\tau\left(m_{\Lambda_c}^2\!-\!m_p^2\right) }{2m_c q^2}	\nonumber\\[0.15cm]
		&\times(f_0 f_++g_0 g_+)\left[4 E m_p q^2+\left(m_\tau^2-q^2\right) \left(m_{\Lambda_c}^2-m_p^2-q^2\right)\right]\,,
		\\[0.2cm]
		\mathcal{A}_{V_{LL}\!-\!S_{LR}}=&\frac{m_\tau \left(q^2\!-\!m_\tau^2\right)}{2m_c q^2} \left[f_0^2 (m_{\Lambda_c}\!-\!m_p)^2 s_+\!-\!g_0^2 (m_{\Lambda_c}\!+\!m_p)^2s_-\right]\!+\!\frac{m_\tau\left(m_{\Lambda_c}^2\!-\!m_p^2\right) }{2m_c q^2}	\nonumber\\[0.15cm]
		&\times(f_0 f_+-g_0 g_+)\left[4 E m_p q^2+\left(m_\tau^2-q^2\right) \left(m_{\Lambda_c}^2-m_p^2-q^2\right)\right]\,,
		\\[0.2cm]
		\mathcal{A}_{V_{LL}\!-\!T_{LL}}=&-\frac{2m_\tau}{q^2}\Big\{\big[m_{\Lambda_c}^2 (m_\tau^2 - q^2) - m_\tau^2 (m_p^2 + q^2) + q^2 (4 E m_p + m_p^2 + q^2)\big]\big[(m_{\Lambda_c}- m_p)  \nonumber\\[0.15cm]
		&\times(f_0\,h_+ +2f_\perp\,\tilde{h}_\perp)+(m_{\Lambda_c} + m_p) (g_0 \,\tilde{h}_+ + 2g_\perp\,h_\perp)\big]-(m_\tau^2 - q^2)\big[(m_{\Lambda_c}+ m_p) \nonumber\\[0.15cm]
		&\times s_-(f_+\, h_+ + 2f_\perp\, h_\perp) +(m_{\Lambda_c}- m_p)s_+(g_+\,\tilde{h}_+ + 2g_\perp\, \tilde{h}_\perp)\big]\Big\}\,,
		\\[0.2cm]
		\mathcal{A}_{V_{LR}\!-\!T_{LL}}=&-\frac{2m_\tau}{q^2}\Big\{\big[m_{\Lambda_c}^2 (m_\tau^2 - q^2) - m_\tau^2 (m_p^2 + q^2) + q^2 (4 E m_p + m_p^2 + q^2)\big]\big[(m_{\Lambda_c}- m_p)  \nonumber\\[0.15cm]
		&\times (f_0\,h_+ +2f_\perp\,\tilde{h}_\perp)-(m_{\Lambda_c} + m_p) (g_0 \,\tilde{h}_+ + 2g_\perp\,h_\perp)\big]-(m_\tau^2 - q^2)\big[(m_{\Lambda_c}+ m_p) \nonumber\\[0.15cm]
		&\times s_-(f_+\, h_+ + 2f_\perp\, h_\perp)-(m_{\Lambda_c}- m_p)s_+(g_+\,\tilde{h}_+ + 2g_\perp\, \tilde{h}_\perp)\big]\Big\}\,,
		\\[0.2cm]
		\mathcal{A}_{S_{LL}\!-\!T_{LL}}=&\frac{2}{m_c}\left[4 E m_p q^2+\left(m_\tau^2-q^2\right) \left(m_{\Lambda_c}^2-m_p^2-q^2\right)\right]\left[f_0 h_+ (m_{\Lambda_c}-m_p)+g_0 \tilde{h}_+ (m_{\Lambda_c}+m_p)\right]\,,	\\[0.2cm]
		\mathcal{A}_{S_{LR}\!-\!T_{LL}}=&\frac{2}{m_c} \left[4 E m_p q^2+\left(m_\tau^2-q^2\right) \left(m_{\Lambda_c}^2-m_p^2-q^2\right)\right]\left[f_0 h_+ (m_{\Lambda_c}-m_p)-g_0 \tilde{h}_+ (m_{\Lambda_c}+m_p)\right]\,,
		\\[0.2cm] 		
		\mathcal{A}_{S_{LL}\!-\!S_{LR}}=&\dfrac{m_\tau^2-q^2 }{2m_c^2}\Big{[}f_0^2 (m_{\Lambda_c}-m_p)^2 s_+-g_0^2 (m_{\Lambda_c}+m_p)^2 s_-\Big{]}\,.
	\end{align}
\end{widetext}
where the various $f$, $g$, and $h$ denote the form factors used to parametrize the hadronic matrix 
elements of the $\Lambda_c\to p$ transition~\cite{Feldmann:2011xf,Meinel:2017ggx,Das:2018sms}.  
Note that $s_\pm=(m_{\Lambda_c}\pm m_p)^2-q^2$, and $m_c$ denotes the $c$-quark running mass.

\section{\boldmath Details of the polarization components of the $\tau$}
\label{app:numerator}

The components of the polarization vector $\mathcal{P}$ are defined in Sec.~\ref{subsec:spin density} as  
\begin{align} 
	P_a=-(\mathcal{P}\cdot s_a)
	=\frac{\Sigma^a_P}{C}
	=\frac{2\Sigma^a_P}{|\mathcal{M}|^2}\,,
\end{align}
where $s_a$ denote the spin four-vectors in the Lab frame. 
Note that the denominator $C$ has been replaced in the last step by $|\mathcal{M}|^2/2$, whose explicit expression 
is given by Eq.~\eqref{eq:amplitude2}. 
In addition, the explicit expression of $\Sigma^a_P$ read
\begin{widetext}
	\begin{align}\label{eq:Aal}
		2v^4\Sigma^a_P=&\left(|g_{V}^{LL}|^2-|g_{V}^{RR}|^2\right)\mathcal{A}^{\tau}_{V_{LL}\!-\!V_{LL}}
		+\left(|g_{V}^{LR}|^2-|g_{V}^{RL}|^2\right)\mathcal{A}^{\tau}_{V_{LR}\!-\!V_{LR}}
		+\left(|g_{T}^{LL}|^2-|g_{T}^{RR}|^2\right)\mathcal{A}^{\tau}_{T_{LL}\!-\!T_{LL}}\nonumber\\[0.15cm]
		&+\left(|g_{S}^{LL}|^2+|g_{S}^{LR}|^2-|g_{S}^{RR}|^2-|g_{S}^{RL}|^2\right)\mathcal{A}^{\tau}_{S_{LL}\!-\!S_{LL}}
		+2\text{Re}\left[\left(g_{V}^{LL}g_{V}^{LR*}-g_{V}^{RR*}g_{V}^{RL}\right)\right]
		\mathcal{A}^{\tau}_{V_{LL}\!-\!V_{LR}}\nonumber\\[0.15cm]
		&+2\text{Re}\left[\left(g_{V}^{LL}g_{S}^{LL*}\!-\!g_{V}^{RR*}g_{S}^{RR}\!
		+\!g_{V}^{LR}g_{S}^{LR*}\!-\!g_{V}^{RL*}g_{S}^{RL}\right)
		\mathcal{A}^{\tau}_{V_{LL}\!-\!S_{LL}}\right]  \nonumber\\[0.15cm]
		&+2\text{Re}\left[\left(g_{V}^{LL}g_{S}^{LR*}\!+\!g_{V}^{LR}g_{S}^{LL*}
		\!-\!g_{V}^{RR*}g_{S}^{RL}\!-\!g_{V}^{RL*}g_{S}^{RR}\right)
		\mathcal{A}^{\tau}_{V_{LL}\!-\!S_{LR}}\right]       \nonumber\\[0.15cm]
		&+2\text{Re}\left[\left(g_{V}^{LL}g_{T}^{LL*}\!-\!g_{V}^{RR*}g_{T}^{RR}\right)
		\mathcal{A}^{\tau}_{V_{LL}\!-\!T_{LL}}\right]
		+2\text{Re}\left[\left(g_{V}^{LR}g_{T}^{LL*}\!-\!g_{V}^{RL*}g_{T}^{RR}\right)
		\mathcal{A}^{\tau}_{V_{LR}\!-\!T_{LL}}\right]  \nonumber\\[0.15cm]
		&+2\text{Re}\left[\left(g_{S}^{LL}g_{S}^{LR*}\!-\!g_{S}^{RR*}g_{S}^{RL}\right)
		\mathcal{A}^{\tau}_{S_{LL}\!-\!S_{LR}}\right]+2\text{Re}\left[\left(g_{S}^{LL}g_{T}^{LL*}\!-\!g_{S}^{RR*}g_{T}^{RR}\right)
		\mathcal{A}^{\tau}_{S_{LL}\!-\!T_{LL}}\right]\nonumber\\[0.15cm]
		&+2\text{Re}\left[\left(g_{S}^{LR}g_{T}^{LL*}\!-\!g_{S}^{RL*}g_{T}^{RR}\right)
		\mathcal{A}^{\tau}_{S_{LR}\!-\!T_{LL}}\right]\,,
	\end{align}
\end{widetext}
where the various patterns discovered in Table~\ref{tab:Numerator} have been also taken into account. 
The explicit expressions of all the $\mathcal{A}^{\tau}$ on the right-hand side of Eq.~\eqref{eq:Aal} are presented as follows:
\begin{widetext}
	\begin{align}
		\mathcal{A}^{\tau}_{V_{LL}\!-\!V_{LL}}=&\frac{m_\tau^3 (s_a\cdot k)}{q^4}\left[f_0^2 (m_{\Lambda_c}-m_p)^2 s_++g_0^2 (m_{\Lambda_c}+m_p)^2s_-\right]-\frac{m_\tau \left(m_{\Lambda_c}^2-m_p^2\right)}{q^4} \nonumber\\[0.15cm]
		& \times (f_0 f_++g_0 g_+)\left\{(s_a\cdot k) \left[4 E m_p q^2+\left(m_\tau^2-q^2\right) \left(2 m_{\Lambda_c}^2-2 m_p^2-q^2\right)\right] 
		\right.\nonumber\\[0.15cm]
		&\left.-q^2\left(m_\tau^2-q^2\right) (s_a\cdot p+s_a\cdot p^\prime)\right\}+m_\tau \left[\frac{ f_+^2 (m_{\Lambda_c}+m_p)^2}{q^4 s_+}+\frac{g_+^2 (m_{\Lambda_c}-m_p)^2}{q^4 s_-}\right]
		\nonumber\\[0.15cm]
		&\times\Big{\{}(s_a\cdot k) \left[\left(m_{\Lambda_c}^2\!-\!m_p^2\right)\Big(m_\tau^2 (m_{\Lambda_c}^2-m_p^2-q^2)\!-\!q^2 (2 m_{\Lambda_c}^2\!-\!4m_p E \!-\!2m_p^2+q^2)\Big) 	\right.\nonumber\\[0.15cm]
		&\left.+q^4\left(4 m_{\Lambda_c}^2-q^2\right)\right]-q^2 (s_a\cdot p+s_a\cdot p^\prime)\left[4 E m_p q^2+\left(m_\tau^2-q^2\right) (m_{\Lambda_c}^2-m_p^2-q^2)\right]\Big{\}} \nonumber\\[0.15cm]
		&-4m_\tau f_\perp g_\perp\left[(s_a\cdot p) \left(m_\tau^2-q^2-2 E m_p\right)+2 E m_p (s_a\cdot p^\prime)\right]
		\nonumber\\[0.15cm]
		&+2 m_\tau \left(\frac{ f_\perp^2}{s_+}+\frac{ g_\perp^2}{s_-}\right)\Big{\{}(s_a\cdot p) \left[2 E m_p \left(m_{\Lambda_c}^2-m_p^2+q^2\right)+\left(m_\tau^2-q^2\right) (m_{\Lambda_c}^2+m_p^2-q^2)\right]\nonumber\\[0.15cm]
		&+2 m_p (s_a\cdot p^\prime) \left[E \left(m_p^2-m_{\Lambda_c}^2+q^2\right)+m_p \left(q^2-m_\tau^2\right)\right]\Big{\}}\,,
		\\[0.2cm]
		\mathcal{A}^{\tau}_{V_{LR}\!-\!V_{LR}}=&\frac{m_\tau^3 (s_a\cdot k)}{q^4}\left[f_0^2 (m_{\Lambda_c}-m_p)^2 s_++g_0^2 (m_{\Lambda_c}+m_p)^2s_-\right]-\frac{m_\tau \left(m_{\Lambda_c}^2-m_p^2\right)}{q^4} \nonumber\\[0.15cm]
		&\times  (f_0 f_++g_0 g_+)\left\{(s_a\cdot k) \left[4 E m_p q^2+\left(m_\tau^2-q^2\right) \left(2 m_{\Lambda_c}^2-2 m_p^2-q^2\right)\right] 
		\right.\nonumber\\[0.15cm]
		&\left.-q^2\left(m_\tau^2-q^2\right) (s_a\cdot p+s_a\cdot p^\prime)\right\}+m_\tau \left[\frac{ f_+^2 (m_{\Lambda_c}+m_p)^2}{q^4 s_+}+\frac{g_+^2 (m_{\Lambda_c}-m_p)^2}{q^4 s_-}\right]
		\nonumber\\[0.15cm]
		&\times \Big{\{}(s_a\cdot k) \left[\left(m_{\Lambda_c}^2\!-\!m_p^2\right)\Big(m_\tau^2 (m_{\Lambda_c}^2-m_p^2-q^2)\!-\!q^2 (2 m_{\Lambda_c}^2\!-\!4m_p E\!-\!2m_p^2+q^2)\Big) 	\right.\nonumber\\[0.15cm]
		&\left.+q^4\left(4 m_{\Lambda_c}^2-q^2\right)\right]-q^2 (s_a\cdot p+s_a\cdot p^\prime)\left[4 E m_p q^2+\left(m_\tau^2-q^2\right) (m_{\Lambda_c}^2-m_p^2-q^2)\right]\Big{\}}
		\nonumber\\[0.15cm]
		&+4m_\tau f_\perp g_\perp\left[(s_a\cdot p) \left(m_\tau^2-q^2-2 E m_p\right)+2 E m_p (s_a\cdot p^\prime)\right]
		\nonumber\\[0.15cm]
		&+2 m_\tau \left(\frac{ f_\perp^2}{s_+}+\frac{ g_\perp^2}{s_-}\right)\Big{\{}(s_a\cdot p) \left[2 E m_p \left(m_{\Lambda_c}^2-m_p^2+q^2\right)+\left(m_\tau^2-q^2\right) (m_{\Lambda_c}^2+m_p^2-q^2)\right]
		\nonumber\\[0.12cm]
		&+2 m_p (s_a\cdot p^\prime) \left[E \left(m_p^2-m_{\Lambda_c}^2+q^2\right)+m_p \left(q^2-m_\tau^2\right)\right]\Big{\}}\,,
		\\[0.2cm]
		\mathcal{A}^{\tau}_{S_{LL}\!-\!S_{LL}}=&\mathcal{A}_{S_{LL}\!-\!S_{LL}}\frac{2m_\tau( k\cdot s_a)}{m_\tau^2-q^2}\label{eq:relation1}\,,
		\\[0.2cm]
		\mathcal{A}^{\tau}_{T_{LL}\!-\!T_{LL}}=&-16 m_\tau\left(\frac{h_+^2}{s_+}+\frac{\tilde{h}_+^2}{s_-}\right)\Big{\{}2 (s_a\cdot p) 
		\big[2 E m_p \left(m_{\Lambda_c}^2\!-\!m_p^2+q^2\right)+\left(m_\tau^2\!-\!q^2\right) (m_{\Lambda_c}^2+m_p^2\!-\!q^2)\big]
		\nonumber\\[0.15cm]
		&+\!4 m_p (s_a\cdot p^\prime) \left[E \left(m_p^2\!-\!m_{\Lambda_c}^2\!+\!q^2\right)\!+\!m_p \left(q^2\!-\!m_\tau^2\right)\right]\!
		+\!(s_a\cdot k) \,s_+ s_- \Big{\}}
		\nonumber\\[0.15cm]
		&-\!16m_\tau\Big\{(s_a\cdot k) \left(m_\tau^2\!-\!q^2\right) s_-s_+ \!-\!\left[4 E m_p q^2\!+\!\left(m_\tau^2\!\!-\!q^2\right) \left(m_{\Lambda_c}^2\!-\!m_p^2\!-\!q^2\right)\right]  
		\nonumber\\[0.15cm]
		&\times\big[(s_a\cdot p)(m_{\Lambda_c}^2-m_p^2+q^2)+(s_a\cdot p^\prime) \left(m_p^2-m_{\Lambda_c}^2+q^2\right)\big]\Big\}
		\left[\frac{(m_{\Lambda_c}-m_p)^2\tilde{h}_\perp^2}{q^4 s_-}
		\right.\nonumber\\[0.15cm]
		&\left.+\frac{(m_{\Lambda_c}+m_p)^2h_\perp^2}{q^4 s_+}\right]-\frac{64m_\tau (m_{\Lambda_c}^2-m_p^2)h_\perp\tilde{h}_\perp}{q^4}\Big\{ (s_a\cdot p) \left[2 E m_p q^2+(m_{\Lambda_c}^2-m_p^2) (m_\tau^2-q^2)\right]
		\nonumber\\[0.15cm]
		&-(s_a\cdot p^\prime) \left[2 E m_p q^2+\left(m_\tau^2-q^2\right) \left(m_{\Lambda_c}^2-m_p^2-q^2\right)\right]\Big\}\,,
		\\[0.2cm]
    	\mathcal{A}^{\tau}_{V_{LL}\!-\!V_{LR}}=&\frac{m_\tau^3 (s_a\cdot k)}{q^4}\left[f_0^2 (m_{\Lambda_c}-m_p)^2 s_+-g_0^2 (m_{\Lambda_c}+m_p)^2s_-\right]-\frac{m_\tau \left(m_{\Lambda_c}^2-m_p^2\right)}{q^4} \nonumber\\[0.15cm]
		&\times (f_0 f_+-g_0 g_+)\left\{(s_a\cdot k) \left[4 E m_p q^2+\left(m_\tau^2-q^2\right) \left(2 m_{\Lambda_c}^2-2 m_p^2-q^2\right)\right]
		\right.\nonumber\\[0.15cm]
		&\left.-q^2 \left(m_\tau^2-q^2\right) (s_a\cdot p+s_a\cdot p^\prime)\right\}+m_\tau \left[\frac{ f_+^2 (m_{\Lambda_c}+m_p)^2}{q^4 s_+}-\frac{g_+^2 (m_{\Lambda_c}-m_p)^2}{q^4 s_-}\right]
		\nonumber\\[0.15cm]
		&\times\Big{\{}(s_a\cdot k) \left[\left(m_{\Lambda_c}^2\!-\!m_p^2\right)\Big(m_\tau^2 (m_{\Lambda_c}^2-m_p^2-q^2)\!-\!q^2 (2 m_{\Lambda_c}^2\!-\!4m_p E\!-\!2m_p^2+q^2)\Big) 	\right.\nonumber\\[0.15cm]
		&\left.+q^4\left(4 m_{\Lambda_c}^2-q^2\right)\right]-q^2 (s_a\cdot p+s_a\cdot p^\prime)\left[4 E m_p q^2+\left(m_\tau^2-q^2\right) (m_{\Lambda_c}^2-m_p^2-q^2)\right]\Big{\}} \nonumber\\[0.15cm]
		&+2m_\tau \left(\frac{ f_\perp^2}{s_+}-\frac{ g_\perp^2}{s_-}\right)\Big{\{}(s_a\cdot p) \left[2 E m_p \left(m_{\Lambda_c}^2-m_p^2+q^2\right)+\left(m_\tau^2-q^2\right)
		\right.\nonumber\\[0.15cm]
		&\left. \times (m_{\Lambda_c}^2+m_p^2-q^2)\right]+2 m_p (s_a\cdot p^\prime) \left[E \left(m_p^2-m_{\Lambda_c}^2+q^2\right)+m_p \left(q^2-m_\tau^2\right)\right]\Big{\}}\,,
		\\[0.2cm]
		\mathcal{A}^{\tau}_{V_{LL}\!-\!S_{LL}}=&-\!\frac{m_\tau^2 (s_a\cdot k)}{m_c q^2}\Big{\{}f_0^2 (m_{\Lambda_c}\!-\!m_p)^2 s_+\!-\!g_0^2 (m_{\Lambda_c}\!+\!m_p)^2 s_-\Big{\}}\!-\!\frac{ \left(m_{\Lambda_c}^2\!-\!m_p^2\right) }{m_c q^2}\nonumber\\[0.15cm]
		&\times(f_0 f_++g_0 g_+)\Big{\{}q^2 \left(m_\tau^2-q^2\right) (s_a\cdot p^\prime)+(s_a\cdot p-s_a\cdot p^\prime) \nonumber\\[0.15cm]
		&\times \left[2 E m_p q^2+\left(m_{\Lambda_c}^2-m_p^2\right) (m_\tau^2-q^2)\right]+2 i q^2 \varepsilon _{\{k\},\{k^\prime\},\{s_a\},\{p\}}\Big{\}}\,,	
		\\[0.2cm]
		\mathcal{A}^{\tau}_{V_{LL}\!-\!S_{LR}}=&-\!\frac{m_\tau^2(s_a\cdot k)}{m_c q^2}\Big{\{}f_0^2 (m_{\Lambda_c}\!-\!m_p)^2 s_+\!-\!g_0^2 (m_{\Lambda_c}\!+\!m_p)^2 s_-\Big{\}}\!-\!\frac{\left(m_{\Lambda_c}^2\!-\!m_p^2\right) }{m_c q^2}\nonumber\\[0.15cm]
		&\times(f_0 f_+-g_0 g_+)\Big{\{}q^2 \left(m_\tau^2-q^2\right) (s_a\cdot p^\prime)+(s_a\cdot p-s_a\cdot p^\prime)\nonumber\\[0.15cm]
		&\times \left[2 E m_p q^2+\left(m_{\Lambda_c}^2-m_p^2\right) (m_\tau^2-q^2)\right]+2 i q^2 \varepsilon _{\{k\},\{k^\prime\},\{s_a\},\{p\}}\Big{\}}\,,	
		\\[0.2cm]
		\mathcal{A}^{\tau}_{V_{LL}\!-\!T_{LL}}=&-4 m_\tau^2\left[\frac{(m_{\Lambda_c}\!-\!m_p)f_0h_+}{q^2}\!+\!\frac{(m_{\Lambda_c}\!+\!m_p)g_0\tilde{h}_+}{q^2}\right]
		\left[ (s_a\cdot p) \left(2 E m_p\!-\!m_\tau^2+q^2\right)
		\right.	\nonumber\\[0.15cm]	
		&\left.-\!2 \left(E m_p (s_a\cdot p^\prime)\!-\!i \varepsilon _{\{k\}\{k^\prime\}\{s_a\}\{p\}}\right)\right]\!\!+\!\!\left[4 E m_p q^2\!+\!\left(m_\tau^2\!-\!q^2\right) \left(m_{\Lambda_c}^2\!\!-\!m_p^2\!-\!q^2\right)\right]
		\nonumber\\[0.15cm]	
		&\times\!\big[-2 i \varepsilon_{\{k\}\{k^\prime\}\{s_a\}\{p\}}\!+\!(s_a\cdot p) \left(m_{\Lambda_c}^2\!\!-\!m_p^2\!+\!m_\tau^2\!-\!2 E m_p\right)\!+\!(s_a\cdot p^\prime)(2 E m_p\!-\!m_{\Lambda_c}^2\!+\!m_p^2\!+\!q^2)\big]
		\nonumber\\[0.15cm]	
		&\times \left[\frac{4(m_{\Lambda_c}\!+\!m_p)f_\perp h_\perp}{s_+q^2}\!+\!\frac{4(m_{\Lambda_c}\!-\!m_p)g_\perp \tilde{h}_\perp}{s_-q^2}\right]\!+\!4 \left(q^2\!-\!m_\tau^2\right)
		\nonumber\\[0.15cm]	
		&\times\left[\frac{(m_{\Lambda_c}\!-\!m_p)f_\perp\tilde{h}_\perp}{q^2}\!+\!\frac{(m_{\Lambda_c}\!+\!m_p)g_\perp h_\perp}{q^2}\right]\left[-2 i \varepsilon _{\{k\}\{k^\prime\}\{s_a\}\{p\}}\!+\!(s_a\cdot p) (m_{\Lambda_c}^2\!-\!m_p^2\!+\!m_\tau^2\!\!-\!2 E m_p)
		\right.	\nonumber\\[0.15cm]	
		&\left.	\!+\!(s_a\cdot p^\prime) \left(2 E  m_p\!-\!m_{\Lambda_c}^2\!+\!m_p^2\!+\!q^2\right)\right]\!\!+\!4\Big{\{}\!\!\left[4 m_p^2 q^2 \left(m_\tau^2\!\!-\!2 E^2\right)
		\right.	\nonumber\\[0.15cm]	
		&\left.+2 E m_p \left(m_\tau^2\!-\!3 q^2\right) \left(m_p^2\!-\!m_{\Lambda_c}^2\!+\!q^2\right)\!-\!q^2 s_- s_+\right](s_a\cdot p\!+\!s_a\cdot p^\prime)\!-\!\left(m_\tau^2\!+\!q^2\right)
		\nonumber\\[0.15cm]	
		&\times\!\left[\left(m_\tau^2\!-\!q^2\right) \left(m_p^2\!-\!m_{\Lambda_c}^2\!+\!q^2\right)\!-\!4 E m_p q^2\right]  (s_a\cdot p)\!+\!2 i \left[4 E m_p q^2\!+\!\left(m_\tau^2\!-\!q^2\right)
		\right.\nonumber\\[0.15cm]	
		& \left.\times\!\left(m_{\Lambda_c}^2\!\!-m_p^2\!-\!q^2\right)\right] \varepsilon _{\{k\}\{k^\prime\}\{s_a\}\{p\}}\Big{\}}
		\left[\frac{ (m_{\Lambda_c}\!+\!m_p)f_+h_+}{s_+q^2}\!+\!\frac{ (m_{\Lambda_c}\!\!-\!m_p)g_+\tilde{h}_+}{s_-q^2}\right]
		\,,
		\\[0.2cm]
		\mathcal{A}^{\tau}_{V_{LR}\!-\!T_{LL}}&=-4 m_\tau^2\left[\frac{(m_{\Lambda_c}\!-\!m_p)f_0h_+}{q^2}\!-\!\frac{(m_{\Lambda_c}\!+\!m_p)g_0\tilde{h}_+}{q^2}\right]\left[ (s_a\cdot p) \left(2 E m_p\!-\!m_\tau^2+q^2\right)
		\right.	\nonumber\\[0.15cm]	
		&\left.-\!2 \left(E m_p (s_a\cdot p^\prime)\!-\!i \varepsilon _{\{k\}\{k^\prime\}\{s_a\}\{p\}}\right)\right]\!\!+\!\!\left[4 E m_p q^2\!+\!\left(m_\tau^2\!-\!q^2\right) \left(m_{\Lambda_c}^2\!\!-\!m_p^2\!-\!q^2\right)\right]
		\nonumber\\[0.15cm]	
		&\times\!\big[-2i \varepsilon_{\{k\}\{k^\prime\}\{s_a\}\{p\}}\!+\!(s_a\cdot p) \left(m_{\Lambda_c}^2\!\!-\!m_p^2\!+\!m_\tau^2\!-\!2 E  m_p\right)\!+\!(s_a\cdot p^\prime)(2 E m_p\!-\!m_{\Lambda_c}^2
		\nonumber\\[0.15cm]	
		&+\!m_p^2\!+\!q^2)\big]\left[\frac{4(m_{\Lambda_c}\!+\!m_p)f_\perp h_\perp}{s_+q^2}\!-\!\frac{4(m_{\Lambda_c}\!-\!m_p)g_\perp \tilde{h}_\perp}{s_-q^2}\right]\!+\!4 \left(q^2\!-\!m_\tau^2\right)
		\nonumber\\[0.15cm]	
		&\times\left[\frac{(m_{\Lambda_c}\!-\!m_p)f_\perp\tilde{h}_\perp}{q^2}\!-\!\frac{(m_{\Lambda_c}\!+\!m_p)g_\perp h_\perp}{q^2}\right]\left[-2 i \varepsilon _{\{k\}\{k^\prime\}\{s_a\}\{p\}}\!+\!(s_a\cdot p) (m_{\Lambda_c}^2\!-\!m_p^2
		\right.	\nonumber\\[0.15cm]	
		&\left.	+m_\tau^2\!\!-\!2 E m_p)\!+\!(s_a\cdot p^\prime) \left(2 E m_p\!-\!m_{\Lambda_c}^2\!+\!m_p^2\!+\!q^2\right)\right]\!\!+\!4\Big{\{}\!\!\left[4 m_p^2 q^2 \left(m_\tau^2\!\!-\!2 E^2\right)
		\right.	\nonumber\\[0.15cm]	
		&\left.+2 E m_p \left(m_\tau^2\!-\!3 q^2\right) \left(m_p^2\!-\!m_{\Lambda_c}^2\!+\!q^2\right)\!-\!q^2 s_- s_+\right](s_a\cdot p\!+\!s_a\cdot p^\prime)\!-\!\left(m_\tau^2\!+\!q^2\right)
		\nonumber\\[0.15cm]	
		& \times\!\left[\left(m_\tau^2\!-\!q^2\right) \left(m_p^2\!-\!m_{\Lambda_c}^2\!+\!q^2\right)\!-\!4 E m_p q^2\right]  (s_a\cdot p)\!+\!2 i \left[4 E m_p q^2\!+\!\left(m_\tau^2\!-\!q^2\right)
		\right.\nonumber\\[0.15cm]	
		& \left.\times\!\left(m_{\Lambda_c}^2\!\!-m_p^2\!-\!q^2\right)\right] \varepsilon _{\{k\}\{k^\prime\}\{s_a\}\{p\}}\Big{\}}
		\left[\frac{ (m_{\Lambda_c}\!+\!m_p)f_+h_+}{s_+q^2}\!-\!\frac{ (m_{\Lambda_c}\!\!-\!m_p)g_+\tilde{h}_+}{s_-q^2}\right]
		\,,
		\\[0.2cm]
		\mathcal{A}^{\tau}_{S_{LL}\!-\!T_{LL}}=&\frac{4 m_\tau}{m_c} \left[-2 i \varepsilon _{\{k\},\{k^\prime\},\{p\},\{s_a\}}+(s_a\cdot p) \left(2 E m_p-m_\tau^2+q^2\right)-2 E m_p (s_a\cdot p^\prime)\right]
		\nonumber\\[0.15cm]
		&\times \left[f_0 h_+ (m_{\Lambda_c}-m_p)+g_0 \tilde{h}_+ (m_{\Lambda_c}+m_p)\right]\,,
		\\[0.2cm]
		\mathcal{A}^{\tau}_{S_{LR}\!-\!T_{LL}}=&\frac{4m_\tau}{m_c} \left[-2 i \varepsilon _{\{k\},\{k^\prime\},\{p\},\{s_a\}}+(s_a\cdot p) \left(2 E  m_p-m_\tau^2+q^2\right)-2 E m_p (s_a\cdot p^\prime)\right]
		\nonumber\\[0.15cm]
		&\times \left[f_0 h_+ (m_{\Lambda_c}-m_p)-g_0 \tilde{h}_+ (m_{\Lambda_c}+m_p)\right]\,,
		\\[0.2cm]
		\mathcal{A}^{\tau}_{S_{LL}\!-\!S_{LR}}=&\mathcal{A}_{S_{LL}\!-\!S_{LR}}\frac{2m_{\tau}(k\cdot s_a)}{m_\tau^2-q^2}\label{eq:relation2}\,,
	\end{align}
\end{widetext}		
where $\varepsilon_{\{k\}\{k^\prime\}\{s_a\}\{p\}}\equiv\varepsilon _{\mu\nu\alpha\beta}k^{\mu}k^{\prime\nu}s_a^{\alpha}p^{\beta}$, with $\varepsilon$ being a totally antisymmetric tensor. 
From the equations above, it is clear that $\mathcal{A}^{\tau}_{V_{LL}\!-\!V_{LR}}$ does not contain the $\varepsilon_{\{k\}\{k^\prime\}\{s_a\}\{p\}}$, resulting in a vanishing $P_P$. 
Interestingly enough, if only $g_S^{LL}$ and $g_S^{LR}$ are activated simultaneously, 
the $P_a$ are independent of those two WCs, due to the relations presented in Eqs.~\eqref{eq:relation1} and \eqref{eq:relation2}.    
Due to the same reason (c.f. Eq.~\eqref{eq:Aal}), the same conclusion also holds, if only the $g_S^{RR}$ and $g_S^{RL}$ are activated at the same time.

\bibliographystyle{apsrev4-2}
\bibliography{reference}

\begin{thebibliography}{100}%
\makeatletter
\providecommand \@ifxundefined [1]{%
 \@ifx{#1\undefined}
}%
\providecommand \@ifnum [1]{%
 \ifnum #1\expandafter \@firstoftwo
 \else \expandafter \@secondoftwo
 \fi
}%
\providecommand \@ifx [1]{%
 \ifx #1\expandafter \@firstoftwo
 \else \expandafter \@secondoftwo
 \fi
}%
\providecommand \natexlab [1]{#1}%
\providecommand \enquote  [1]{``#1''}%
\providecommand \bibnamefont  [1]{#1}%
\providecommand \bibfnamefont [1]{#1}%
\providecommand \citenamefont [1]{#1}%
\providecommand \href@noop [0]{\@secondoftwo}%
\providecommand \href [0]{\begingroup \@sanitize@url \@href}%
\providecommand \@href[1]{\@@startlink{#1}\@@href}%
\providecommand \@@href[1]{\endgroup#1\@@endlink}%
\providecommand \@sanitize@url [0]{\catcode `\\12\catcode `\$12\catcode
  `\&12\catcode `\#12\catcode `\^12\catcode `\_12\catcode `\%12\relax}%
\providecommand \@@startlink[1]{}%
\providecommand \@@endlink[0]{}%
\providecommand \url  [0]{\begingroup\@sanitize@url \@url }%
\providecommand \@url [1]{\endgroup\@href {#1}{\urlprefix }}%
\providecommand \urlprefix  [0]{URL }%
\providecommand \Eprint [0]{\href }%
\providecommand \doibase [0]{https://doi.org/}%
\providecommand \selectlanguage [0]{\@gobble}%
\providecommand \bibinfo  [0]{\@secondoftwo}%
\providecommand \bibfield  [0]{\@secondoftwo}%
\providecommand \translation [1]{[#1]}%
\providecommand \BibitemOpen [0]{}%
\providecommand \bibitemStop [0]{}%
\providecommand \bibitemNoStop [0]{.\EOS\space}%
\providecommand \EOS [0]{\spacefactor3000\relax}%
\providecommand \BibitemShut  [1]{\csname bibitem#1\endcsname}%
\let\auto@bib@innerbib\@empty
\bibitem [{\citenamefont {Calibbi}\ and\ \citenamefont
  {Signorelli}(2018)}]{Calibbi:2017uvl}%
  \BibitemOpen
  \bibfield  {author} {\bibinfo {author} {\bibfnamefont {L.}~\bibnamefont
  {Calibbi}}\ and\ \bibinfo {author} {\bibfnamefont {G.}~\bibnamefont
  {Signorelli}},\ }\href {https://doi.org/10.1393/ncr/i2018-10144-0} {\bibfield
   {journal} {\bibinfo  {journal} {Riv. Nuovo Cim.}\ }\textbf {\bibinfo
  {volume} {41}},\ \bibinfo {pages} {71} (\bibinfo {year} {2018})},\ \Eprint
  {https://arxiv.org/abs/1709.00294} {arXiv:1709.00294 [hep-ph]} \BibitemShut
  {NoStop}%
\bibitem [{\citenamefont {Ardu}\ and\ \citenamefont
  {Pezzullo}(2022)}]{Ardu:2022sbt}%
  \BibitemOpen
  \bibfield  {author} {\bibinfo {author} {\bibfnamefont {M.}~\bibnamefont
  {Ardu}}\ and\ \bibinfo {author} {\bibfnamefont {G.}~\bibnamefont
  {Pezzullo}},\ }\href {https://doi.org/10.3390/universe8060299} {\bibfield
  {journal} {\bibinfo  {journal} {Universe}\ }\textbf {\bibinfo {volume} {8}},\
  \bibinfo {pages} {299} (\bibinfo {year} {2022})},\ \Eprint
  {https://arxiv.org/abs/2204.08220} {arXiv:2204.08220 [hep-ph]} \BibitemShut
  {NoStop}%
\bibitem [{\citenamefont {Altmannshofer}\ and\ \citenamefont
  {Archilli}(2022)}]{Altmannshofer:2022hfs}%
  \BibitemOpen
  \bibfield  {author} {\bibinfo {author} {\bibfnamefont {W.}~\bibnamefont
  {Altmannshofer}}\ and\ \bibinfo {author} {\bibfnamefont {F.}~\bibnamefont
  {Archilli}},\ }in\ \href@noop {} {\emph {\bibinfo {booktitle} {{Snowmass
  2021}}}}\ (\bibinfo {year} {2022})\ \Eprint
  {https://arxiv.org/abs/2206.11331} {arXiv:2206.11331 [hep-ph]} \BibitemShut
  {NoStop}%
\bibitem [{\citenamefont {Guadagnoli}\ and\ \citenamefont
  {Koppenburg}(2023)}]{Guadagnoli:2022oxk}%
  \BibitemOpen
  \bibfield  {author} {\bibinfo {author} {\bibfnamefont {D.}~\bibnamefont
  {Guadagnoli}}\ and\ \bibinfo {author} {\bibfnamefont {P.}~\bibnamefont
  {Koppenburg}},\ }\href {https://doi.org/10.1146/annurev-nucl-101122-045442}
  {\bibfield  {journal} {\bibinfo  {journal} {Ann. Rev. Nucl. Part. Sci.}\
  }\textbf {\bibinfo {volume} {73}},\ \bibinfo {pages} {1} (\bibinfo {year}
  {2023})},\ \Eprint {https://arxiv.org/abs/2207.01851} {arXiv:2207.01851
  [hep-ph]} \BibitemShut {NoStop}%
\bibitem [{\citenamefont {Davidson}\ \emph {et~al.}(2022)\citenamefont
  {Davidson}, \citenamefont {Echenard}, \citenamefont {Bernstein},
  \citenamefont {Heeck},\ and\ \citenamefont {Hitlin}}]{Davidson:2022jai}%
  \BibitemOpen
  \bibfield  {author} {\bibinfo {author} {\bibfnamefont {S.}~\bibnamefont
  {Davidson}}, \bibinfo {author} {\bibfnamefont {B.}~\bibnamefont {Echenard}},
  \bibinfo {author} {\bibfnamefont {R.~H.}\ \bibnamefont {Bernstein}}, \bibinfo
  {author} {\bibfnamefont {J.}~\bibnamefont {Heeck}},\ and\ \bibinfo {author}
  {\bibfnamefont {D.~G.}\ \bibnamefont {Hitlin}},\ }\href@noop {} {\  (\bibinfo
  {year} {2022})},\ \Eprint {https://arxiv.org/abs/2209.00142}
  {arXiv:2209.00142 [hep-ex]} \BibitemShut {NoStop}%
\bibitem [{\citenamefont {Kodama}\ \emph {et~al.}(1995)\citenamefont {Kodama}
  \emph {et~al.}}]{E653:1995rpz}%
  \BibitemOpen
  \bibfield  {author} {\bibinfo {author} {\bibfnamefont {K.}~\bibnamefont
  {Kodama}} \emph {et~al.} (\bibinfo {collaboration} {E653}),\ }\href
  {https://doi.org/10.1016/0370-2693(94)01610-O} {\bibfield  {journal}
  {\bibinfo  {journal} {Phys. Lett. B}\ }\textbf {\bibinfo {volume} {345}},\
  \bibinfo {pages} {85} (\bibinfo {year} {1995})}\BibitemShut {NoStop}%
\bibitem [{\citenamefont {Lees}\ \emph {et~al.}(2011)\citenamefont {Lees} \emph
  {et~al.}}]{Lees:2011hb}%
  \BibitemOpen
  \bibfield  {author} {\bibinfo {author} {\bibfnamefont {J.}~\bibnamefont
  {Lees}} \emph {et~al.} (\bibinfo {collaboration} {BaBar}),\ }\href
  {https://doi.org/10.1103/PhysRevD.84.072006} {\bibfield  {journal} {\bibinfo
  {journal} {Phys. Rev. D}\ }\textbf {\bibinfo {volume} {84}},\ \bibinfo
  {pages} {072006} (\bibinfo {year} {2011})},\ \Eprint
  {https://arxiv.org/abs/1107.4465} {arXiv:1107.4465 [hep-ex]} \BibitemShut
  {NoStop}%
\bibitem [{\citenamefont {de~Boer}\ and\ \citenamefont
  {Hiller}(2016)}]{deBoer:2015boa}%
  \BibitemOpen
  \bibfield  {author} {\bibinfo {author} {\bibfnamefont {S.}~\bibnamefont
  {de~Boer}}\ and\ \bibinfo {author} {\bibfnamefont {G.}~\bibnamefont
  {Hiller}},\ }\href {https://doi.org/10.1103/PhysRevD.93.074001} {\bibfield
  {journal} {\bibinfo  {journal} {Phys. Rev.}\ }\textbf {\bibinfo {volume}
  {D93}},\ \bibinfo {pages} {074001} (\bibinfo {year} {2016})},\ \Eprint
  {https://arxiv.org/abs/1510.00311} {arXiv:1510.00311 [hep-ph]} \BibitemShut
  {NoStop}%
\bibitem [{\citenamefont {Aaij}\ \emph {et~al.}(2016)\citenamefont {Aaij} \emph
  {et~al.}}]{Aaij:2015qmj}%
  \BibitemOpen
  \bibfield  {author} {\bibinfo {author} {\bibfnamefont {R.}~\bibnamefont
  {Aaij}} \emph {et~al.} (\bibinfo {collaboration} {LHCb}),\ }\href
  {https://doi.org/10.1016/j.physletb.2016.01.029} {\bibfield  {journal}
  {\bibinfo  {journal} {Phys. Lett. B}\ }\textbf {\bibinfo {volume} {754}},\
  \bibinfo {pages} {167} (\bibinfo {year} {2016})},\ \Eprint
  {https://arxiv.org/abs/1512.00322} {arXiv:1512.00322 [hep-ex]} \BibitemShut
  {NoStop}%
\bibitem [{\citenamefont {Aaij}\ \emph {et~al.}(2018)\citenamefont {Aaij} \emph
  {et~al.}}]{LHCb:2017yqf}%
  \BibitemOpen
  \bibfield  {author} {\bibinfo {author} {\bibfnamefont {R.}~\bibnamefont
  {Aaij}} \emph {et~al.} (\bibinfo {collaboration} {LHCb}),\ }\href
  {https://doi.org/10.1103/PhysRevD.97.091101} {\bibfield  {journal} {\bibinfo
  {journal} {Phys. Rev. D}\ }\textbf {\bibinfo {volume} {97}},\ \bibinfo
  {pages} {091101} (\bibinfo {year} {2018})},\ \Eprint
  {https://arxiv.org/abs/1712.07938} {arXiv:1712.07938 [hep-ex]} \BibitemShut
  {NoStop}%
\bibitem [{\citenamefont {De~Boer}\ and\ \citenamefont
  {Hiller}(2018)}]{DeBoer:2018pdx}%
  \BibitemOpen
  \bibfield  {author} {\bibinfo {author} {\bibfnamefont {S.}~\bibnamefont
  {De~Boer}}\ and\ \bibinfo {author} {\bibfnamefont {G.}~\bibnamefont
  {Hiller}},\ }\href {https://doi.org/10.1103/PhysRevD.98.035041} {\bibfield
  {journal} {\bibinfo  {journal} {Phys. Rev. D}\ }\textbf {\bibinfo {volume}
  {98}},\ \bibinfo {pages} {035041} (\bibinfo {year} {2018})},\ \Eprint
  {https://arxiv.org/abs/1805.08516} {arXiv:1805.08516 [hep-ph]} \BibitemShut
  {NoStop}%
\bibitem [{\citenamefont {Bause}\ \emph {et~al.}(2020)\citenamefont {Bause},
  \citenamefont {Golz}, \citenamefont {Hiller},\ and\ \citenamefont
  {Tayduganov}}]{Bause:2019vpr}%
  \BibitemOpen
  \bibfield  {author} {\bibinfo {author} {\bibfnamefont {R.}~\bibnamefont
  {Bause}}, \bibinfo {author} {\bibfnamefont {M.}~\bibnamefont {Golz}},
  \bibinfo {author} {\bibfnamefont {G.}~\bibnamefont {Hiller}},\ and\ \bibinfo
  {author} {\bibfnamefont {A.}~\bibnamefont {Tayduganov}},\ }\href
  {https://doi.org/10.1140/epjc/s10052-020-7621-7} {\bibfield  {journal}
  {\bibinfo  {journal} {Eur. Phys. J. C}\ }\textbf {\bibinfo {volume} {80}},\
  \bibinfo {pages} {65} (\bibinfo {year} {2020})},\ \Eprint
  {https://arxiv.org/abs/1909.11108} {arXiv:1909.11108 [hep-ph]} \BibitemShut
  {NoStop}%
\bibitem [{\citenamefont {Lees}\ \emph {et~al.}(2020)\citenamefont {Lees} \emph
  {et~al.}}]{BaBar:2020faa}%
  \BibitemOpen
  \bibfield  {author} {\bibinfo {author} {\bibfnamefont {J.~P.}\ \bibnamefont
  {Lees}} \emph {et~al.} (\bibinfo {collaboration} {BaBar}),\ }\href
  {https://doi.org/10.1103/PhysRevD.101.112003} {\bibfield  {journal} {\bibinfo
   {journal} {Phys. Rev. D}\ }\textbf {\bibinfo {volume} {101}},\ \bibinfo
  {pages} {112003} (\bibinfo {year} {2020})},\ \Eprint
  {https://arxiv.org/abs/2004.09457} {arXiv:2004.09457 [hep-ex]} \BibitemShut
  {NoStop}%
\bibitem [{\citenamefont {Bause}\ \emph {et~al.}(2021)\citenamefont {Bause},
  \citenamefont {Gisbert}, \citenamefont {Golz},\ and\ \citenamefont
  {Hiller}}]{Bause:2020xzj}%
  \BibitemOpen
  \bibfield  {author} {\bibinfo {author} {\bibfnamefont {R.}~\bibnamefont
  {Bause}}, \bibinfo {author} {\bibfnamefont {H.}~\bibnamefont {Gisbert}},
  \bibinfo {author} {\bibfnamefont {M.}~\bibnamefont {Golz}},\ and\ \bibinfo
  {author} {\bibfnamefont {G.}~\bibnamefont {Hiller}},\ }\href
  {https://doi.org/10.1103/PhysRevD.103.015033} {\bibfield  {journal} {\bibinfo
   {journal} {Phys. Rev. D}\ }\textbf {\bibinfo {volume} {103}},\ \bibinfo
  {pages} {015033} (\bibinfo {year} {2021})},\ \Eprint
  {https://arxiv.org/abs/2010.02225} {arXiv:2010.02225 [hep-ph]} \BibitemShut
  {NoStop}%
\bibitem [{\citenamefont {Aaij}\ \emph {et~al.}(2021)\citenamefont {Aaij} \emph
  {et~al.}}]{LHCb:2020car}%
  \BibitemOpen
  \bibfield  {author} {\bibinfo {author} {\bibfnamefont {R.}~\bibnamefont
  {Aaij}} \emph {et~al.} (\bibinfo {collaboration} {LHCb}),\ }\href
  {https://doi.org/10.1007/JHEP06(2021)044} {\bibfield  {journal} {\bibinfo
  {journal} {JHEP}\ }\textbf {\bibinfo {volume} {06}},\ \bibinfo {pages}
  {044}},\ \Eprint {https://arxiv.org/abs/2011.00217} {arXiv:2011.00217
  [hep-ex]} \BibitemShut {NoStop}%
\bibitem [{\citenamefont {Gisbert}\ \emph {et~al.}(2021)\citenamefont
  {Gisbert}, \citenamefont {Golz},\ and\ \citenamefont
  {Mitzel}}]{Gisbert:2020vjx}%
  \BibitemOpen
  \bibfield  {author} {\bibinfo {author} {\bibfnamefont {H.}~\bibnamefont
  {Gisbert}}, \bibinfo {author} {\bibfnamefont {M.}~\bibnamefont {Golz}},\ and\
  \bibinfo {author} {\bibfnamefont {D.~S.}\ \bibnamefont {Mitzel}},\ }\href
  {https://doi.org/10.1142/S0217732321300020} {\bibfield  {journal} {\bibinfo
  {journal} {Mod. Phys. Lett. A}\ }\textbf {\bibinfo {volume} {36}},\ \bibinfo
  {pages} {2130002} (\bibinfo {year} {2021})},\ \Eprint
  {https://arxiv.org/abs/2011.09478} {arXiv:2011.09478 [hep-ph]} \BibitemShut
  {NoStop}%
\bibitem [{\citenamefont {Golz}\ \emph {et~al.}(2021)\citenamefont {Golz},
  \citenamefont {Hiller},\ and\ \citenamefont {Magorsch}}]{Golz:2021imq}%
  \BibitemOpen
  \bibfield  {author} {\bibinfo {author} {\bibfnamefont {M.}~\bibnamefont
  {Golz}}, \bibinfo {author} {\bibfnamefont {G.}~\bibnamefont {Hiller}},\ and\
  \bibinfo {author} {\bibfnamefont {T.}~\bibnamefont {Magorsch}},\ }\href
  {https://doi.org/10.1007/JHEP09(2021)208} {\bibfield  {journal} {\bibinfo
  {journal} {JHEP}\ }\textbf {\bibinfo {volume} {09}},\ \bibinfo {pages}
  {208}},\ \Eprint {https://arxiv.org/abs/2107.13010} {arXiv:2107.13010
  [hep-ph]} \BibitemShut {NoStop}%
\bibitem [{\citenamefont {Aaboud}\ \emph {et~al.}(2018)\citenamefont {Aaboud}
  \emph {et~al.}}]{ATLAS:2018mrn}%
  \BibitemOpen
  \bibfield  {author} {\bibinfo {author} {\bibfnamefont {M.}~\bibnamefont
  {Aaboud}} \emph {et~al.} (\bibinfo {collaboration} {ATLAS}),\ }\href
  {https://doi.org/10.1103/PhysRevD.98.092008} {\bibfield  {journal} {\bibinfo
  {journal} {Phys. Rev. D}\ }\textbf {\bibinfo {volume} {98}},\ \bibinfo
  {pages} {092008} (\bibinfo {year} {2018})},\ \Eprint
  {https://arxiv.org/abs/1807.06573} {arXiv:1807.06573 [hep-ex]} \BibitemShut
  {NoStop}%
\bibitem [{\citenamefont {Angelescu}\ \emph {et~al.}(2020)\citenamefont
  {Angelescu}, \citenamefont {Faroughy},\ and\ \citenamefont
  {Sumensari}}]{Angelescu:2020uug}%
  \BibitemOpen
  \bibfield  {author} {\bibinfo {author} {\bibfnamefont {A.}~\bibnamefont
  {Angelescu}}, \bibinfo {author} {\bibfnamefont {D.~A.}\ \bibnamefont
  {Faroughy}},\ and\ \bibinfo {author} {\bibfnamefont {O.}~\bibnamefont
  {Sumensari}},\ }\href {https://doi.org/10.1140/epjc/s10052-020-8210-5}
  {\bibfield  {journal} {\bibinfo  {journal} {Eur. Phys. J. C}\ }\textbf
  {\bibinfo {volume} {80}},\ \bibinfo {pages} {641} (\bibinfo {year} {2020})},\
  \Eprint {https://arxiv.org/abs/2002.05684} {arXiv:2002.05684 [hep-ph]}
  \BibitemShut {NoStop}%
\bibitem [{\citenamefont {Descotes-Genon}\ \emph {et~al.}(2023)\citenamefont
  {Descotes-Genon}, \citenamefont {Faroughy}, \citenamefont {Plakias},\ and\
  \citenamefont {Sumensari}}]{Descotes-Genon:2023pen}%
  \BibitemOpen
  \bibfield  {author} {\bibinfo {author} {\bibfnamefont {S.}~\bibnamefont
  {Descotes-Genon}}, \bibinfo {author} {\bibfnamefont {D.~A.}\ \bibnamefont
  {Faroughy}}, \bibinfo {author} {\bibfnamefont {I.}~\bibnamefont {Plakias}},\
  and\ \bibinfo {author} {\bibfnamefont {O.}~\bibnamefont {Sumensari}},\ }\href
  {https://doi.org/10.1140/epjc/s10052-023-11860-w} {\bibfield  {journal}
  {\bibinfo  {journal} {Eur. Phys. J. C}\ }\textbf {\bibinfo {volume} {83}},\
  \bibinfo {pages} {753} (\bibinfo {year} {2023})},\ \Eprint
  {https://arxiv.org/abs/2303.07521} {arXiv:2303.07521 [hep-ph]} \BibitemShut
  {NoStop}%
\bibitem [{\citenamefont {Lai}\ \emph {et~al.}(2022{\natexlab{a}})\citenamefont
  {Lai}, \citenamefont {Li}, \citenamefont {Yan},\ and\ \citenamefont
  {Yang}}]{Lai:2021sww}%
  \BibitemOpen
  \bibfield  {author} {\bibinfo {author} {\bibfnamefont {L.-F.}\ \bibnamefont
  {Lai}}, \bibinfo {author} {\bibfnamefont {X.-Q.}\ \bibnamefont {Li}},
  \bibinfo {author} {\bibfnamefont {X.-S.}\ \bibnamefont {Yan}},\ and\ \bibinfo
  {author} {\bibfnamefont {Y.-D.}\ \bibnamefont {Yang}},\ }\href
  {https://doi.org/10.1103/PhysRevD.105.035007} {\bibfield  {journal} {\bibinfo
   {journal} {Phys. Rev. D}\ }\textbf {\bibinfo {volume} {105}},\ \bibinfo
  {pages} {035007} (\bibinfo {year} {2022}{\natexlab{a}})},\ \Eprint
  {https://arxiv.org/abs/2111.01463} {arXiv:2111.01463 [hep-ph]} \BibitemShut
  {NoStop}%
\bibitem [{\citenamefont {Fleischer}\ \emph {et~al.}(2020)\citenamefont
  {Fleischer}, \citenamefont {Jaarsma},\ and\ \citenamefont
  {Koole}}]{Fleischer:2019wlx}%
  \BibitemOpen
  \bibfield  {author} {\bibinfo {author} {\bibfnamefont {R.}~\bibnamefont
  {Fleischer}}, \bibinfo {author} {\bibfnamefont {R.}~\bibnamefont {Jaarsma}},\
  and\ \bibinfo {author} {\bibfnamefont {G.}~\bibnamefont {Koole}},\ }\href
  {https://doi.org/10.1140/epjc/s10052-020-7702-7} {\bibfield  {journal}
  {\bibinfo  {journal} {Eur. Phys. J. C}\ }\textbf {\bibinfo {volume} {80}},\
  \bibinfo {pages} {153} (\bibinfo {year} {2020})},\ \Eprint
  {https://arxiv.org/abs/1912.08641} {arXiv:1912.08641 [hep-ph]} \BibitemShut
  {NoStop}%
\bibitem [{\citenamefont {Be\v{c}irevi\'c}\ \emph {et~al.}(2021)\citenamefont
  {Be\v{c}irevi\'c}, \citenamefont {Jaffredo}, \citenamefont {Pe\~nuelas},\
  and\ \citenamefont {Sumensari}}]{Becirevic:2020rzi}%
  \BibitemOpen
  \bibfield  {author} {\bibinfo {author} {\bibfnamefont {D.}~\bibnamefont
  {Be\v{c}irevi\'c}}, \bibinfo {author} {\bibfnamefont {F.}~\bibnamefont
  {Jaffredo}}, \bibinfo {author} {\bibfnamefont {A.}~\bibnamefont
  {Pe\~nuelas}},\ and\ \bibinfo {author} {\bibfnamefont {O.}~\bibnamefont
  {Sumensari}},\ }\href {https://doi.org/10.1007/JHEP05(2021)175} {\bibfield
  {journal} {\bibinfo  {journal} {JHEP}\ }\textbf {\bibinfo {volume} {05}},\
  \bibinfo {pages} {175}},\ \Eprint {https://arxiv.org/abs/2012.09872}
  {arXiv:2012.09872 [hep-ph]} \BibitemShut {NoStop}%
\bibitem [{\citenamefont {Leng}\ \emph {et~al.}(2021)\citenamefont {Leng},
  \citenamefont {Mu}, \citenamefont {Zou},\ and\ \citenamefont
  {Li}}]{Leng:2020fei}%
  \BibitemOpen
  \bibfield  {author} {\bibinfo {author} {\bibfnamefont {X.}~\bibnamefont
  {Leng}}, \bibinfo {author} {\bibfnamefont {X.-L.}\ \bibnamefont {Mu}},
  \bibinfo {author} {\bibfnamefont {Z.-T.}\ \bibnamefont {Zou}},\ and\ \bibinfo
  {author} {\bibfnamefont {Y.}~\bibnamefont {Li}},\ }\href
  {https://doi.org/10.1088/1674-1137/abf489} {\bibfield  {journal} {\bibinfo
  {journal} {Chin. Phys. C}\ }\textbf {\bibinfo {volume} {45}},\ \bibinfo
  {pages} {063107} (\bibinfo {year} {2021})},\ \Eprint
  {https://arxiv.org/abs/2011.01061} {arXiv:2011.01061 [hep-ph]} \BibitemShut
  {NoStop}%
\bibitem [{\citenamefont {Colangelo}\ \emph {et~al.}(2021)\citenamefont
  {Colangelo}, \citenamefont {De~Fazio},\ and\ \citenamefont
  {Loparco}}]{Colangelo:2021dnv}%
  \BibitemOpen
  \bibfield  {author} {\bibinfo {author} {\bibfnamefont {P.}~\bibnamefont
  {Colangelo}}, \bibinfo {author} {\bibfnamefont {F.}~\bibnamefont
  {De~Fazio}},\ and\ \bibinfo {author} {\bibfnamefont {F.}~\bibnamefont
  {Loparco}},\ }\href {https://doi.org/10.1103/PhysRevD.103.075019} {\bibfield
  {journal} {\bibinfo  {journal} {Phys. Rev. D}\ }\textbf {\bibinfo {volume}
  {103}},\ \bibinfo {pages} {075019} (\bibinfo {year} {2021})},\ \Eprint
  {https://arxiv.org/abs/2102.05365} {arXiv:2102.05365 [hep-ph]} \BibitemShut
  {NoStop}%
\bibitem [{\citenamefont {Workman}\ and\ \citenamefont
  {Others}(2022)}]{Workman:2022ynf}%
  \BibitemOpen
  \bibfield  {author} {\bibinfo {author} {\bibfnamefont {R.~L.}\ \bibnamefont
  {Workman}}\ and\ \bibinfo {author} {\bibnamefont {Others}} (\bibinfo
  {collaboration} {Particle Data Group}),\ }\href
  {https://doi.org/10.1093/ptep/ptac097} {\bibfield  {journal} {\bibinfo
  {journal} {PTEP}\ }\textbf {\bibinfo {volume} {2022}},\ \bibinfo {pages}
  {083C01} (\bibinfo {year} {2022})}\BibitemShut {NoStop}%
\bibitem [{\citenamefont {Arrington}\ \emph {et~al.}(2022)\citenamefont
  {Arrington} \emph {et~al.}}]{Arrington:2021alx}%
  \BibitemOpen
  \bibfield  {author} {\bibinfo {author} {\bibfnamefont {J.}~\bibnamefont
  {Arrington}} \emph {et~al.},\ }\href
  {https://doi.org/10.1016/j.ppnp.2022.103985} {\bibfield  {journal} {\bibinfo
  {journal} {Prog. Part. Nucl. Phys.}\ }\textbf {\bibinfo {volume} {127}},\
  \bibinfo {pages} {103985} (\bibinfo {year} {2022})},\ \Eprint
  {https://arxiv.org/abs/2112.00060} {arXiv:2112.00060 [nucl-ex]} \BibitemShut
  {NoStop}%
\bibitem [{\citenamefont {Vergados}(1986)}]{Vergados:1985pq}%
  \BibitemOpen
  \bibfield  {author} {\bibinfo {author} {\bibfnamefont {J.~D.}\ \bibnamefont
  {Vergados}},\ }\href {https://doi.org/10.1016/0370-1573(86)90088-8}
  {\bibfield  {journal} {\bibinfo  {journal} {Phys. Rept.}\ }\textbf {\bibinfo
  {volume} {133}},\ \bibinfo {pages} {1} (\bibinfo {year} {1986})}\BibitemShut
  {NoStop}%
\bibitem [{\citenamefont {Bernabeu}\ \emph {et~al.}(1993)\citenamefont
  {Bernabeu}, \citenamefont {Nardi},\ and\ \citenamefont
  {Tommasini}}]{Bernabeu:1993ta}%
  \BibitemOpen
  \bibfield  {author} {\bibinfo {author} {\bibfnamefont {J.}~\bibnamefont
  {Bernabeu}}, \bibinfo {author} {\bibfnamefont {E.}~\bibnamefont {Nardi}},\
  and\ \bibinfo {author} {\bibfnamefont {D.}~\bibnamefont {Tommasini}},\ }\href
  {https://doi.org/10.1016/0550-3213(93)90446-V} {\bibfield  {journal}
  {\bibinfo  {journal} {Nucl. Phys. B}\ }\textbf {\bibinfo {volume} {409}},\
  \bibinfo {pages} {69} (\bibinfo {year} {1993})},\ \Eprint
  {https://arxiv.org/abs/hep-ph/9306251} {arXiv:hep-ph/9306251} \BibitemShut
  {NoStop}%
\bibitem [{\citenamefont {Hirsch}\ \emph {et~al.}(1995)\citenamefont {Hirsch},
  \citenamefont {Klapdor-Kleingrothaus},\ and\ \citenamefont
  {Kovalenko}}]{Hirsch:1995zi}%
  \BibitemOpen
  \bibfield  {author} {\bibinfo {author} {\bibfnamefont {M.}~\bibnamefont
  {Hirsch}}, \bibinfo {author} {\bibfnamefont {H.~V.}\ \bibnamefont
  {Klapdor-Kleingrothaus}},\ and\ \bibinfo {author} {\bibfnamefont {S.~G.}\
  \bibnamefont {Kovalenko}},\ }\href
  {https://doi.org/10.1103/PhysRevLett.75.17} {\bibfield  {journal} {\bibinfo
  {journal} {Phys. Rev. Lett.}\ }\textbf {\bibinfo {volume} {75}},\ \bibinfo
  {pages} {17} (\bibinfo {year} {1995})}\BibitemShut {NoStop}%
\bibitem [{\citenamefont {Faessler}\ \emph
  {et~al.}(1998{\natexlab{a}})\citenamefont {Faessler}, \citenamefont
  {Kovalenko},\ and\ \citenamefont {Simkovic}}]{Faessler:1997db}%
  \BibitemOpen
  \bibfield  {author} {\bibinfo {author} {\bibfnamefont {A.}~\bibnamefont
  {Faessler}}, \bibinfo {author} {\bibfnamefont {S.}~\bibnamefont
  {Kovalenko}},\ and\ \bibinfo {author} {\bibfnamefont {F.}~\bibnamefont
  {Simkovic}},\ }\href {https://doi.org/10.1103/PhysRevD.58.055004} {\bibfield
  {journal} {\bibinfo  {journal} {Phys. Rev. D}\ }\textbf {\bibinfo {volume}
  {58}},\ \bibinfo {pages} {055004} (\bibinfo {year} {1998}{\natexlab{a}})},\
  \Eprint {https://arxiv.org/abs/hep-ph/9712535} {arXiv:hep-ph/9712535}
  \BibitemShut {NoStop}%
\bibitem [{\citenamefont {Hisano}\ \emph {et~al.}(1996)\citenamefont {Hisano},
  \citenamefont {Moroi}, \citenamefont {Tobe},\ and\ \citenamefont
  {Yamaguchi}}]{Hisano:1995cp}%
  \BibitemOpen
  \bibfield  {author} {\bibinfo {author} {\bibfnamefont {J.}~\bibnamefont
  {Hisano}}, \bibinfo {author} {\bibfnamefont {T.}~\bibnamefont {Moroi}},
  \bibinfo {author} {\bibfnamefont {K.}~\bibnamefont {Tobe}},\ and\ \bibinfo
  {author} {\bibfnamefont {M.}~\bibnamefont {Yamaguchi}},\ }\href
  {https://doi.org/10.1103/PhysRevD.53.2442} {\bibfield  {journal} {\bibinfo
  {journal} {Phys. Rev. D}\ }\textbf {\bibinfo {volume} {53}},\ \bibinfo
  {pages} {2442} (\bibinfo {year} {1996})},\ \Eprint
  {https://arxiv.org/abs/hep-ph/9510309} {arXiv:hep-ph/9510309} \BibitemShut
  {NoStop}%
\bibitem [{\citenamefont {Faessler}\ \emph {et~al.}(1997)\citenamefont
  {Faessler}, \citenamefont {Kovalenko}, \citenamefont {Simkovic},\ and\
  \citenamefont {Schwieger}}]{Faessler:1996ph}%
  \BibitemOpen
  \bibfield  {author} {\bibinfo {author} {\bibfnamefont {A.}~\bibnamefont
  {Faessler}}, \bibinfo {author} {\bibfnamefont {S.}~\bibnamefont {Kovalenko}},
  \bibinfo {author} {\bibfnamefont {F.}~\bibnamefont {Simkovic}},\ and\
  \bibinfo {author} {\bibfnamefont {J.}~\bibnamefont {Schwieger}},\ }\href
  {https://doi.org/10.1103/PhysRevLett.78.183} {\bibfield  {journal} {\bibinfo
  {journal} {Phys. Rev. Lett.}\ }\textbf {\bibinfo {volume} {78}},\ \bibinfo
  {pages} {183} (\bibinfo {year} {1997})},\ \Eprint
  {https://arxiv.org/abs/hep-ph/9612357} {arXiv:hep-ph/9612357} \BibitemShut
  {NoStop}%
\bibitem [{\citenamefont {Faessler}\ \emph
  {et~al.}(1998{\natexlab{b}})\citenamefont {Faessler}, \citenamefont
  {Kovalenko},\ and\ \citenamefont {Simkovic}}]{Faessler:1998qv}%
  \BibitemOpen
  \bibfield  {author} {\bibinfo {author} {\bibfnamefont {A.}~\bibnamefont
  {Faessler}}, \bibinfo {author} {\bibfnamefont {S.}~\bibnamefont
  {Kovalenko}},\ and\ \bibinfo {author} {\bibfnamefont {F.}~\bibnamefont
  {Simkovic}},\ }\href {https://doi.org/10.1103/PhysRevD.58.115004} {\bibfield
  {journal} {\bibinfo  {journal} {Phys. Rev. D}\ }\textbf {\bibinfo {volume}
  {58}},\ \bibinfo {pages} {115004} (\bibinfo {year} {1998}{\natexlab{b}})},\
  \Eprint {https://arxiv.org/abs/hep-ph/9803253} {arXiv:hep-ph/9803253}
  \BibitemShut {NoStop}%
\bibitem [{\citenamefont {Faessler}\ \emph {et~al.}(2000)\citenamefont
  {Faessler}, \citenamefont {Kosmas}, \citenamefont {Kovalenko},\ and\
  \citenamefont {Vergados}}]{Faessler:1999jf}%
  \BibitemOpen
  \bibfield  {author} {\bibinfo {author} {\bibfnamefont {A.}~\bibnamefont
  {Faessler}}, \bibinfo {author} {\bibfnamefont {T.~S.}\ \bibnamefont
  {Kosmas}}, \bibinfo {author} {\bibfnamefont {S.}~\bibnamefont {Kovalenko}},\
  and\ \bibinfo {author} {\bibfnamefont {J.~D.}\ \bibnamefont {Vergados}},\
  }\href {https://doi.org/10.1016/S0550-3213(00)00446-6} {\bibfield  {journal}
  {\bibinfo  {journal} {Nucl. Phys. B}\ }\textbf {\bibinfo {volume} {587}},\
  \bibinfo {pages} {25} (\bibinfo {year} {2000})},\ \Eprint
  {https://arxiv.org/abs/hep-ph/9904335} {arXiv:hep-ph/9904335} \BibitemShut
  {NoStop}%
\bibitem [{\citenamefont {Kosmas}\ \emph
  {et~al.}(2001{\natexlab{a}})\citenamefont {Kosmas}, \citenamefont
  {Kovalenko},\ and\ \citenamefont {Schmidt}}]{Kosmas:2001mv}%
  \BibitemOpen
  \bibfield  {author} {\bibinfo {author} {\bibfnamefont {T.~S.}\ \bibnamefont
  {Kosmas}}, \bibinfo {author} {\bibfnamefont {S.}~\bibnamefont {Kovalenko}},\
  and\ \bibinfo {author} {\bibfnamefont {I.}~\bibnamefont {Schmidt}},\ }\href
  {https://doi.org/10.1016/S0370-2693(01)00657-8} {\bibfield  {journal}
  {\bibinfo  {journal} {Phys. Lett. B}\ }\textbf {\bibinfo {volume} {511}},\
  \bibinfo {pages} {203} (\bibinfo {year} {2001}{\natexlab{a}})},\ \Eprint
  {https://arxiv.org/abs/hep-ph/0102101} {arXiv:hep-ph/0102101} \BibitemShut
  {NoStop}%
\bibitem [{\citenamefont {Kosmas}\ \emph
  {et~al.}(2001{\natexlab{b}})\citenamefont {Kosmas}, \citenamefont
  {Kovalenko},\ and\ \citenamefont {Schmidt}}]{Kosmas:2001sx}%
  \BibitemOpen
  \bibfield  {author} {\bibinfo {author} {\bibfnamefont {T.~S.}\ \bibnamefont
  {Kosmas}}, \bibinfo {author} {\bibfnamefont {S.}~\bibnamefont {Kovalenko}},\
  and\ \bibinfo {author} {\bibfnamefont {I.}~\bibnamefont {Schmidt}},\ }\href
  {https://doi.org/10.1016/S0370-2693(01)01096-6} {\bibfield  {journal}
  {\bibinfo  {journal} {Phys. Lett. B}\ }\textbf {\bibinfo {volume} {519}},\
  \bibinfo {pages} {78} (\bibinfo {year} {2001}{\natexlab{b}})},\ \Eprint
  {https://arxiv.org/abs/hep-ph/0107292} {arXiv:hep-ph/0107292} \BibitemShut
  {NoStop}%
\bibitem [{\citenamefont {Kuno}\ and\ \citenamefont
  {Okada}(2001)}]{Kuno:1999jp}%
  \BibitemOpen
  \bibfield  {author} {\bibinfo {author} {\bibfnamefont {Y.}~\bibnamefont
  {Kuno}}\ and\ \bibinfo {author} {\bibfnamefont {Y.}~\bibnamefont {Okada}},\
  }\href {https://doi.org/10.1103/RevModPhys.73.151} {\bibfield  {journal}
  {\bibinfo  {journal} {Rev. Mod. Phys.}\ }\textbf {\bibinfo {volume} {73}},\
  \bibinfo {pages} {151} (\bibinfo {year} {2001})},\ \Eprint
  {https://arxiv.org/abs/hep-ph/9909265} {arXiv:hep-ph/9909265} \BibitemShut
  {NoStop}%
\bibitem [{\citenamefont {Faessler}\ \emph
  {et~al.}(2004{\natexlab{a}})\citenamefont {Faessler}, \citenamefont
  {Gutsche}, \citenamefont {Kovalenko}, \citenamefont {Lyubovitskij},
  \citenamefont {Schmidt},\ and\ \citenamefont {Simkovic}}]{Faessler:2004jt}%
  \BibitemOpen
  \bibfield  {author} {\bibinfo {author} {\bibfnamefont {A.}~\bibnamefont
  {Faessler}}, \bibinfo {author} {\bibfnamefont {T.}~\bibnamefont {Gutsche}},
  \bibinfo {author} {\bibfnamefont {S.}~\bibnamefont {Kovalenko}}, \bibinfo
  {author} {\bibfnamefont {V.~E.}\ \bibnamefont {Lyubovitskij}}, \bibinfo
  {author} {\bibfnamefont {I.}~\bibnamefont {Schmidt}},\ and\ \bibinfo {author}
  {\bibfnamefont {F.}~\bibnamefont {Simkovic}},\ }\href
  {https://doi.org/10.1016/j.physletb.2004.03.068} {\bibfield  {journal}
  {\bibinfo  {journal} {Phys. Lett. B}\ }\textbf {\bibinfo {volume} {590}},\
  \bibinfo {pages} {57} (\bibinfo {year} {2004}{\natexlab{a}})},\ \Eprint
  {https://arxiv.org/abs/hep-ph/0403033} {arXiv:hep-ph/0403033} \BibitemShut
  {NoStop}%
\bibitem [{\citenamefont {Faessler}\ \emph
  {et~al.}(2004{\natexlab{b}})\citenamefont {Faessler}, \citenamefont
  {Gutsche}, \citenamefont {Kovalenko}, \citenamefont {Lyubovitskij},
  \citenamefont {Schmidt},\ and\ \citenamefont {Simkovic}}]{Faessler:2004ea}%
  \BibitemOpen
  \bibfield  {author} {\bibinfo {author} {\bibfnamefont {A.}~\bibnamefont
  {Faessler}}, \bibinfo {author} {\bibfnamefont {T.}~\bibnamefont {Gutsche}},
  \bibinfo {author} {\bibfnamefont {S.}~\bibnamefont {Kovalenko}}, \bibinfo
  {author} {\bibfnamefont {V.~E.}\ \bibnamefont {Lyubovitskij}}, \bibinfo
  {author} {\bibfnamefont {I.}~\bibnamefont {Schmidt}},\ and\ \bibinfo {author}
  {\bibfnamefont {F.}~\bibnamefont {Simkovic}},\ }\href
  {https://doi.org/10.1103/PhysRevD.70.055008} {\bibfield  {journal} {\bibinfo
  {journal} {Phys. Rev. D}\ }\textbf {\bibinfo {volume} {70}},\ \bibinfo
  {pages} {055008} (\bibinfo {year} {2004}{\natexlab{b}})},\ \Eprint
  {https://arxiv.org/abs/hep-ph/0405164} {arXiv:hep-ph/0405164} \BibitemShut
  {NoStop}%
\bibitem [{\citenamefont {Faessler}\ \emph {et~al.}(2005)\citenamefont
  {Faessler}, \citenamefont {Gutsche}, \citenamefont {Kovalenko}, \citenamefont
  {Lyubovitskij},\ and\ \citenamefont {Schmidt}}]{Faessler:2005hx}%
  \BibitemOpen
  \bibfield  {author} {\bibinfo {author} {\bibfnamefont {A.}~\bibnamefont
  {Faessler}}, \bibinfo {author} {\bibfnamefont {T.}~\bibnamefont {Gutsche}},
  \bibinfo {author} {\bibfnamefont {S.}~\bibnamefont {Kovalenko}}, \bibinfo
  {author} {\bibfnamefont {V.~E.}\ \bibnamefont {Lyubovitskij}},\ and\ \bibinfo
  {author} {\bibfnamefont {I.}~\bibnamefont {Schmidt}},\ }\href
  {https://doi.org/10.1103/PhysRevD.72.075006} {\bibfield  {journal} {\bibinfo
  {journal} {Phys. Rev. D}\ }\textbf {\bibinfo {volume} {72}},\ \bibinfo
  {pages} {075006} (\bibinfo {year} {2005})},\ \Eprint
  {https://arxiv.org/abs/hep-ph/0507033} {arXiv:hep-ph/0507033} \BibitemShut
  {NoStop}%
\bibitem [{\citenamefont {Gutsche}\ \emph {et~al.}(2010)\citenamefont
  {Gutsche}, \citenamefont {Helo}, \citenamefont {Kovalenko},\ and\
  \citenamefont {Lyubovitskij}}]{Gutsche:2009vp}%
  \BibitemOpen
  \bibfield  {author} {\bibinfo {author} {\bibfnamefont {T.}~\bibnamefont
  {Gutsche}}, \bibinfo {author} {\bibfnamefont {J.~C.}\ \bibnamefont {Helo}},
  \bibinfo {author} {\bibfnamefont {S.}~\bibnamefont {Kovalenko}},\ and\
  \bibinfo {author} {\bibfnamefont {V.~E.}\ \bibnamefont {Lyubovitskij}},\
  }\href {https://doi.org/10.1103/PhysRevD.81.037702} {\bibfield  {journal}
  {\bibinfo  {journal} {Phys. Rev. D}\ }\textbf {\bibinfo {volume} {81}},\
  \bibinfo {pages} {037702} (\bibinfo {year} {2010})},\ \Eprint
  {https://arxiv.org/abs/0912.4562} {arXiv:0912.4562 [hep-ph]} \BibitemShut
  {NoStop}%
\bibitem [{\citenamefont {Gutsche}\ \emph {et~al.}(2011)\citenamefont
  {Gutsche}, \citenamefont {Helo}, \citenamefont {Kovalenko},\ and\
  \citenamefont {Lyubovitskij}}]{Gutsche:2011bi}%
  \BibitemOpen
  \bibfield  {author} {\bibinfo {author} {\bibfnamefont {T.}~\bibnamefont
  {Gutsche}}, \bibinfo {author} {\bibfnamefont {J.~C.}\ \bibnamefont {Helo}},
  \bibinfo {author} {\bibfnamefont {S.}~\bibnamefont {Kovalenko}},\ and\
  \bibinfo {author} {\bibfnamefont {V.~E.}\ \bibnamefont {Lyubovitskij}},\
  }\href {https://doi.org/10.1103/PhysRevD.83.115015} {\bibfield  {journal}
  {\bibinfo  {journal} {Phys. Rev. D}\ }\textbf {\bibinfo {volume} {83}},\
  \bibinfo {pages} {115015} (\bibinfo {year} {2011})},\ \Eprint
  {https://arxiv.org/abs/1103.1317} {arXiv:1103.1317 [hep-ph]} \BibitemShut
  {NoStop}%
\bibitem [{\citenamefont {Gonzalez}\ \emph {et~al.}(2013)\citenamefont
  {Gonzalez}, \citenamefont {Gutsche}, \citenamefont {Helo}, \citenamefont
  {Kovalenko}, \citenamefont {Lyubovitskij},\ and\ \citenamefont
  {Schmidt}}]{Gonzalez:2013rea}%
  \BibitemOpen
  \bibfield  {author} {\bibinfo {author} {\bibfnamefont {M.}~\bibnamefont
  {Gonzalez}}, \bibinfo {author} {\bibfnamefont {T.}~\bibnamefont {Gutsche}},
  \bibinfo {author} {\bibfnamefont {J.~C.}\ \bibnamefont {Helo}}, \bibinfo
  {author} {\bibfnamefont {S.}~\bibnamefont {Kovalenko}}, \bibinfo {author}
  {\bibfnamefont {V.~E.}\ \bibnamefont {Lyubovitskij}},\ and\ \bibinfo {author}
  {\bibfnamefont {I.}~\bibnamefont {Schmidt}},\ }\href
  {https://doi.org/10.1103/PhysRevD.87.096020} {\bibfield  {journal} {\bibinfo
  {journal} {Phys. Rev. D}\ }\textbf {\bibinfo {volume} {87}},\ \bibinfo
  {pages} {096020} (\bibinfo {year} {2013})},\ \Eprint
  {https://arxiv.org/abs/1303.0596} {arXiv:1303.0596 [hep-ph]} \BibitemShut
  {NoStop}%
\bibitem [{\citenamefont {Black}\ \emph {et~al.}(2002)\citenamefont {Black},
  \citenamefont {Han}, \citenamefont {He},\ and\ \citenamefont
  {Sher}}]{Black:2002wh}%
  \BibitemOpen
  \bibfield  {author} {\bibinfo {author} {\bibfnamefont {D.}~\bibnamefont
  {Black}}, \bibinfo {author} {\bibfnamefont {T.}~\bibnamefont {Han}}, \bibinfo
  {author} {\bibfnamefont {H.-J.}\ \bibnamefont {He}},\ and\ \bibinfo {author}
  {\bibfnamefont {M.}~\bibnamefont {Sher}},\ }\href
  {https://doi.org/10.1103/PhysRevD.66.053002} {\bibfield  {journal} {\bibinfo
  {journal} {Phys. Rev. D}\ }\textbf {\bibinfo {volume} {66}},\ \bibinfo
  {pages} {053002} (\bibinfo {year} {2002})},\ \Eprint
  {https://arxiv.org/abs/hep-ph/0206056} {arXiv:hep-ph/0206056} \BibitemShut
  {NoStop}%
\bibitem [{\citenamefont {Kitano}\ \emph {et~al.}(2002)\citenamefont {Kitano},
  \citenamefont {Koike},\ and\ \citenamefont {Okada}}]{Kitano:2002mt}%
  \BibitemOpen
  \bibfield  {author} {\bibinfo {author} {\bibfnamefont {R.}~\bibnamefont
  {Kitano}}, \bibinfo {author} {\bibfnamefont {M.}~\bibnamefont {Koike}},\ and\
  \bibinfo {author} {\bibfnamefont {Y.}~\bibnamefont {Okada}},\ }\href
  {https://doi.org/10.1103/PhysRevD.76.059902} {\bibfield  {journal} {\bibinfo
  {journal} {Phys. Rev. D}\ }\textbf {\bibinfo {volume} {66}},\ \bibinfo
  {pages} {096002} (\bibinfo {year} {2002})},\ \bibinfo {note} {[Erratum:
  Phys.Rev.D 76, 059902 (2007)]},\ \Eprint
  {https://arxiv.org/abs/hep-ph/0203110} {arXiv:hep-ph/0203110} \BibitemShut
  {NoStop}%
\bibitem [{\citenamefont {Cirigliano}\ \emph {et~al.}(2009)\citenamefont
  {Cirigliano}, \citenamefont {Kitano}, \citenamefont {Okada},\ and\
  \citenamefont {Tuzon}}]{Cirigliano:2009bz}%
  \BibitemOpen
  \bibfield  {author} {\bibinfo {author} {\bibfnamefont {V.}~\bibnamefont
  {Cirigliano}}, \bibinfo {author} {\bibfnamefont {R.}~\bibnamefont {Kitano}},
  \bibinfo {author} {\bibfnamefont {Y.}~\bibnamefont {Okada}},\ and\ \bibinfo
  {author} {\bibfnamefont {P.}~\bibnamefont {Tuzon}},\ }\href
  {https://doi.org/10.1103/PhysRevD.80.013002} {\bibfield  {journal} {\bibinfo
  {journal} {Phys. Rev. D}\ }\textbf {\bibinfo {volume} {80}},\ \bibinfo
  {pages} {013002} (\bibinfo {year} {2009})},\ \Eprint
  {https://arxiv.org/abs/0904.0957} {arXiv:0904.0957 [hep-ph]} \BibitemShut
  {NoStop}%
\bibitem [{\citenamefont {Abada}\ \emph {et~al.}(2014)\citenamefont {Abada},
  \citenamefont {Krauss}, \citenamefont {Porod}, \citenamefont {Staub},
  \citenamefont {Vicente},\ and\ \citenamefont {Weiland}}]{Abada:2014kba}%
  \BibitemOpen
  \bibfield  {author} {\bibinfo {author} {\bibfnamefont {A.}~\bibnamefont
  {Abada}}, \bibinfo {author} {\bibfnamefont {M.~E.}\ \bibnamefont {Krauss}},
  \bibinfo {author} {\bibfnamefont {W.}~\bibnamefont {Porod}}, \bibinfo
  {author} {\bibfnamefont {F.}~\bibnamefont {Staub}}, \bibinfo {author}
  {\bibfnamefont {A.}~\bibnamefont {Vicente}},\ and\ \bibinfo {author}
  {\bibfnamefont {C.}~\bibnamefont {Weiland}},\ }\href
  {https://doi.org/10.1007/JHEP11(2014)048} {\bibfield  {journal} {\bibinfo
  {journal} {JHEP}\ }\textbf {\bibinfo {volume} {11}},\ \bibinfo {pages}
  {048}},\ \Eprint {https://arxiv.org/abs/1408.0138} {arXiv:1408.0138 [hep-ph]}
  \BibitemShut {NoStop}%
\bibitem [{\citenamefont {Crivellin}\ \emph {et~al.}(2017)\citenamefont
  {Crivellin}, \citenamefont {Davidson}, \citenamefont {Pruna},\ and\
  \citenamefont {Signer}}]{Crivellin:2017rmk}%
  \BibitemOpen
  \bibfield  {author} {\bibinfo {author} {\bibfnamefont {A.}~\bibnamefont
  {Crivellin}}, \bibinfo {author} {\bibfnamefont {S.}~\bibnamefont {Davidson}},
  \bibinfo {author} {\bibfnamefont {G.~M.}\ \bibnamefont {Pruna}},\ and\
  \bibinfo {author} {\bibfnamefont {A.}~\bibnamefont {Signer}},\ }\href
  {https://doi.org/10.1007/JHEP05(2017)117} {\bibfield  {journal} {\bibinfo
  {journal} {JHEP}\ }\textbf {\bibinfo {volume} {05}},\ \bibinfo {pages}
  {117}},\ \Eprint {https://arxiv.org/abs/1702.03020} {arXiv:1702.03020
  [hep-ph]} \BibitemShut {NoStop}%
\bibitem [{\citenamefont {Davidson}\ \emph {et~al.}(2018)\citenamefont
  {Davidson}, \citenamefont {Gorbahn},\ and\ \citenamefont
  {Leak}}]{Davidson:2018zuo}%
  \BibitemOpen
  \bibfield  {author} {\bibinfo {author} {\bibfnamefont {S.}~\bibnamefont
  {Davidson}}, \bibinfo {author} {\bibfnamefont {M.}~\bibnamefont {Gorbahn}},\
  and\ \bibinfo {author} {\bibfnamefont {M.}~\bibnamefont {Leak}},\ }\href
  {https://doi.org/10.1103/PhysRevD.98.095014} {\bibfield  {journal} {\bibinfo
  {journal} {Phys. Rev. D}\ }\textbf {\bibinfo {volume} {98}},\ \bibinfo
  {pages} {095014} (\bibinfo {year} {2018})},\ \Eprint
  {https://arxiv.org/abs/1807.04283} {arXiv:1807.04283 [hep-ph]} \BibitemShut
  {NoStop}%
\bibitem [{\citenamefont {Gninenko}\ \emph {et~al.}(2018)\citenamefont
  {Gninenko}, \citenamefont {Kovalenko}, \citenamefont {Kuleshov},
  \citenamefont {Lyubovitskij},\ and\ \citenamefont
  {Zhevlakov}}]{Gninenko:2018num}%
  \BibitemOpen
  \bibfield  {author} {\bibinfo {author} {\bibfnamefont {S.}~\bibnamefont
  {Gninenko}}, \bibinfo {author} {\bibfnamefont {S.}~\bibnamefont {Kovalenko}},
  \bibinfo {author} {\bibfnamefont {S.}~\bibnamefont {Kuleshov}}, \bibinfo
  {author} {\bibfnamefont {V.~E.}\ \bibnamefont {Lyubovitskij}},\ and\ \bibinfo
  {author} {\bibfnamefont {A.~S.}\ \bibnamefont {Zhevlakov}},\ }\href
  {https://doi.org/10.1103/PhysRevD.98.015007} {\bibfield  {journal} {\bibinfo
  {journal} {Phys. Rev. D}\ }\textbf {\bibinfo {volume} {98}},\ \bibinfo
  {pages} {015007} (\bibinfo {year} {2018})},\ \Eprint
  {https://arxiv.org/abs/1804.05550} {arXiv:1804.05550 [hep-ph]} \BibitemShut
  {NoStop}%
\bibitem [{\citenamefont {Davidson}\ \emph {et~al.}(2019)\citenamefont
  {Davidson}, \citenamefont {Kuno},\ and\ \citenamefont
  {Yamanaka}}]{Davidson:2018kud}%
  \BibitemOpen
  \bibfield  {author} {\bibinfo {author} {\bibfnamefont {S.}~\bibnamefont
  {Davidson}}, \bibinfo {author} {\bibfnamefont {Y.}~\bibnamefont {Kuno}},\
  and\ \bibinfo {author} {\bibfnamefont {M.}~\bibnamefont {Yamanaka}},\ }\href
  {https://doi.org/10.1016/j.physletb.2019.01.042} {\bibfield  {journal}
  {\bibinfo  {journal} {Phys. Lett. B}\ }\textbf {\bibinfo {volume} {790}},\
  \bibinfo {pages} {380} (\bibinfo {year} {2019})},\ \Eprint
  {https://arxiv.org/abs/1810.01884} {arXiv:1810.01884 [hep-ph]} \BibitemShut
  {NoStop}%
\bibitem [{\citenamefont {Dib}\ \emph {et~al.}(2019)\citenamefont {Dib},
  \citenamefont {Gutsche}, \citenamefont {Kovalenko}, \citenamefont
  {Lyubovitskij},\ and\ \citenamefont {Schmidt}}]{Dib:2018rpy}%
  \BibitemOpen
  \bibfield  {author} {\bibinfo {author} {\bibfnamefont {C.~O.}\ \bibnamefont
  {Dib}}, \bibinfo {author} {\bibfnamefont {T.}~\bibnamefont {Gutsche}},
  \bibinfo {author} {\bibfnamefont {S.~G.}\ \bibnamefont {Kovalenko}}, \bibinfo
  {author} {\bibfnamefont {V.~E.}\ \bibnamefont {Lyubovitskij}},\ and\ \bibinfo
  {author} {\bibfnamefont {I.}~\bibnamefont {Schmidt}},\ }\href
  {https://doi.org/10.1103/PhysRevD.99.035020} {\bibfield  {journal} {\bibinfo
  {journal} {Phys. Rev. D}\ }\textbf {\bibinfo {volume} {99}},\ \bibinfo
  {pages} {035020} (\bibinfo {year} {2019})},\ \Eprint
  {https://arxiv.org/abs/1812.02638} {arXiv:1812.02638 [hep-ph]} \BibitemShut
  {NoStop}%
\bibitem [{\citenamefont {Fuentes-Martin}\ \emph {et~al.}(2020)\citenamefont
  {Fuentes-Martin}, \citenamefont {Greljo}, \citenamefont {Martin~Camalich},\
  and\ \citenamefont {Ruiz-Alvarez}}]{Fuentes-Martin:2020lea}%
  \BibitemOpen
  \bibfield  {author} {\bibinfo {author} {\bibfnamefont {J.}~\bibnamefont
  {Fuentes-Martin}}, \bibinfo {author} {\bibfnamefont {A.}~\bibnamefont
  {Greljo}}, \bibinfo {author} {\bibfnamefont {J.}~\bibnamefont
  {Martin~Camalich}},\ and\ \bibinfo {author} {\bibfnamefont {J.~D.}\
  \bibnamefont {Ruiz-Alvarez}},\ }\href
  {https://doi.org/10.1007/JHEP11(2020)080} {\bibfield  {journal} {\bibinfo
  {journal} {JHEP}\ }\textbf {\bibinfo {volume} {11}},\ \bibinfo {pages}
  {080}},\ \Eprint {https://arxiv.org/abs/2003.12421} {arXiv:2003.12421
  [hep-ph]} \BibitemShut {NoStop}%
\bibitem [{\citenamefont {Dor\v{s}ner}\ \emph {et~al.}(2016)\citenamefont
  {Dor\v{s}ner}, \citenamefont {Fajfer}, \citenamefont {Greljo}, \citenamefont
  {Kamenik},\ and\ \citenamefont {Ko\v{s}nik}}]{Dorsner:2016wpm}%
  \BibitemOpen
  \bibfield  {author} {\bibinfo {author} {\bibfnamefont {I.}~\bibnamefont
  {Dor\v{s}ner}}, \bibinfo {author} {\bibfnamefont {S.}~\bibnamefont {Fajfer}},
  \bibinfo {author} {\bibfnamefont {A.}~\bibnamefont {Greljo}}, \bibinfo
  {author} {\bibfnamefont {J.~F.}\ \bibnamefont {Kamenik}},\ and\ \bibinfo
  {author} {\bibfnamefont {N.}~\bibnamefont {Ko\v{s}nik}},\ }\href
  {https://doi.org/10.1016/j.physrep.2016.06.001} {\bibfield  {journal}
  {\bibinfo  {journal} {Phys. Rept.}\ }\textbf {\bibinfo {volume} {641}},\
  \bibinfo {pages} {1} (\bibinfo {year} {2016})},\ \Eprint
  {https://arxiv.org/abs/1603.04993} {arXiv:1603.04993 [hep-ph]} \BibitemShut
  {NoStop}%
\bibitem [{\citenamefont {Mandal}\ and\ \citenamefont
  {Pich}(2019)}]{Mandal:2019gff}%
  \BibitemOpen
  \bibfield  {author} {\bibinfo {author} {\bibfnamefont {R.}~\bibnamefont
  {Mandal}}\ and\ \bibinfo {author} {\bibfnamefont {A.}~\bibnamefont {Pich}},\
  }\href {https://doi.org/10.1007/JHEP12(2019)089} {\bibfield  {journal}
  {\bibinfo  {journal} {JHEP}\ }\textbf {\bibinfo {volume} {12}},\ \bibinfo
  {pages} {089}},\ \Eprint {https://arxiv.org/abs/1908.11155} {arXiv:1908.11155
  [hep-ph]} \BibitemShut {NoStop}%
\bibitem [{\citenamefont {Godbole}\ \emph {et~al.}(2006)\citenamefont
  {Godbole}, \citenamefont {Rindani},\ and\ \citenamefont
  {Singh}}]{Godbole:2006tq}%
  \BibitemOpen
  \bibfield  {author} {\bibinfo {author} {\bibfnamefont {R.~M.}\ \bibnamefont
  {Godbole}}, \bibinfo {author} {\bibfnamefont {S.~D.}\ \bibnamefont
  {Rindani}},\ and\ \bibinfo {author} {\bibfnamefont {R.~K.}\ \bibnamefont
  {Singh}},\ }\href {https://doi.org/10.1088/1126-6708/2006/12/021} {\bibfield
  {journal} {\bibinfo  {journal} {JHEP}\ }\textbf {\bibinfo {volume} {12}},\
  \bibinfo {pages} {021}},\ \Eprint {https://arxiv.org/abs/hep-ph/0605100}
  {arXiv:hep-ph/0605100} \BibitemShut {NoStop}%
\bibitem [{\citenamefont {Haber}(1994)}]{Haber:1994pe}%
  \BibitemOpen
  \bibfield  {author} {\bibinfo {author} {\bibfnamefont {H.~E.}\ \bibnamefont
  {Haber}},\ }in\ \href@noop {} {\emph {\bibinfo {booktitle} {{21st Annual SLAC
  Summer Institute on Particle Physics: Spin Structure in High-energy Processes
  (School: 26 Jul - 3 Aug, Topical Conference: 4-6 Aug) (SSI 93)}}}}\ (\bibinfo
  {year} {1994})\ pp.\ \bibinfo {pages} {231--272},\ \Eprint
  {https://arxiv.org/abs/hep-ph/9405376} {arXiv:hep-ph/9405376} \BibitemShut
  {NoStop}%
\bibitem [{\citenamefont {Kong}\ \emph {et~al.}(2023)\citenamefont {Kong},
  \citenamefont {Lai}, \citenamefont {Li}, \citenamefont {Yan}, \citenamefont
  {Yang},\ and\ \citenamefont {Zheng}}]{Kong:2023kkd}%
  \BibitemOpen
  \bibfield  {author} {\bibinfo {author} {\bibfnamefont {Y.-R.}\ \bibnamefont
  {Kong}}, \bibinfo {author} {\bibfnamefont {L.-F.}\ \bibnamefont {Lai}},
  \bibinfo {author} {\bibfnamefont {X.-Q.}\ \bibnamefont {Li}}, \bibinfo
  {author} {\bibfnamefont {X.-S.}\ \bibnamefont {Yan}}, \bibinfo {author}
  {\bibfnamefont {Y.-D.}\ \bibnamefont {Yang}},\ and\ \bibinfo {author}
  {\bibfnamefont {D.-H.}\ \bibnamefont {Zheng}},\ }\href
  {https://doi.org/10.1103/PhysRevD.108.095036} {\bibfield  {journal} {\bibinfo
   {journal} {Phys. Rev. D}\ }\textbf {\bibinfo {volume} {108}},\ \bibinfo
  {pages} {095036} (\bibinfo {year} {2023})},\ \Eprint
  {https://arxiv.org/abs/2307.07239} {arXiv:2307.07239 [hep-ph]} \BibitemShut
  {NoStop}%
\bibitem [{\citenamefont {Hagiwara}\ \emph {et~al.}(2014)\citenamefont
  {Hagiwara}, \citenamefont {Nojiri},\ and\ \citenamefont
  {Sakaki}}]{Hagiwara:2014tsa}%
  \BibitemOpen
  \bibfield  {author} {\bibinfo {author} {\bibfnamefont {K.}~\bibnamefont
  {Hagiwara}}, \bibinfo {author} {\bibfnamefont {M.~M.}\ \bibnamefont
  {Nojiri}},\ and\ \bibinfo {author} {\bibfnamefont {Y.}~\bibnamefont
  {Sakaki}},\ }\href {https://doi.org/10.1103/PhysRevD.89.094009} {\bibfield
  {journal} {\bibinfo  {journal} {Phys. Rev. D}\ }\textbf {\bibinfo {volume}
  {89}},\ \bibinfo {pages} {094009} (\bibinfo {year} {2014})},\ \Eprint
  {https://arxiv.org/abs/1403.5892} {arXiv:1403.5892 [hep-ph]} \BibitemShut
  {NoStop}%
\bibitem [{\citenamefont {Bordone}\ \emph {et~al.}(2016)\citenamefont
  {Bordone}, \citenamefont {Isidori},\ and\ \citenamefont {van
  Dyk}}]{Bordone:2016tex}%
  \BibitemOpen
  \bibfield  {author} {\bibinfo {author} {\bibfnamefont {M.}~\bibnamefont
  {Bordone}}, \bibinfo {author} {\bibfnamefont {G.}~\bibnamefont {Isidori}},\
  and\ \bibinfo {author} {\bibfnamefont {D.}~\bibnamefont {van Dyk}},\ }\href
  {https://doi.org/10.1140/epjc/s10052-016-4202-x} {\bibfield  {journal}
  {\bibinfo  {journal} {Eur. Phys. J. C}\ }\textbf {\bibinfo {volume} {76}},\
  \bibinfo {pages} {360} (\bibinfo {year} {2016})},\ \Eprint
  {https://arxiv.org/abs/1602.06143} {arXiv:1602.06143 [hep-ph]} \BibitemShut
  {NoStop}%
\bibitem [{\citenamefont {Alonso}\ \emph {et~al.}(2016)\citenamefont {Alonso},
  \citenamefont {Kobach},\ and\ \citenamefont
  {Martin~Camalich}}]{Alonso:2016gym}%
  \BibitemOpen
  \bibfield  {author} {\bibinfo {author} {\bibfnamefont {R.}~\bibnamefont
  {Alonso}}, \bibinfo {author} {\bibfnamefont {A.}~\bibnamefont {Kobach}},\
  and\ \bibinfo {author} {\bibfnamefont {J.}~\bibnamefont {Martin~Camalich}},\
  }\href {https://doi.org/10.1103/PhysRevD.94.094021} {\bibfield  {journal}
  {\bibinfo  {journal} {Phys. Rev. D}\ }\textbf {\bibinfo {volume} {94}},\
  \bibinfo {pages} {094021} (\bibinfo {year} {2016})},\ \Eprint
  {https://arxiv.org/abs/1602.07671} {arXiv:1602.07671 [hep-ph]} \BibitemShut
  {NoStop}%
\bibitem [{\citenamefont {Ligeti}\ \emph {et~al.}(2017)\citenamefont {Ligeti},
  \citenamefont {Papucci},\ and\ \citenamefont {Robinson}}]{Ligeti:2016npd}%
  \BibitemOpen
  \bibfield  {author} {\bibinfo {author} {\bibfnamefont {Z.}~\bibnamefont
  {Ligeti}}, \bibinfo {author} {\bibfnamefont {M.}~\bibnamefont {Papucci}},\
  and\ \bibinfo {author} {\bibfnamefont {D.~J.}\ \bibnamefont {Robinson}},\
  }\href {https://doi.org/10.1007/JHEP01(2017)083} {\bibfield  {journal}
  {\bibinfo  {journal} {JHEP}\ }\textbf {\bibinfo {volume} {01}},\ \bibinfo
  {pages} {083}},\ \Eprint {https://arxiv.org/abs/1610.02045} {arXiv:1610.02045
  [hep-ph]} \BibitemShut {NoStop}%
\bibitem [{\citenamefont {Asadi}\ \emph {et~al.}(2019)\citenamefont {Asadi},
  \citenamefont {Buckley},\ and\ \citenamefont {Shih}}]{Asadi:2018sym}%
  \BibitemOpen
  \bibfield  {author} {\bibinfo {author} {\bibfnamefont {P.}~\bibnamefont
  {Asadi}}, \bibinfo {author} {\bibfnamefont {M.~R.}\ \bibnamefont {Buckley}},\
  and\ \bibinfo {author} {\bibfnamefont {D.}~\bibnamefont {Shih}},\ }\href
  {https://doi.org/10.1103/PhysRevD.99.035015} {\bibfield  {journal} {\bibinfo
  {journal} {Phys. Rev. D}\ }\textbf {\bibinfo {volume} {99}},\ \bibinfo
  {pages} {035015} (\bibinfo {year} {2019})},\ \Eprint
  {https://arxiv.org/abs/1810.06597} {arXiv:1810.06597 [hep-ph]} \BibitemShut
  {NoStop}%
\bibitem [{\citenamefont {Alonso}\ \emph {et~al.}(2019)\citenamefont {Alonso},
  \citenamefont {Martin~Camalich},\ and\ \citenamefont
  {Westhoff}}]{Alonso:2018vwa}%
  \BibitemOpen
  \bibfield  {author} {\bibinfo {author} {\bibfnamefont {R.}~\bibnamefont
  {Alonso}}, \bibinfo {author} {\bibfnamefont {J.}~\bibnamefont
  {Martin~Camalich}},\ and\ \bibinfo {author} {\bibfnamefont {S.}~\bibnamefont
  {Westhoff}},\ }\href {https://doi.org/10.21468/SciPostPhysProc.1.012}
  {\bibfield  {journal} {\bibinfo  {journal} {SciPost Phys. Proc.}\ }\textbf
  {\bibinfo {volume} {1}},\ \bibinfo {pages} {012} (\bibinfo {year} {2019})},\
  \Eprint {https://arxiv.org/abs/1811.05664} {arXiv:1811.05664 [hep-ph]}
  \BibitemShut {NoStop}%
\bibitem [{\citenamefont {Bhattacharya}\ \emph {et~al.}(2020)\citenamefont
  {Bhattacharya}, \citenamefont {Datta}, \citenamefont {Kamali},\ and\
  \citenamefont {London}}]{Bhattacharya:2020lfm}%
  \BibitemOpen
  \bibfield  {author} {\bibinfo {author} {\bibfnamefont {B.}~\bibnamefont
  {Bhattacharya}}, \bibinfo {author} {\bibfnamefont {A.}~\bibnamefont {Datta}},
  \bibinfo {author} {\bibfnamefont {S.}~\bibnamefont {Kamali}},\ and\ \bibinfo
  {author} {\bibfnamefont {D.}~\bibnamefont {London}},\ }\href
  {https://doi.org/10.1007/JHEP07(2020)194} {\bibfield  {journal} {\bibinfo
  {journal} {JHEP}\ }\textbf {\bibinfo {volume} {07}}\bibfield  {number}
  {\bibinfo  {number} { (07)},\ \bibinfo {pages} {194}},\ }\Eprint
  {https://arxiv.org/abs/2005.03032} {arXiv:2005.03032 [hep-ph]} \BibitemShut
  {NoStop}%
\bibitem [{\citenamefont {Hu}\ \emph {et~al.}(2021{\natexlab{a}})\citenamefont
  {Hu}, \citenamefont {Li}, \citenamefont {Yang},\ and\ \citenamefont
  {Zheng}}]{Hu:2020axt}%
  \BibitemOpen
  \bibfield  {author} {\bibinfo {author} {\bibfnamefont {Q.-Y.}\ \bibnamefont
  {Hu}}, \bibinfo {author} {\bibfnamefont {X.-Q.}\ \bibnamefont {Li}}, \bibinfo
  {author} {\bibfnamefont {Y.-D.}\ \bibnamefont {Yang}},\ and\ \bibinfo
  {author} {\bibfnamefont {D.-H.}\ \bibnamefont {Zheng}},\ }\href
  {https://doi.org/10.1007/JHEP02(2021)183} {\bibfield  {journal} {\bibinfo
  {journal} {JHEP}\ }\textbf {\bibinfo {volume} {02}},\ \bibinfo {pages}
  {183}},\ \Eprint {https://arxiv.org/abs/2011.05912} {arXiv:2011.05912
  [hep-ph]} \BibitemShut {NoStop}%
\bibitem [{\citenamefont {Nierste}\ \emph {et~al.}(2008)\citenamefont
  {Nierste}, \citenamefont {Trine},\ and\ \citenamefont
  {Westhoff}}]{Nierste:2008qe}%
  \BibitemOpen
  \bibfield  {author} {\bibinfo {author} {\bibfnamefont {U.}~\bibnamefont
  {Nierste}}, \bibinfo {author} {\bibfnamefont {S.}~\bibnamefont {Trine}},\
  and\ \bibinfo {author} {\bibfnamefont {S.}~\bibnamefont {Westhoff}},\ }\href
  {https://doi.org/10.1103/PhysRevD.78.015006} {\bibfield  {journal} {\bibinfo
  {journal} {Phys. Rev. D}\ }\textbf {\bibinfo {volume} {78}},\ \bibinfo
  {pages} {015006} (\bibinfo {year} {2008})},\ \Eprint
  {https://arxiv.org/abs/0801.4938} {arXiv:0801.4938 [hep-ph]} \BibitemShut
  {NoStop}%
\bibitem [{\citenamefont {Tanaka}\ and\ \citenamefont
  {Watanabe}(2010)}]{Tanaka:2010se}%
  \BibitemOpen
  \bibfield  {author} {\bibinfo {author} {\bibfnamefont {M.}~\bibnamefont
  {Tanaka}}\ and\ \bibinfo {author} {\bibfnamefont {R.}~\bibnamefont
  {Watanabe}},\ }\href {https://doi.org/10.1103/PhysRevD.82.034027} {\bibfield
  {journal} {\bibinfo  {journal} {Phys. Rev. D}\ }\textbf {\bibinfo {volume}
  {82}},\ \bibinfo {pages} {034027} (\bibinfo {year} {2010})},\ \Eprint
  {https://arxiv.org/abs/1005.4306} {arXiv:1005.4306 [hep-ph]} \BibitemShut
  {NoStop}%
\bibitem [{\citenamefont {Alonso}\ \emph {et~al.}(2017)\citenamefont {Alonso},
  \citenamefont {Martin~Camalich},\ and\ \citenamefont
  {Westhoff}}]{Alonso:2017ktd}%
  \BibitemOpen
  \bibfield  {author} {\bibinfo {author} {\bibfnamefont {R.}~\bibnamefont
  {Alonso}}, \bibinfo {author} {\bibfnamefont {J.}~\bibnamefont
  {Martin~Camalich}},\ and\ \bibinfo {author} {\bibfnamefont {S.}~\bibnamefont
  {Westhoff}},\ }\href {https://doi.org/10.1103/PhysRevD.95.093006} {\bibfield
  {journal} {\bibinfo  {journal} {Phys. Rev. D}\ }\textbf {\bibinfo {volume}
  {95}},\ \bibinfo {pages} {093006} (\bibinfo {year} {2017})},\ \Eprint
  {https://arxiv.org/abs/1702.02773} {arXiv:1702.02773 [hep-ph]} \BibitemShut
  {NoStop}%
\bibitem [{\citenamefont {Asadi}\ \emph {et~al.}(2020)\citenamefont {Asadi},
  \citenamefont {Hallin}, \citenamefont {Martin~Camalich}, \citenamefont
  {Shih},\ and\ \citenamefont {Westhoff}}]{Asadi:2020fdo}%
  \BibitemOpen
  \bibfield  {author} {\bibinfo {author} {\bibfnamefont {P.}~\bibnamefont
  {Asadi}}, \bibinfo {author} {\bibfnamefont {A.}~\bibnamefont {Hallin}},
  \bibinfo {author} {\bibfnamefont {J.}~\bibnamefont {Martin~Camalich}},
  \bibinfo {author} {\bibfnamefont {D.}~\bibnamefont {Shih}},\ and\ \bibinfo
  {author} {\bibfnamefont {S.}~\bibnamefont {Westhoff}},\ }\href
  {https://doi.org/10.1103/PhysRevD.102.095028} {\bibfield  {journal} {\bibinfo
   {journal} {Phys. Rev. D}\ }\textbf {\bibinfo {volume} {102}},\ \bibinfo
  {pages} {095028} (\bibinfo {year} {2020})},\ \Eprint
  {https://arxiv.org/abs/2006.16416} {arXiv:2006.16416 [hep-ph]} \BibitemShut
  {NoStop}%
\bibitem [{\citenamefont {Penalva}\ \emph
  {et~al.}(2021{\natexlab{a}})\citenamefont {Penalva}, \citenamefont
  {Hern\'andez},\ and\ \citenamefont {Nieves}}]{Penalva:2021gef}%
  \BibitemOpen
  \bibfield  {author} {\bibinfo {author} {\bibfnamefont {N.}~\bibnamefont
  {Penalva}}, \bibinfo {author} {\bibfnamefont {E.}~\bibnamefont
  {Hern\'andez}},\ and\ \bibinfo {author} {\bibfnamefont {J.}~\bibnamefont
  {Nieves}},\ }\href {https://doi.org/10.1007/JHEP06(2021)118} {\bibfield
  {journal} {\bibinfo  {journal} {JHEP}\ }\textbf {\bibinfo {volume} {06}},\
  \bibinfo {pages} {118}},\ \Eprint {https://arxiv.org/abs/2103.01857}
  {arXiv:2103.01857 [hep-ph]} \BibitemShut {NoStop}%
\bibitem [{\citenamefont {Hu}\ \emph {et~al.}(2021{\natexlab{b}})\citenamefont
  {Hu}, \citenamefont {Li}, \citenamefont {Mu}, \citenamefont {Yang},\ and\
  \citenamefont {Zheng}}]{Hu:2021emb}%
  \BibitemOpen
  \bibfield  {author} {\bibinfo {author} {\bibfnamefont {Q.-Y.}\ \bibnamefont
  {Hu}}, \bibinfo {author} {\bibfnamefont {X.-Q.}\ \bibnamefont {Li}}, \bibinfo
  {author} {\bibfnamefont {X.-L.}\ \bibnamefont {Mu}}, \bibinfo {author}
  {\bibfnamefont {Y.-D.}\ \bibnamefont {Yang}},\ and\ \bibinfo {author}
  {\bibfnamefont {D.-H.}\ \bibnamefont {Zheng}},\ }\href
  {https://doi.org/10.1007/JHEP06(2021)075} {\bibfield  {journal} {\bibinfo
  {journal} {JHEP}\ }\textbf {\bibinfo {volume} {06}},\ \bibinfo {pages}
  {075}},\ \Eprint {https://arxiv.org/abs/2104.04942} {arXiv:2104.04942
  [hep-ph]} \BibitemShut {NoStop}%
\bibitem [{\citenamefont {Penalva}\ \emph
  {et~al.}(2021{\natexlab{b}})\citenamefont {Penalva}, \citenamefont
  {Hern\'andez},\ and\ \citenamefont {Nieves}}]{Penalva:2021wye}%
  \BibitemOpen
  \bibfield  {author} {\bibinfo {author} {\bibfnamefont {N.}~\bibnamefont
  {Penalva}}, \bibinfo {author} {\bibfnamefont {E.}~\bibnamefont
  {Hern\'andez}},\ and\ \bibinfo {author} {\bibfnamefont {J.}~\bibnamefont
  {Nieves}},\ }\href {https://doi.org/10.1007/JHEP10(2021)122} {\bibfield
  {journal} {\bibinfo  {journal} {JHEP}\ }\textbf {\bibinfo {volume} {10}},\
  \bibinfo {pages} {122}},\ \Eprint {https://arxiv.org/abs/2107.13406}
  {arXiv:2107.13406 [hep-ph]} \BibitemShut {NoStop}%
\bibitem [{\citenamefont {Penalva}\ \emph {et~al.}(2022)\citenamefont
  {Penalva}, \citenamefont {Penalva}, \citenamefont {Hern\'andez},
  \citenamefont {Hern\'andez}, \citenamefont {Nieves},\ and\ \citenamefont
  {Nieves}}]{Penalva:2022vxy}%
  \BibitemOpen
  \bibfield  {author} {\bibinfo {author} {\bibfnamefont {N.}~\bibnamefont
  {Penalva}}, \bibinfo {author} {\bibfnamefont {N.}~\bibnamefont {Penalva}},
  \bibinfo {author} {\bibfnamefont {E.}~\bibnamefont {Hern\'andez}}, \bibinfo
  {author} {\bibfnamefont {E.}~\bibnamefont {Hern\'andez}}, \bibinfo {author}
  {\bibfnamefont {J.}~\bibnamefont {Nieves}},\ and\ \bibinfo {author}
  {\bibfnamefont {J.}~\bibnamefont {Nieves}},\ }\href
  {https://doi.org/10.1007/JHEP04(2022)026} {\bibfield  {journal} {\bibinfo
  {journal} {JHEP}\ }\textbf {\bibinfo {volume} {04}},\ \bibinfo {pages}
  {026}},\ \bibinfo {note} {[Erratum: JHEP 03, 011 (2023)]},\ \Eprint
  {https://arxiv.org/abs/2201.05537} {arXiv:2201.05537 [hep-ph]} \BibitemShut
  {NoStop}%
\bibitem [{\citenamefont {Li}\ \emph {et~al.}(2023)\citenamefont {Li},
  \citenamefont {Xu}, \citenamefont {Yang},\ and\ \citenamefont
  {Zheng}}]{Li:2023gev}%
  \BibitemOpen
  \bibfield  {author} {\bibinfo {author} {\bibfnamefont {X.-Q.}\ \bibnamefont
  {Li}}, \bibinfo {author} {\bibfnamefont {X.}~\bibnamefont {Xu}}, \bibinfo
  {author} {\bibfnamefont {Y.-D.}\ \bibnamefont {Yang}},\ and\ \bibinfo
  {author} {\bibfnamefont {D.-H.}\ \bibnamefont {Zheng}},\ }\href
  {https://doi.org/10.1007/JHEP05(2023)173} {\bibfield  {journal} {\bibinfo
  {journal} {JHEP}\ }\textbf {\bibinfo {volume} {05}},\ \bibinfo {pages}
  {173}},\ \Eprint {https://arxiv.org/abs/2302.13743} {arXiv:2302.13743
  [hep-ph]} \BibitemShut {NoStop}%
\bibitem [{\citenamefont {Hern\'andez}\ \emph {et~al.}(2022)\citenamefont
  {Hern\'andez}, \citenamefont {Nieves}, \citenamefont {S\'anchez},\ and\
  \citenamefont {Sobczyk}}]{Hernandez:2022nmp}%
  \BibitemOpen
  \bibfield  {author} {\bibinfo {author} {\bibfnamefont {E.}~\bibnamefont
  {Hern\'andez}}, \bibinfo {author} {\bibfnamefont {J.}~\bibnamefont {Nieves}},
  \bibinfo {author} {\bibfnamefont {F.}~\bibnamefont {S\'anchez}},\ and\
  \bibinfo {author} {\bibfnamefont {J.~E.}\ \bibnamefont {Sobczyk}},\ }\href
  {https://doi.org/10.1016/j.physletb.2022.137046} {\bibfield  {journal}
  {\bibinfo  {journal} {Phys. Lett. B}\ }\textbf {\bibinfo {volume} {829}},\
  \bibinfo {pages} {137046} (\bibinfo {year} {2022})},\ \Eprint
  {https://arxiv.org/abs/2202.07539} {arXiv:2202.07539 [hep-ph]} \BibitemShut
  {NoStop}%
\bibitem [{\citenamefont {Isaacson}\ \emph {et~al.}(2023)\citenamefont
  {Isaacson}, \citenamefont {H\"oche}, \citenamefont {Siegert},\ and\
  \citenamefont {Wang}}]{Isaacson:2023gwp}%
  \BibitemOpen
  \bibfield  {author} {\bibinfo {author} {\bibfnamefont {J.}~\bibnamefont
  {Isaacson}}, \bibinfo {author} {\bibfnamefont {S.}~\bibnamefont {H\"oche}},
  \bibinfo {author} {\bibfnamefont {F.}~\bibnamefont {Siegert}},\ and\ \bibinfo
  {author} {\bibfnamefont {S.}~\bibnamefont {Wang}},\ }\href@noop {} {\
  (\bibinfo {year} {2023})},\ \Eprint {https://arxiv.org/abs/2303.08104}
  {arXiv:2303.08104 [hep-ph]} \BibitemShut {NoStop}%
\bibitem [{\citenamefont {Gonderinger}\ and\ \citenamefont
  {Ramsey-Musolf}(2010)}]{Gonderinger:2010yn}%
  \BibitemOpen
  \bibfield  {author} {\bibinfo {author} {\bibfnamefont {M.}~\bibnamefont
  {Gonderinger}}\ and\ \bibinfo {author} {\bibfnamefont {M.~J.}\ \bibnamefont
  {Ramsey-Musolf}},\ }\href {https://doi.org/10.1007/JHEP11(2010)045}
  {\bibfield  {journal} {\bibinfo  {journal} {JHEP}\ }\textbf {\bibinfo
  {volume} {11}},\ \bibinfo {pages} {045}},\ \bibinfo {note} {[Erratum: JHEP
  05, 047 (2012)]},\ \Eprint {https://arxiv.org/abs/1006.5063} {arXiv:1006.5063
  [hep-ph]} \BibitemShut {NoStop}%
\bibitem [{\citenamefont {Cirigliano}\ \emph {et~al.}(2021)\citenamefont
  {Cirigliano}, \citenamefont {Fuyuto}, \citenamefont {Lee}, \citenamefont
  {Mereghetti},\ and\ \citenamefont {Yan}}]{Cirigliano:2021img}%
  \BibitemOpen
  \bibfield  {author} {\bibinfo {author} {\bibfnamefont {V.}~\bibnamefont
  {Cirigliano}}, \bibinfo {author} {\bibfnamefont {K.}~\bibnamefont {Fuyuto}},
  \bibinfo {author} {\bibfnamefont {C.}~\bibnamefont {Lee}}, \bibinfo {author}
  {\bibfnamefont {E.}~\bibnamefont {Mereghetti}},\ and\ \bibinfo {author}
  {\bibfnamefont {B.}~\bibnamefont {Yan}},\ }\href
  {https://doi.org/10.1007/JHEP03(2021)256} {\bibfield  {journal} {\bibinfo
  {journal} {JHEP}\ }\textbf {\bibinfo {volume} {03}},\ \bibinfo {pages}
  {256}},\ \Eprint {https://arxiv.org/abs/2102.06176} {arXiv:2102.06176
  [hep-ph]} \BibitemShut {NoStop}%
\bibitem [{\citenamefont {Sobczyk}\ \emph {et~al.}(2019)\citenamefont
  {Sobczyk}, \citenamefont {Rocco}, \citenamefont {Lovato},\ and\ \citenamefont
  {Nieves}}]{Sobczyk:2019uej}%
  \BibitemOpen
  \bibfield  {author} {\bibinfo {author} {\bibfnamefont {J.~E.}\ \bibnamefont
  {Sobczyk}}, \bibinfo {author} {\bibfnamefont {N.}~\bibnamefont {Rocco}},
  \bibinfo {author} {\bibfnamefont {A.}~\bibnamefont {Lovato}},\ and\ \bibinfo
  {author} {\bibfnamefont {J.}~\bibnamefont {Nieves}},\ }\href
  {https://doi.org/10.1103/PhysRevC.99.065503} {\bibfield  {journal} {\bibinfo
  {journal} {Phys. Rev. C}\ }\textbf {\bibinfo {volume} {99}},\ \bibinfo
  {pages} {065503} (\bibinfo {year} {2019})},\ \Eprint
  {https://arxiv.org/abs/1901.10192} {arXiv:1901.10192 [nucl-th]} \BibitemShut
  {NoStop}%
\bibitem [{\citenamefont {Lai}\ \emph {et~al.}(2022{\natexlab{b}})\citenamefont
  {Lai}, \citenamefont {Li}, \citenamefont {Yan},\ and\ \citenamefont
  {Yang}}]{Lai:2022ekw}%
  \BibitemOpen
  \bibfield  {author} {\bibinfo {author} {\bibfnamefont {L.-F.}\ \bibnamefont
  {Lai}}, \bibinfo {author} {\bibfnamefont {X.-Q.}\ \bibnamefont {Li}},
  \bibinfo {author} {\bibfnamefont {X.-S.}\ \bibnamefont {Yan}},\ and\ \bibinfo
  {author} {\bibfnamefont {Y.-D.}\ \bibnamefont {Yang}},\ }\href
  {https://doi.org/10.1103/PhysRevD.105.115025} {\bibfield  {journal} {\bibinfo
   {journal} {Phys. Rev. D}\ }\textbf {\bibinfo {volume} {105}},\ \bibinfo
  {pages} {115025} (\bibinfo {year} {2022}{\natexlab{b}})},\ \Eprint
  {https://arxiv.org/abs/2203.17104} {arXiv:2203.17104 [hep-ph]} \BibitemShut
  {NoStop}%
\bibitem [{\citenamefont {Feldmann}\ and\ \citenamefont
  {Yip}(2012)}]{Feldmann:2011xf}%
  \BibitemOpen
  \bibfield  {author} {\bibinfo {author} {\bibfnamefont {T.}~\bibnamefont
  {Feldmann}}\ and\ \bibinfo {author} {\bibfnamefont {M.~W.~Y.}\ \bibnamefont
  {Yip}},\ }\href {https://doi.org/10.1103/PhysRevD.85.014035,
  10.1103/physrevd.86.079901} {\bibfield  {journal} {\bibinfo  {journal} {Phys.
  Rev.}\ }\textbf {\bibinfo {volume} {D85}},\ \bibinfo {pages} {014035}
  (\bibinfo {year} {2012})},\ \bibinfo {note} {[Erratum: Phys.
  Rev.D86,079901(2012)]},\ \Eprint {https://arxiv.org/abs/1111.1844}
  {arXiv:1111.1844 [hep-ph]} \BibitemShut {NoStop}%
\bibitem [{\citenamefont {Meinel}(2018)}]{Meinel:2017ggx}%
  \BibitemOpen
  \bibfield  {author} {\bibinfo {author} {\bibfnamefont {S.}~\bibnamefont
  {Meinel}},\ }\href {https://doi.org/10.1103/PhysRevD.97.034511} {\bibfield
  {journal} {\bibinfo  {journal} {Phys. Rev.}\ }\textbf {\bibinfo {volume}
  {D97}},\ \bibinfo {pages} {034511} (\bibinfo {year} {2018})},\ \Eprint
  {https://arxiv.org/abs/1712.05783} {arXiv:1712.05783 [hep-lat]} \BibitemShut
  {NoStop}%
\bibitem [{\citenamefont {Das}(2018)}]{Das:2018sms}%
  \BibitemOpen
  \bibfield  {author} {\bibinfo {author} {\bibfnamefont {D.}~\bibnamefont
  {Das}},\ }\href {https://doi.org/10.1140/epjc/s10052-018-5731-2} {\bibfield
  {journal} {\bibinfo  {journal} {Eur. Phys. J. C}\ }\textbf {\bibinfo {volume}
  {78}},\ \bibinfo {pages} {230} (\bibinfo {year} {2018})},\ \Eprint
  {https://arxiv.org/abs/1802.09404} {arXiv:1802.09404 [hep-ph]} \BibitemShut
  {NoStop}%
\bibitem [{\citenamefont {Chodos}\ \emph
  {et~al.}(1974{\natexlab{a}})\citenamefont {Chodos}, \citenamefont {Jaffe},
  \citenamefont {Johnson}, \citenamefont {Thorn},\ and\ \citenamefont
  {Weisskopf}}]{Chodos:1974je}%
  \BibitemOpen
  \bibfield  {author} {\bibinfo {author} {\bibfnamefont {A.}~\bibnamefont
  {Chodos}}, \bibinfo {author} {\bibfnamefont {R.~L.}\ \bibnamefont {Jaffe}},
  \bibinfo {author} {\bibfnamefont {K.}~\bibnamefont {Johnson}}, \bibinfo
  {author} {\bibfnamefont {C.~B.}\ \bibnamefont {Thorn}},\ and\ \bibinfo
  {author} {\bibfnamefont {V.~F.}\ \bibnamefont {Weisskopf}},\ }\href
  {https://doi.org/10.1103/PhysRevD.9.3471} {\bibfield  {journal} {\bibinfo
  {journal} {Phys. Rev. D}\ }\textbf {\bibinfo {volume} {9}},\ \bibinfo {pages}
  {3471} (\bibinfo {year} {1974}{\natexlab{a}})}\BibitemShut {NoStop}%
\bibitem [{\citenamefont {Chodos}\ \emph
  {et~al.}(1974{\natexlab{b}})\citenamefont {Chodos}, \citenamefont {Jaffe},
  \citenamefont {Johnson},\ and\ \citenamefont {Thorn}}]{Chodos:1974pn}%
  \BibitemOpen
  \bibfield  {author} {\bibinfo {author} {\bibfnamefont {A.}~\bibnamefont
  {Chodos}}, \bibinfo {author} {\bibfnamefont {R.~L.}\ \bibnamefont {Jaffe}},
  \bibinfo {author} {\bibfnamefont {K.}~\bibnamefont {Johnson}},\ and\ \bibinfo
  {author} {\bibfnamefont {C.~B.}\ \bibnamefont {Thorn}},\ }\href
  {https://doi.org/10.1103/PhysRevD.10.2599} {\bibfield  {journal} {\bibinfo
  {journal} {Phys. Rev. D}\ }\textbf {\bibinfo {volume} {10}},\ \bibinfo
  {pages} {2599} (\bibinfo {year} {1974}{\natexlab{b}})}\BibitemShut {NoStop}%
\bibitem [{\citenamefont {Kokkedee}(1969)}]{Kokkedee1969}%
  \BibitemOpen
  \bibfield  {author} {\bibinfo {author} {\bibfnamefont {J.~J.~J.}\
  \bibnamefont {Kokkedee}},\ }\href@noop {} {\emph {\bibinfo {title} {{The
  quark model}}}}\ (\bibinfo  {publisher} {W. A. Benjamin},\ \bibinfo {year}
  {1969})\BibitemShut {NoStop}%
\bibitem [{\citenamefont {Ivanov}\ \emph {et~al.}(1997)\citenamefont {Ivanov},
  \citenamefont {Lyubovitskij}, \citenamefont {Korner},\ and\ \citenamefont
  {Kroll}}]{Ivanov:1996fj}%
  \BibitemOpen
  \bibfield  {author} {\bibinfo {author} {\bibfnamefont {M.~A.}\ \bibnamefont
  {Ivanov}}, \bibinfo {author} {\bibfnamefont {V.~E.}\ \bibnamefont
  {Lyubovitskij}}, \bibinfo {author} {\bibfnamefont {J.}~\bibnamefont
  {Korner}},\ and\ \bibinfo {author} {\bibfnamefont {P.}~\bibnamefont
  {Kroll}},\ }\href {https://doi.org/10.1103/PhysRevD.56.348} {\bibfield
  {journal} {\bibinfo  {journal} {Phys. Rev. D}\ }\textbf {\bibinfo {volume}
  {56}},\ \bibinfo {pages} {348} (\bibinfo {year} {1997})},\ \Eprint
  {https://arxiv.org/abs/hep-ph/9612463} {arXiv:hep-ph/9612463} \BibitemShut
  {NoStop}%
\bibitem [{\citenamefont {Branz}\ \emph {et~al.}(2010)\citenamefont {Branz},
  \citenamefont {Faessler}, \citenamefont {Gutsche}, \citenamefont {Ivanov},
  \citenamefont {Korner},\ and\ \citenamefont {Lyubovitskij}}]{Branz:2009cd}%
  \BibitemOpen
  \bibfield  {author} {\bibinfo {author} {\bibfnamefont {T.}~\bibnamefont
  {Branz}}, \bibinfo {author} {\bibfnamefont {A.}~\bibnamefont {Faessler}},
  \bibinfo {author} {\bibfnamefont {T.}~\bibnamefont {Gutsche}}, \bibinfo
  {author} {\bibfnamefont {M.~A.}\ \bibnamefont {Ivanov}}, \bibinfo {author}
  {\bibfnamefont {J.~G.}\ \bibnamefont {Korner}},\ and\ \bibinfo {author}
  {\bibfnamefont {V.~E.}\ \bibnamefont {Lyubovitskij}},\ }\href
  {https://doi.org/10.1103/PhysRevD.81.034010} {\bibfield  {journal} {\bibinfo
  {journal} {Phys. Rev. D}\ }\textbf {\bibinfo {volume} {81}},\ \bibinfo
  {pages} {034010} (\bibinfo {year} {2010})},\ \Eprint
  {https://arxiv.org/abs/0912.3710} {arXiv:0912.3710 [hep-ph]} \BibitemShut
  {NoStop}%
\bibitem [{\citenamefont {Bourrely}\ \emph {et~al.}(2009)\citenamefont
  {Bourrely}, \citenamefont {Caprini},\ and\ \citenamefont
  {Lellouch}}]{Bourrely:2008za}%
  \BibitemOpen
  \bibfield  {author} {\bibinfo {author} {\bibfnamefont {C.}~\bibnamefont
  {Bourrely}}, \bibinfo {author} {\bibfnamefont {I.}~\bibnamefont {Caprini}},\
  and\ \bibinfo {author} {\bibfnamefont {L.}~\bibnamefont {Lellouch}},\ }\href
  {https://doi.org/10.1103/PhysRevD.82.099902} {\bibfield  {journal} {\bibinfo
  {journal} {Phys. Rev. D}\ }\textbf {\bibinfo {volume} {79}},\ \bibinfo
  {pages} {013008} (\bibinfo {year} {2009})},\ \bibinfo {note} {[Erratum:
  Phys.Rev.D 82, 099902 (2010)]},\ \Eprint {https://arxiv.org/abs/0807.2722}
  {arXiv:0807.2722 [hep-ph]} \BibitemShut {NoStop}%
\bibitem [{\citenamefont {De~Lellis}\ \emph {et~al.}(2004)\citenamefont
  {De~Lellis}, \citenamefont {Migliozzi},\ and\ \citenamefont
  {Santorelli}}]{DeLellis:2004ovi}%
  \BibitemOpen
  \bibfield  {author} {\bibinfo {author} {\bibfnamefont {G.}~\bibnamefont
  {De~Lellis}}, \bibinfo {author} {\bibfnamefont {P.}~\bibnamefont
  {Migliozzi}},\ and\ \bibinfo {author} {\bibfnamefont {P.}~\bibnamefont
  {Santorelli}},\ }\href {https://doi.org/10.1016/j.physrep.2005.02.001}
  {\bibfield  {journal} {\bibinfo  {journal} {Phys. Rept.}\ }\textbf {\bibinfo
  {volume} {399}},\ \bibinfo {pages} {227} (\bibinfo {year} {2004})},\ \bibinfo
  {note} {[Erratum: Phys.Rept. 411, 323--324 (2005)]}\BibitemShut {NoStop}%
\bibitem [{\citenamefont {Petric}\ \emph {et~al.}(2010)\citenamefont {Petric}
  \emph {et~al.}}]{Belle:2010ouj}%
  \BibitemOpen
  \bibfield  {author} {\bibinfo {author} {\bibfnamefont {M.}~\bibnamefont
  {Petric}} \emph {et~al.} (\bibinfo {collaboration} {Belle}),\ }\href
  {https://doi.org/10.1103/PhysRevD.81.091102} {\bibfield  {journal} {\bibinfo
  {journal} {Phys. Rev. D}\ }\textbf {\bibinfo {volume} {81}},\ \bibinfo
  {pages} {091102} (\bibinfo {year} {2010})},\ \Eprint
  {https://arxiv.org/abs/1003.2345} {arXiv:1003.2345 [hep-ex]} \BibitemShut
  {NoStop}%
\bibitem [{\citenamefont {Perez-Marcial}\ \emph {et~al.}(1989)\citenamefont
  {Perez-Marcial}, \citenamefont {Huerta}, \citenamefont {Garcia},\ and\
  \citenamefont {Avila-Aoki}}]{Perez-Marcial:1989sch}%
  \BibitemOpen
  \bibfield  {author} {\bibinfo {author} {\bibfnamefont {R.}~\bibnamefont
  {Perez-Marcial}}, \bibinfo {author} {\bibfnamefont {R.}~\bibnamefont
  {Huerta}}, \bibinfo {author} {\bibfnamefont {A.}~\bibnamefont {Garcia}},\
  and\ \bibinfo {author} {\bibfnamefont {M.}~\bibnamefont {Avila-Aoki}},\
  }\href {https://doi.org/10.1103/PhysRevD.44.2203} {\bibfield  {journal}
  {\bibinfo  {journal} {Phys. Rev. D}\ }\textbf {\bibinfo {volume} {40}},\
  \bibinfo {pages} {2955} (\bibinfo {year} {1989})},\ \bibinfo {note}
  {[Erratum: Phys.Rev.D 44, 2203 (1991)]}\BibitemShut {NoStop}%
\bibitem [{\citenamefont {Avila-Aoki}\ \emph {et~al.}(1989)\citenamefont
  {Avila-Aoki}, \citenamefont {Garcia}, \citenamefont {Huerta},\ and\
  \citenamefont {Perez-Marcial}}]{Avila-Aoki:1989arc}%
  \BibitemOpen
  \bibfield  {author} {\bibinfo {author} {\bibfnamefont {M.}~\bibnamefont
  {Avila-Aoki}}, \bibinfo {author} {\bibfnamefont {A.}~\bibnamefont {Garcia}},
  \bibinfo {author} {\bibfnamefont {R.}~\bibnamefont {Huerta}},\ and\ \bibinfo
  {author} {\bibfnamefont {R.}~\bibnamefont {Perez-Marcial}},\ }\href
  {https://doi.org/10.1103/PhysRevD.40.2944} {\bibfield  {journal} {\bibinfo
  {journal} {Phys. Rev. D}\ }\textbf {\bibinfo {volume} {40}},\ \bibinfo
  {pages} {2944} (\bibinfo {year} {1989})}\BibitemShut {NoStop}%
\bibitem [{\citenamefont {Arrington}\ \emph {et~al.}(2023)\citenamefont
  {Arrington} \emph {et~al.}}]{JeffersonLabSoLID:2022iod}%
  \BibitemOpen
  \bibfield  {author} {\bibinfo {author} {\bibfnamefont {J.}~\bibnamefont
  {Arrington}} \emph {et~al.} (\bibinfo {collaboration} {Jefferson Lab
  SoLID}),\ }\href {https://doi.org/10.1088/1361-6471/acda21} {\bibfield
  {journal} {\bibinfo  {journal} {J. Phys. G}\ }\textbf {\bibinfo {volume}
  {50}},\ \bibinfo {pages} {110501} (\bibinfo {year} {2023})},\ \Eprint
  {https://arxiv.org/abs/2209.13357} {arXiv:2209.13357 [nucl-ex]} \BibitemShut
  {NoStop}%
\bibitem [{\citenamefont {Hafidi}\ \emph {et~al.}(2012)\citenamefont {Hafidi}
  \emph {et~al.}}]{K.Hafidi}%
  \BibitemOpen
  \bibfield  {author} {\bibinfo {author} {\bibfnamefont {K.}~\bibnamefont
  {Hafidi}} \emph {et~al.} (\bibinfo {collaboration} {E12-12-006}),\ }\href
  {https://www.jlab.org/exp_prog/proposals/12/PR12-12-006.pdf} {\bibinfo
  {title} {{Near Threshold Electroproduction of $J/\Psi$ at 11 GeV}}} (\bibinfo
  {year} {2012})\BibitemShut {NoStop}%
\bibitem [{\citenamefont {Liu}\ \emph {et~al.}(2023)\citenamefont {Liu},
  \citenamefont {Zhao}, \citenamefont {Cai}, \citenamefont {Byer},\ and\
  \citenamefont {Gao}}]{Liu:2023htv}%
  \BibitemOpen
  \bibfield  {author} {\bibinfo {author} {\bibfnamefont {T.}~\bibnamefont
  {Liu}}, \bibinfo {author} {\bibfnamefont {Z.~W.}\ \bibnamefont {Zhao}},
  \bibinfo {author} {\bibfnamefont {M.}~\bibnamefont {Cai}}, \bibinfo {author}
  {\bibfnamefont {D.}~\bibnamefont {Byer}},\ and\ \bibinfo {author}
  {\bibfnamefont {H.}~\bibnamefont {Gao}},\ }\href@noop {} {\  (\bibinfo {year}
  {2023})},\ \Eprint {https://arxiv.org/abs/2310.05405} {arXiv:2310.05405
  [nucl-ex]} \BibitemShut {NoStop}%
\bibitem [{\citenamefont {Bouchiat}\ and\ \citenamefont
  {Michel}(1958)}]{Bouchiat:1958yui}%
  \BibitemOpen
  \bibfield  {author} {\bibinfo {author} {\bibfnamefont {C.}~\bibnamefont
  {Bouchiat}}\ and\ \bibinfo {author} {\bibfnamefont {L.}~\bibnamefont
  {Michel}},\ }\href {https://doi.org/10.1016/0029-5582(58)90046-4} {\bibfield
  {journal} {\bibinfo  {journal} {Nucl. Phys.}\ }\textbf {\bibinfo {volume}
  {5}},\ \bibinfo {pages} {416} (\bibinfo {year} {1958})}\BibitemShut {NoStop}%
\bibitem [{\citenamefont {Michel}(1959)}]{Michel:1959dvg}%
  \BibitemOpen
  \bibfield  {author} {\bibinfo {author} {\bibfnamefont {L.}~\bibnamefont
  {Michel}},\ }\href {https://doi.org/10.1007/BF02724842} {\bibfield  {journal}
  {\bibinfo  {journal} {Nuovo Cim.}\ }\textbf {\bibinfo {volume} {14}},\
  \bibinfo {pages} {95} (\bibinfo {year} {1959})}\BibitemShut {NoStop}%
\end{thebibliography}%

\end{document}